\documentclass[letterpaper,titlepage,11pt]{article}
\usepackage{amsmath,amssymb,mathtools,bm}
\usepackage{physics}
\usepackage{cancel}
\usepackage{slashed}
\usepackage{mathrsfs}
\usepackage{color}
\usepackage[svgnames]{xcolor}
\usepackage{enumerate}
\usepackage{graphicx}
\usepackage{tikz}
\usepackage{caption}
\usepackage[symbol]{footmisc}
\usepackage{enumitem}

\usepackage{multirow}
\usepackage{booktabs}

\usetikzlibrary{arrows.meta, positioning, calc, decorations.pathmorphing, shapes.geometric}
\usepackage{tikzsymbols}
\usepackage{fontawesome5}
\usepackage{pifont}
\usetikzlibrary{shadows}

\usepackage[numbers,sort&compress]{natbib}
\allowdisplaybreaks
\usepackage[
colorlinks=true,
linkcolor=DarkBlue,
anchorcolor=Black,
urlcolor=DarkRed,
filecolor=green,
citecolor=DarkRed,
linktoc=page,
pdfstartview=FitV,
pdftitle={Stochastic Survival near Swampland Boundaries},
pdfauthor={Omer Guleryuz},
pdfsubject={},
pdfkeywords={},
bookmarksopen=true
]{hyperref}

\usepackage[nameinlink,capitalise]{cleveref}

\renewcommand{\dd}{\mathrm d}

\newcommand{\revdel}[1]{{\color{red}#1}}
\begin{document}

 \hypersetup{pageanchor=false}
 \begin{titlepage}
			%%%%%%%%%%%%%%%%%%%%%%%%%%%%%%%%%%
			
			%%%%%%%%%%%%%%%%%%%%%%%%%%%%%%%%%%
			\bigskip
			
			\begin{center}
				{\LARGE\bfseries Stochastic Survival near Swampland Boundaries}
				\\[10mm]
				\textbf{Omer Guleryuz}\\[5mm]
				\vskip 25pt

				{\em \hskip -.1truecm Department of Physics, Istanbul Technical University, \\
					Maslak 34469 Istanbul, Türkiye \vskip 5pt }

				{{\tt \href{mailto:omerguleryuz@itu.edu.tr}{omerguleryuz@itu.edu.tr}}
 
 }

			\end{center}
			
			\vspace{3ex}

			\begin{center}
				{\bfseries Abstract}
			\end{center}
\begin{quotation}
Swampland and compactification data tell us where a chosen EFT description can lose parametric control; stochastic cosmology asks which histories survive near that edge. We turn this question into a survival problem for fluctuating moduli over cosmological time scales. Given a valid stochastic generator, hard loss surfaces, soft degradation profiles, and finite horizons define a survival probability, whose negative logarithm is the survival action. The Doob transform then converts this logarithmic survival cost into the drift of the ensemble conditioned to remain on the controlled side. Near a regular hard boundary with nonzero normal diffusion, the inward normal component of the conditioned response is universal: it is fixed at leading order by the proper distance to the wall and the normal diffusion coefficient. The same finite-horizon construction also determines loss probabilities, first-exit statistics, survival hazards, and the control margin required to retain a prescribed fraction of histories. In this way, tower/species cutoffs, weak-coupling limits, string and Kaluza–Klein thresholds, and carefully qualified potential-based diagnostics acquire stochastic boundary layers without becoming microscopic forces. The inverse map tests when a conditioned drift is compatible with a scalar operational loss surface and reconstructs its boundary-normal Doob class. The construction therefore gives a stochastic survival interface between quantum-gravity control data and the cosmological histories that remain inside the prescribed EFT domain.
\end{quotation}
			
			\vfill
			
			%%%%%%%
 \begin{center}{\flushleft{\today}}\end{center}
			%%%%%%
		\end{titlepage}
		\hypersetup{pageanchor=true}
		\setcounter{page}{1}
		\tableofcontents

		%%%%%%%%%%%%%%%%%%%%%%%%%%%%%%%%%%%%%%%%%%%%%%%%%%%%%%%%%%%%%%%%%
\newpage

\section{Introduction}
\label{sec:intro}

The Swampland program~\cite{Vafa:2005ui,Palti:2019pca,Brennan:2017rbf,vanBeest:2021lhn,Agmon:2022thq,Lehnert:2025izp} organizes proposed quantum-gravity restrictions on low-energy effective field theories. It is important, however, to distinguish two logically different statements. Some criteria aim to separate EFTs that can admit a quantum-gravity completion from EFTs that cannot, whereas tower, species, weak-coupling, string, and Kaluza--Klein data often diagnose the loss of parametric control of a particular low-energy description~\cite{Ooguri:2006in,Klaewer:2016kiy,Grimm:2018ohb,Lee:2019wij,Heidenreich:2018kpg,Harlow:2015lma}. Reaching a cutoff/Hubble threshold can require additional degrees of freedom or a new EFT without implying that the underlying theory is inconsistent with quantum gravity~\cite{Dvali:2007hz,Dvali:2007wp,Dvali:2010vm,Arkani-Hamed:2006emk,Heidenreich:2015nta,Bedroya:2019snp,Bedroya:2019tba,Brandenberger:2021pzy,Cagan:2025rbc}. In this paper, the phrase ``Swampland boundary'' refers to an operational boundary of EFT control motivated by Swampland or compactification data, not to a sharp ontological frontier between the landscape and the Swampland~\cite{Andriot:2020lea,Lehnert:2025izp}. The controlled side is the domain in which the chosen stochastic EFT is assumed to be valid; crossing its boundary means only that this chosen description should no longer be extrapolated.

An important additional requirement is that the stochastic description itself remain valid up to the proposed loss surface.  A compactification-derived threshold may define a candidate boundary, but it becomes a physically admissible first-passage surface for a given stochastic EFT only if the degrees of freedom retained in that EFT, together with their drift and noise, remain controlled throughout the intervening region.  In particular, the appearance of additional light tower states can require an enlarged stochastic description before a formal species-scale surface is reached.  We therefore distinguish throughout between candidate control surfaces supplied by compactification data and loss surfaces that are actually accessible within the specified stochastic model.

In cosmology, however, the relevant degrees of freedom are not fixed points in field space. During inflation, sufficiently light coarse-grained scalar degrees of freedom, including moduli in an appropriate regime, receive stochastic kicks sourced by sub-horizon quantum fluctuations~\cite{Starobinsky:1986fx,Starobinsky:1994bd,Rigopoulos:2005xx,
Vennin:2015hra,Assadullahi:2016gkk,Guleryuz:2024vix}. Over cosmological time scales, these small kicks can accumulate, and the relevant object is no longer a single point in moduli space but an ensemble of histories. The natural question is therefore not only where the boundaries of EFT control lie, but what it means for fluctuating cosmological histories to survive near them.

Conditioned stochastic dynamics has already proved useful in inflationary settings. Doob transforms have been applied to histories conditioned on their inflationary duration, with the associated inflationary-meander construction exhibiting the half-line survival, first-passage, and conditioned-drift structures relevant to the local normal problem considered
below~\cite{Tokeshi:2023swe}.  The purpose of the present work is different: the conditioning event is tied to compactification- or EFT-control data.  Rather than conditioning a history only on how long it inflates, we ask whether a specified EFT can continue to describe that history without crossing a prescribed loss surface during the required interval.  The new physical input is therefore the field-dependent geometry of EFT loss, while the Doob transform supplies the corresponding conditioned ensemble once that input and a valid stochastic generator have been specified.

We formulate this question as a survival problem for moduli subject to cosmological stochastic fluctuations. The organizing step is to impose EFT control on histories rather than only on instantaneous field values. Once EFT control is represented by a controlled domain in field space, selected Swampland-motivated and compactification control criteria can then be encoded as survival data for histories required to remain inside that domain for a prescribed time. A hard loss of EFT control is encoded by a boundary function \(F_A(\phi)=0\), with \(F_A>0\) on the controlled side. A gradual loss is encoded by a killing profile \(\kappa_A(\phi)\), and a finite-duration restriction by a horizon \(\tau\). Together with a stochastic generator \((b^i,D^{ij})\), these data determine a survival probability \(h(\phi,\tau)\).

These ingredients belong to two distinct levels of modeling.  Compactification physics specifies candidate control data such as \(M_{\rm KK}(\phi)\), \(M_s(\phi)\), or a species scale, together with their threshold conventions.  The stochastic model separately specifies which fields are retained, the background evolution, the diffusion tensor, and the domain over which the corresponding Markovian coarse-grained description is valid.  Quantitative first-passage probabilities are meaningful only after both sets of assumptions have been fixed.

The central object is the survival action
\begin{equation}
\mathcal S_{\rm surv}(\phi,\tau) \equiv -\ln h(\phi,\tau),
\label{eq:intro_survival_action}
\end{equation}
whose gradient gives the Doob-transformed drift of the conditioned ensemble~\cite{Doob1957,Risken1996,Redner2001,Collet:2013,ChetriteTouchette:2014,Chetrite:2015},
\begin{equation}
\Delta b^i_{\rm Doob} = -2D^{ij}\nabla_j\mathcal S_{\rm surv}.
\label{eq:intro_doob_drift}
\end{equation}
Thus the survival action is not a scalar potential and does not introduce a new fundamental force term. It is the logarithmic cost of remaining in the EFT-controlled ensemble; its gradient is the statistical drift induced by conditioning on survival~\cite{Touchette:2009,ChetriteTouchette:2014}.

The leading local consequence has a universal normal component. Near a regular absorbing boundary, let \(s\) be the inward proper distance, \(n_i=\nabla_i s\) the inward unit normal covector, and \(D_\perp=n_iD^{ij}n_j\) the normal diffusion coefficient. At fixed positive remaining time, regularity and nonvanishing normal diffusion imply that the survival probability vanishes linearly at the absorbing wall, \(h\sim s\). Hence
\begin{equation}
 \Delta b_{\perp}^{\rm Doob} \equiv n_i\Delta b_{\rm Doob}^{i} \simeq \frac{2D_\perp}{s}.
 \label{eq:intro_wall_law}
\end{equation}
The full leading singular vector is
\begin{equation}
 \Delta b_{\rm Doob}^{i} \simeq \frac{2D^{ij}n_j}{s}.
 \label{eq:intro_wall_vector_law}
\end{equation}
Therefore, anisotropic or mixed diffusion can generate singular components that are not aligned with the metric normal, while the invariant normal projection remains \(2D_\perp/s\).

Equivalently, for a regular boundary function \(F(\phi)>0\),
\begin{equation}
 \Delta b^i_{\rm Doob} \simeq 2D^{ij} \frac{\nabla_jF}{F}.
 \label{eq:intro_boundary_function_law}
\end{equation}
This is the local bridge from operational EFT-validity data to stochastic response: compactification data supply \(F\), \(\kappa_A\), or \(\tau\), while the survival problem supplies the conditioned drift.

The universal \(1/s\) behavior is a local statement and does not by itself establish the accuracy of a finite-duration calculation with a stationary boundary.  Global survival probabilities depend on the motion and softness of the loss surface, the retained degrees of freedom, the stochastic correlation time, and the complete drift and diffusion data.  These dependencies are kept explicit below.

\begin{figure}[!htb]
\centering
\includegraphics[width=0.89\textwidth]{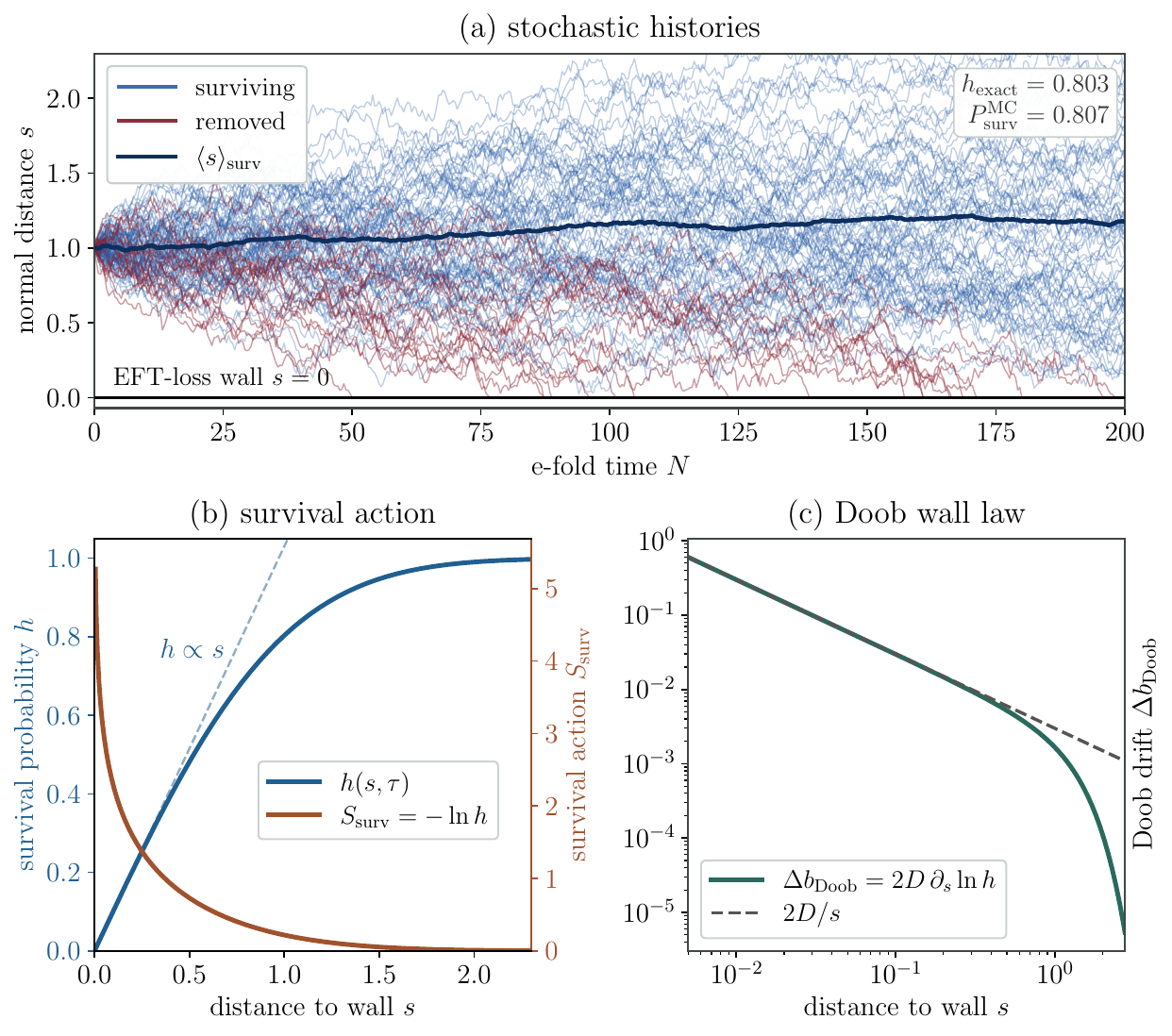}
\caption{Local hard-wall survival problem for the driftless half-line model \(\mathcal L=D\partial_s^2\) with absorbing wall \(s=0\). Panel (a) shows a Brownian-bridge-corrected Monte Carlo ensemble of removed and surviving histories; the quoted survival fraction is compared with the exact finite-horizon value \(h(s_0,\tau=N)=\operatorname{erf}[s_0/(2\sqrt{DN})]\), where \(N\) is the displayed remaining horizon. Panel (b) shows the analytic survival probability \(h(s,\tau)=\operatorname{erf}[s/(2\sqrt{D\tau})]\) and the survival action \(\mathcal S_{\rm surv}=-\ln h\). Panel (c) compares the exact conditioned drift \(2D\,\partial_s\ln h\) with its near-wall limit \(2D/s\).}
\label{fig:survivalEnsembleWallLaw}
\end{figure}

An explicitly solvable normal-coordinate example of the local hard-wall problem is shown in \cref{fig:survivalEnsembleWallLaw}.

The microscopic control data and the stochastic dynamics play distinct roles. A boundary function, killing profile, or finite horizon specifies what counts as loss of control, while the stochastic generator specifies how the retained degrees of freedom fluctuate before that loss occurs. Neither input determines the other. The construction developed below is therefore conditional on the pair
\begin{equation}
 \bigl\{F_A,\,\kappa_A,\,\tau\bigr\} \quad\text{and}\quad \mathcal L =b^i\nabla_i+D^{ij}\nabla_i\nabla_j,
 \label{eq:input_separation_revision}
\end{equation}
and should not be read as a derivation of stochastic dynamics directly from a Swampland conjecture.

Once these inputs are specified, the formalism answers finite-time physical questions rather than only assigning a local drift. In particular, for autonomous hard-wall loss without soft killing, it gives
\begin{equation}
 P_{\rm surv}(\phi,\tau)=h(\phi,\tau), \qquad
 P_{\rm loss}(\phi,\tau)=1-h(\phi,\tau), \qquad
 f_{\rm exit}(\phi,\tau)=-\partial_\tau h(\phi,\tau),
 \label{eq:physical_outputs_intro}
\end{equation}
together with the drift and distribution of histories conditioned on survival. Thus, the operational question is: for a prescribed EFT-control domain and a justified stochastic generator, what fraction of histories remain controlled for the required duration, when do histories that fail first exit the controlled domain, and how is the surviving ensemble statistically biased?

With soft killing, \(-\partial_\tau h\) instead measures total loss, including both boundary exits and killing in the interior.

In string compactifications, the stochastic variables are macroscopic moduli. Tower spectra, species counting, direct string and Kaluza--Klein thresholds, weak-coupling cutoffs, local Hubble profiles, potential-based diagnostics, and finite-duration inputs become survival data once they are written as operational loss conditions.

Not all such formal surfaces are simultaneously accessible within a given reduced EFT.  For example, a tower may become lighter than the Hubble scale before the corresponding species scale reaches the Hubble scale.  In that case, a stochastic model that retains only the original moduli cannot be continued unchanged to the species surface.  The ordering of thresholds and the associated retained field content must therefore be checked before a compactification surface is used in a quantitative first-passage calculation.

A typical hard-wall input is a cutoff/Hubble ratio,
\begin{equation}
F_{\rm QG/H}^{(q_{\rm ctrl})}(\phi)
=
\ln\frac{\Lambda_{\rm QG}(\phi)}
{q_{\rm ctrl}H(\phi)},
\qquad q_{\rm ctrl}\geq 1,
\label{eq:intro_general_cutoff_ratio}
\end{equation}
where \(\Lambda_{\rm QG}\) may be a species scale, string scale, KK scale, or another channel-dependent cutoff. The choice \(q_{\rm ctrl}>1\) places the operational wall on the controlled side of the nominal cutoff/Hubble crossing, while \(q_{\rm ctrl}=1\) gives the limiting threshold.

The factor \(q_{\rm ctrl}\) is part of the loss prescription rather than a fundamental compactification parameter.  Survival probabilities can therefore depend sensitively on it when the initial state is deliberately placed within only a few stochastic displacements of the prescribed surface.  Such near-wall examples are useful for resolving first-passage physics, but their numerical probabilities are conditional on the chosen threshold.  Below we distinguish these prescription-sensitive quantities from the local hard-wall asymptotics that are independent of the threshold normalization.

This separation is what makes the construction an operational map rather than a collection of examples with the same near-wall formula. Different quantum-gravity inputs define genuinely different survival problems.

A species scale may be determined by a single tower, by genuinely separate tower sectors, or by a higher-dimensional lattice of jointly charged states; these cases need not have the same parametric field dependence. Independent microscopic loss conditions require a multi-boundary problem, gradual degradation is described by killing rather than a Dirichlet surface, and finite-duration constraints require a path event that specifies what must occur before the available duration expires.

Potential-based diagnostics require the additional qualification described below and are not automatically hard EFT-loss surfaces.

We show that, for regular hard loss surfaces, the leading local answer is universal: the inward normal component of the conditioned drift is fixed by the proper distance to EFT loss and by the normal diffusion coefficient, while microscopic data, background evolution, threshold motion, and the retained stochastic degrees of freedom determine the global survival problem.

The novelty of the construction is therefore not the absorbing-wall asymptotics by itself, nor the use of a Doob transform by itself, but the physical dictionary that turns compactification-dependent EFT-control data into explicit first-loss events for stochastic cosmological histories.

When the corresponding stochastic EFT remains valid up to the chosen surface, cutoff ratios such as \(\Lambda_{\rm sp}/H\), \(M_s/H\), and \(M_{\rm KK}/H\), as well as carefully qualified potential diagnostics such as \(F_\nabla\) and \(F_{\rm Hess}\), can be used as loss data for a conditioned path ensemble.

The microscopic loss criterion \(F_A=0\), the survival probability \(h_A\), the survival action, the induced drift \(2D\nabla\ln h_A\), and the inverse-reconstructed boundary-normal class are then distinct objects connected by the survival map.

The construction can be summarized as the operational map in \cref{fig:operational_chain}.

\begin{figure}[!htb]
\centering
\footnotesize
\begin{tikzpicture}[
 node distance=1.55cm,
 box/.style={
 rounded corners=2pt,
 draw=DarkBlue,
 line width=0.45pt,
 align=center,
 inner xsep=7pt,
 inner ysep=5pt,
 text width=2.8cm
 },
 arrow/.style={
 -{Latex[length=2.2mm]},
 line width=0.45pt,
 draw=DarkBlue
 }
]
\node[box] (qg) {EFT-control\\ data};
\node[box, right=of qg] (inputs) {Survival inputs\\
\(F_A,\kappa_A,\tau;\,b,D\)};
\node[box, right=of inputs] (h) {Survival\\ probability\\ \(h_A\)};
\node[box, below=1.05cm of h] (S) {Survival action\\
\(\mathcal S_A=-\ln h_A\)};
\node[box, left=of S] (doob) {Conditioned drift\\
\(\Delta b_{\rm Doob}\)};
\node[box, left=of doob] (out) {Boundary layers\\
and Doob class};

\draw[arrow] (qg) -- (inputs);
\draw[arrow] (inputs) -- (h);
\draw[arrow] (h) -- (S);
\draw[arrow] (S) -- (doob);
\draw[arrow] (doob) -- (out);

\end{tikzpicture}
\vspace{3mm}
\caption{Operational survival map. Compactification, Swampland, and diagnostic control data specify candidate loss surfaces, killing profiles, and finite duration data, while the stochastic description supplies the generator. A quantitative first-passage problem additionally requires that the chosen generator remain valid up to the prescribed loss surface.
The survival problem determines the surviving ensemble, its conditioned drift, and the corresponding boundary-layer and Doob-equivalence data.}
\label{fig:operational_chain}
\end{figure}

The paper is organized as follows. \Cref{sec:covariant_doob} develops the finite-horizon survival equation, the associated Doob transform, the universal near-wall law, first-exit statistics, and the domain of validity of the stochastic description.

\Cref{sec:string_towers} converts tower/species, weak-coupling, direct string/KK, soft-loss, and finite-horizon inputs into operational survival data, clarifies when those surfaces are accessible within the retained
stochastic EFT.

\Cref{sec:swampland_layers} analyzes the resulting stochastic boundary layers and one-e-fold stochastic displacement scales, together with carefully qualified potential-based diagnostics. It also studies a light spectator modulus in a quadratic potential with a field-dependent KK threshold.

\Cref{sec:inverse_reconstruction} develops the inverse reconstruction problem, the associated integrability obstruction, and the boundary-normal Doob-equivalence class.

Finally, \Cref{sec:discussion} summarizes the physical implications, limitations, and possible extensions, while technical variants and supporting derivations are collected in the appendices.
\section{Survival Actions and Local Wall Laws}
\label{sec:covariant_doob}

We first formulate the survival problem for a covariant diffusion on moduli space. The backward equation determines the survival probability, while the Doob transform converts its logarithmic weight into the drift of the survival-conditioned ensemble. The near-boundary limit then gives the local wall law and its boundary-function form. Technical variants involving independent multi-boundary problems, soft killing, and finite horizons are collected in the appendix.

Throughout this section, a prescribed surface becomes a physical first-passage boundary only after the stochastic description has been shown to remain valid up to that surface.  This distinction between a candidate control surface and an admissible stopping surface will be important for the compactification applications below.

\subsection{Backward survival equation and Doob transform}

Let \(\mathcal M_{\rm EFT}\) be the field-space domain in which the low-energy description is under control. We equip it with metric \(G_{ij}(\phi)\) and Levi-Civita connection \(\nabla_i\). The unconditioned stochastic dynamics is specified by the backward generator
\begin{equation}
\mathcal L f
=
b^i\nabla_i f
+
D^{ij}\nabla_i\nabla_j f ,
\label{eq:backward_generator}
\end{equation}
where \(D^{ij}=D^{ji}\) is a smooth positive-semidefinite diffusion tensor and \(b^i\) is the covariant drift appearing in the generator. In a local coordinate chart, the corresponding It\^o realization has coordinate drift \(\widetilde b^i\), related to \(b^i\) by
\begin{equation}
 \widetilde b^k
 =
 b^k
 -
 D^{ij}\Gamma^k{}_{ij}.
 \label{eq:coordinate_ito_drift_main}
\end{equation}
Consequently,
\begin{equation}
 \left\langle
 \Delta\phi^i
 \right\rangle
 =
 \widetilde b^i\Delta t,
 \qquad
 \left\langle
 \Delta\phi^i\Delta\phi^j
 \right\rangle
 =
 2D^{ij}\Delta t,
 \label{eq:local_ito_moments_main}
\end{equation}
up to higher orders in the time step. Here \(\Delta t>0\) is a forward elapsed increment of the stochastic time (\(\Delta N\) in the e-fold convention), not a forward increase of the remaining horizon \(\tau\) defined below. We take the covariant generator \(\mathcal L\), rather than a coordinate-dependent stochastic differential equation, as the primary object. Once this generator is fixed, the conditioning calculation is coordinate covariant.

Let \(T_{\partial\mathcal M_{\rm EFT}}\) denote the first exit time from the controlled EFT domain. The finite-horizon survival probability is
\begin{equation}
h(\phi,\tau)
=
\Pr_\phi\!\left[
T_{\partial\mathcal M_{\rm EFT}}>\tau
\right],
\label{eq:survival_probability}
\end{equation}
where \(\tau\) is the remaining time, or the remaining number of e-folds in the cosmological applications.

Throughout the main text, the Doob transform associated with \(h(\phi,\tau)\) is a finite-horizon transform. The parameter \(\tau\) is the chosen remaining survival time, not a unique cosmological value of the construction. Different physical questions correspond to different path events and horizons.  Survival until the end of inflation, survival for a prescribed interval, and completion before a separate finite-duration bound are not automatically the same event; the relevant stopping or completion rule must be specified explicitly.

The survival probability obeys
\begin{equation}
h|_{\partial\mathcal M_{\rm EFT}}=0,
\qquad
h(\phi,0)=1
\quad
(\phi\in\mathcal M_{\rm EFT}),
\label{eq:survival_boundary_conditions}
\end{equation}
and satisfies the backward equation
\begin{equation}
\partial_\tau h
=
\mathcal Lh .
\label{eq:backward_survival}
\end{equation}
The first condition removes histories that start on the EFT-loss surface; the second states that a history already inside the controlled domain has survived when no time remains.

Conditioning on survival weights short-time transitions by the future survival probability of the point they reach. The resulting finite-horizon Doob transform is
\begin{equation}
\mathcal L^h f
=
h^{-1}\mathcal L(hf)
-
h^{-1}f\,\mathcal Lh .
\label{eq:doob_generator}
\end{equation}
Using
\begin{equation}
\mathcal L(hf)
=
h\,\mathcal L f
+
f\,\mathcal L h
+
2D^{ij}(\nabla_i h)(\nabla_j f),
\end{equation}
one obtains
\begin{equation}
\mathcal L^h f
=
\left(
b^i+2D^{ij}\nabla_j\ln h
\right)\nabla_i f
+
D^{ij}\nabla_i\nabla_j f .
\label{eq:doob_transformed_generator}
\end{equation}
Thus, the conditioned process has the same diffusion tensor as the original one, but its drift is shifted by
\begin{equation}
\Delta b^i_{\rm Doob}
=
2D^{ij}\nabla_j\ln h .
\label{eq:doob_drift_logh}
\end{equation}
This is the statistical drift induced by conditioning the path ensemble on future survival.

It is useful to write the survival probability as a dimensionless survival action,
\begin{equation}
\mathcal S_{\rm surv}(\phi,\tau)
\equiv
-\ln h(\phi,\tau).
\label{eq:survival_action}
\end{equation}
Then
\begin{equation}
\Delta b^i_{\rm Doob}
=
-2D^{ij}\nabla_j\mathcal S_{\rm surv}.
\label{eq:doob_drift_survival_action}
\end{equation}
The word ``action'' refers here to the logarithmic cost of survival~\cite{Touchette:2009}. It is not a scalar potential and does not introduce a fundamental force term; it is the object whose gradient generates the drift of the survival-conditioned ensemble.

\subsection{Local wall law and finite-horizon normal model}

Let \(\Sigma\) be a smooth absorbing boundary with nonzero normal diffusion, and let \(s(\phi)\) be the inward proper distance from \(\Sigma\). Thus \(s=0\) on the wall and
\begin{equation}
n_i
=
\nabla_i s
\end{equation}
is the inward unit normal. Since the survival probability vanishes on the absorbing boundary, a regular positive solution has the leading near-wall form
\begin{equation}
h(\phi,\tau)
=
C(\sigma,\tau)\,s
+
\mathcal O(s^2),
\qquad
C(\sigma,\tau)>0,
\label{eq:h_linear_boundary}
\end{equation}
where \(\sigma\) denotes coordinates along \(\Sigma\). Therefore
\begin{equation}
\nabla_i\ln h
=
\frac{n_i}{s}
+
\mathcal O(1),
\label{eq:logh_boundary_gradient}
\end{equation}
and the inward normal component of the conditioned drift is
\begin{equation}
\Delta b_\perp^{\rm Doob}
\equiv
n_i\Delta b^i_{\rm Doob}
\simeq
\frac{2D_\perp}{s},
\qquad
D_\perp
\equiv
n_iD^{ij}n_j.
\label{eq:single_boundary_law}
\end{equation}
This is the local wall law. It fixes the leading normal response near a regular hard boundary; the global conditioned process still depends on the full boundary-value problem for \(h\).

In particular, Eq.~\eqref{eq:single_boundary_law} is an asymptotic statement and does not by itself justify treating a physically moving or finite-width control transition as a static absorbing surface over a long duration.  The finite-time accuracy of such an approximation must be checked against the motion, width, and stochastic time scales of the specific problem.

The same result can be written directly in terms of a boundary function. Let the controlled side be described locally by
\begin{equation}
F(\phi)>0,
\qquad
\Sigma:\ F(\phi)=0,
\end{equation}
with \(\nabla_iF\neq0\) on \(\Sigma\). Near the wall,
\begin{equation}
s
=
\frac{F}{|\nabla F|}
+
\mathcal O(F^2),
\qquad
n_i
=
\frac{\nabla_iF}{|\nabla F|}
+
\mathcal O(F),
\end{equation}
where
\begin{equation}
|\nabla F|
\equiv
\sqrt{
G^{ij}\nabla_iF\nabla_jF
}.
\end{equation}
Hence
\begin{equation}
\nabla_i\ln h
=
\frac{\nabla_iF}{F}
+
\mathcal O(1),
\label{eq:survival_boundary_form}
\end{equation}
and the Doob drift has the local boundary-function form
\begin{equation}
\Delta b^i_{\rm Doob}
\simeq
2D^{ij}\frac{\nabla_jF}{F}
+
\mathcal O(1).
\label{eq:doob_boundary_function}
\end{equation}
Thus, once an EFT-loss condition is written as a regular function \(F(\phi)\), its leading boundary structure directly determines the singular statistical drift of the conditioned ensemble.

For an explicitly time-dependent control function \(F(\phi,N)\), however, the same level set can move through field space. Writing the controlled domain as
\begin{equation}
 {
 \Omega(N)
 =
 \left\{
 \phi:
 F(\phi,N)>0
 \right\},
 }
 \label{eq:moving_level_set_domain_revision}
\end{equation}
the normal velocity of the surface \(F=0\), measured along the inward unit normal \(n_i=\nabla_iF/|\nabla F|\), is
\begin{equation}
 {
 v_{\Sigma,\perp}
 =
 -\frac{\partial_NF}{|\nabla F|}
 \qquad
 \text{on}
 \qquad
 F=0.
 }
 \label{eq:moving_level_set_velocity_revision}
\end{equation}
The first-passage problem must then be formulated in the moving domain \(\Omega(N)\).  In a coordinate measuring distance relative to the moving surface, this boundary velocity contributes to the relative drift.  This effect is included automatically by the nonautonomous formulation below and will be quantified for the KK example in
\cref{subsec:two_field_survival_benchmark}.

The local origin of this singular term can be seen by writing the generator in boundary-adapted coordinates \((s,\sigma^a)\). Locally,
\begin{equation}
\mathcal L
=
D_\perp\partial_s^2
+
2D^{sa}\partial_s\nabla_a
+
D^{ab}\nabla_a\nabla_b
+
B_\perp\partial_s
+
B^a\nabla_a
+\cdots ,
\label{eq:boundary_adapted_generator}
\end{equation}
where \(D_\perp=D^{ss}=n_iD^{ij}n_j\), the effective first-order coefficients \(B_\perp,B^a\) include the original drift and connection terms, and all coefficients are smooth at the wall. The Dirichlet condition forces any regular positive survival solution to vanish linearly in \(s\), provided the boundary is regular and \(D_\perp\neq0\). Consequently,
\begin{equation}
 \Delta b_{\rm Doob}^{i}
 \simeq
 \frac{2D^{ij}n_j}{s},
 \qquad
 n_i\Delta b_{\rm Doob}^{i}
 \simeq
 \frac{2D_\perp}{s}.
 \label{eq:anisotropic_wall_vector_revision}
\end{equation}
Mixed normal--tangential diffusion can therefore produce singular tangential components of the full drift vector. The invariant universal coefficient is the inward normal projection \(2D_\perp/s\). The original drift, boundary curvature, tangential variation, and spatial variation of the diffusion tensor enter the regular terms, while the complete conditioned process still depends on the global boundary-value problem.

To explicitly display this universal singular layer, we now introduce a solvable finite-horizon normal model. Freeze the coefficients at the wall and keep only the leading normal diffusion operator. This gives the driftless half-line generator \(\mathcal L=D\partial_s^2\) on \(s\ge0\), with \(D=D_\perp\) evaluated locally at the wall. Global compactification data determine the boundary function and the full generator; the half-line problem is a local normal model that isolates the singular survival layer common to regular hard walls~\cite{Pinsky:1995,Collet:2013,2022PhRvE.106d4117M}.

Closely related half-line meander and first-passage formulae arise in duration-conditioned stochastic-inflation constructions~\cite{Tokeshi:2023swe}.  Here the same mathematics is used with a different physical stopping event: loss of a prescribed EFT-control domain.

The survival problem is
\begin{equation}
\partial_\tau h
=
D\,\partial_s^2 h,
\qquad
h(0,\tau)=0,
\qquad
h(s,0)=1 .
\label{eq:half_line_survival_problem}
\end{equation}
The absorbing heat kernel on the half-line is obtained by the image method,
\begin{equation}
K_{\rm abs}(s,s';\tau)
=
\frac{1}{\sqrt{4\pi D\tau}}
\left[
\exp\!\left(
-\frac{(s-s')^2}{4D\tau}
\right)
-
\exp\!\left(
-\frac{(s+s')^2}{4D\tau}
\right)
\right],
\label{eq:half_line_absorbing_kernel}
\end{equation}
which vanishes at \(s=0\). Since the initial survival probability is unity in the interior, the finite-horizon solution is
\begin{equation}
h(s,\tau)
=
\int_0^\infty
K_{\rm abs}(s,s';\tau)\,ds'
=
\frac{1}{\sqrt{\pi}}
\int_{-s/(2\sqrt{D\tau})}^{s/(2\sqrt{D\tau})}
e^{-u^2}\,du
=
\operatorname{erf}\!\left(
\frac{s}{2\sqrt{D\tau}}
\right).
\label{eq:half_line_survival_solution}
\end{equation}
Thus
\begin{equation}
\mathcal S_{\rm surv}(s,\tau)
=
-\ln h(s,\tau)
=
-\ln
\operatorname{erf}\!\left(
\frac{s}{2\sqrt{D\tau}}
\right).
\label{eq:half_line_survival_action}
\end{equation}
The corresponding Doob drift follows from
\begin{equation}
\Delta b_{\rm Doob}(s,\tau)
=
2D\,\partial_s\ln h .
\label{eq:half_line_doob_start}
\end{equation}
Therefore
\begin{equation}
\Delta b_{\rm Doob}(s,\tau)
=
2D\,\frac{\partial_s h}{h}
=
\frac{2\sqrt{D}}{\sqrt{\pi\tau}}
\,
\frac{
\exp[-s^2/(4D\tau)]
}{
\operatorname{erf}[s/(2\sqrt{D\tau})]
}.
\label{eq:half_line_doob_exact}
\end{equation}
Near the wall, as \(s\to0\),
\begin{equation}
h(s,\tau)
\simeq
\frac{s}{\sqrt{\pi D\tau}},
\qquad
\Delta b_{\rm Doob}(s,\tau)
\simeq
\frac{2D}{s}.
\label{eq:half_line_near_wall_limit}
\end{equation}

The same solution provides direct first-passage observables. The probability that the wall has been reached by the remaining time \(\tau\) is
\begin{equation}
 P_{\rm loss}(s,\tau)
 =
 1-h(s,\tau)
 =
 \operatorname{erfc}\!\left(
 \frac{s}{2\sqrt{D\tau}}
 \right).
 \label{eq:halfline_loss_probability_revision}
\end{equation}
The corresponding first-exit-time density is
\begin{equation}
 f_{\rm exit}(s,\tau)
 \equiv
 -\partial_\tau h(s,\tau)
 =
 \frac{s}{\sqrt{4\pi D\tau^3}}
 \exp\!\left[
 -\frac{s^2}{4D\tau}
 \right],
 \label{eq:halfline_exit_density_revision}
\end{equation}
and the finite-time survival hazard is
\begin{equation}
 \Gamma_{\rm exit}(s,\tau)
 \equiv
 \frac{f_{\rm exit}(s,\tau)}{h(s,\tau)}
 =
 -\partial_\tau\ln h(s,\tau)
 =
 \partial_\tau\mathcal S_{\rm surv}(s,\tau).
 \label{eq:halfline_hazard_revision}
\end{equation}
Conversely, requiring a target survival probability \(p\in(0,1)\) fixes the minimum initial proper distance from the wall,
\begin{equation}
 h(s_p,\tau)=p
 \qquad\Longrightarrow\qquad
 s_p(\tau)
 =
 2\sqrt{D\tau}\,
 \operatorname{erf}^{-1}(p).
 \label{eq:halfline_survival_quantile_revision}
\end{equation}

Equations~\eqref{eq:halfline_loss_probability_revision}--\eqref{eq:halfline_survival_quantile_revision} give, directly, the fraction of histories that lose EFT control, the distribution of their first-loss times, and the initial margin required to retain a prescribed fraction of histories for a specified duration.  Their numerical use for a microscopic EFT boundary remains conditional on the static hard-wall approximation being valid over that duration.

The finite-horizon solution therefore reproduces the local wall law derived above. \Cref{fig:survivalEnsembleWallLaw} illustrates this local model in two complementary ways: panel (a) samples the killed diffusion by a Brownian-bridge-corrected Monte Carlo ensemble, while panels (b) and (c) plot the analytic functions in \cref{eq:half_line_survival_solution,eq:half_line_survival_action,eq:half_line_doob_exact}. The Monte Carlo panel is only a numerical visualization of the same half-line survival problem; the survival action and Doob drift used in the analysis are the analytic expressions above.

\subsection{Scope and validity of the stochastic description}
\label{sec:stochastic_validity_revision}

The survival construction developed in this paper is a statement about a specified stochastic process; it is not, by itself, a derivation of that process from an ultraviolet completion. The stochastic state variables, their drift and diffusion data, the Markov approximation, and the prescribed loss rule must therefore be justified independently in every application.

A further consistency requirement concerns the ordering of physical thresholds.  Let \(\Omega_{\rm stoch}(N)\) denote the region in which the chosen stochastic state variables and generator provide a controlled description.  A proposed loss surface \(\partial\Omega_{\rm loss}(N)\) can be used with that generator only if the relevant histories can reach it without first leaving \(\Omega_{\rm stoch}(N)\).  If additional degrees of freedom become light or the coarse-grained description otherwise fails earlier, one must either enlarge the stochastic state space or stop the reduced process at that earlier threshold.

Operationally, one may define
\begin{equation}
 {
 T_{\rm valid}
 \equiv
 \inf\left\{
 N'>N:
 \phi(N')\notin\Omega_{\rm stoch}(N')
 \right\},
 }
 \label{eq:stochastic_validity_exit_time_revision}
\end{equation}
in addition to the specified EFT-loss time \(T_{\rm exit}\).  A quantitative first-passage calculation based on the reduced generator is self-consistent only over histories for which the prescribed stopping event occurs no later than the loss of validity of that generator, unless the latter is included explicitly as an additional stopping surface.  This criterion will be important for species-scale applications, where tower states may become dynamical before the formal species/Hubble equality is reached.

Let \(\phi^i\), with \(i=1,\ldots,n_\phi\), denote local coordinates on a field-space manifold equipped with metric \(G_{ij}(\phi)\), inverse metric \(G^{ij}(\phi)\), Levi--Civita connection \(\Gamma^i{}_{jk}\), and covariant derivative \(\nabla_i\). We use the number of e-folds,
\begin{equation}
 N
 \equiv
 \ln a,
 \label{eq:efold_time_definition_revision}
\end{equation}
as the stochastic time variable, where \(a\) is the cosmological scale factor.

In a local coordinate chart, an It\^o realization of the coarse-grained field dynamics may be written as
\begin{equation}
 \mathrm d\phi^i
 =
 \widetilde b_N^i(\phi,N)\,\mathrm dN
 +
 \Sigma^i{}_{a}(\phi,N)\,\mathrm dW_N^a,
 \qquad
 D_N^{ij}
 \equiv
 \frac12
 \sum_{a=1}^{n_{\rm noise}}
 \Sigma^i{}_{a}\Sigma^j{}_{a}.
 \label{eq:general_fieldspace_sde_revision}
\end{equation}
Here \(\widetilde b_N^i\) is the coordinate It\^o drift per e-fold, \(\Sigma^i{}_{a}\) is the noise-amplitude matrix, and \(a=1,\ldots,n_{\rm noise}\) labels statistically independent noise channels. The Wiener increments are normalized by
\begin{equation}
 \bigl\langle
 \mathrm dW_N^a
 \bigr\rangle
 =
 0,
 \qquad
 \bigl\langle
 \mathrm dW_N^a
 \mathrm dW_N^b
 \bigr\rangle
 =
 \delta^{ab}\,\mathrm dN.
 \label{eq:wiener_normalization_revision}
\end{equation}
The symmetric positive-semidefinite tensor \(D_N^{ij}\) is the diffusion tensor per e-fold.

The coordinate It\^o equation \eqref{eq:general_fieldspace_sde_revision} generates
\begin{equation}
 \mathcal L_N f
 =
 \widetilde b_N^i\partial_i f
 +
 D_N^{ij}\partial_i\partial_j f
 \label{eq:coordinate_ito_generator_revision}
\end{equation}
on sufficiently regular scalar test functions \(f(\phi,N)\). Equivalently, the same generator can be written in covariant form as
\begin{equation}
 \mathcal L_N f
 =
 b_N^i\nabla_i f
 +
 D_N^{ij}\nabla_i\nabla_j f,
 \label{eq:fieldspace_generator_validity_revision}
\end{equation}
provided the coordinate and covariant drifts are related by
\begin{equation}
 \widetilde b_N^k
 =
 b_N^k
 -
 D_N^{ij}\Gamma^k{}_{ij}.
 \label{eq:ito_covariant_drift_relation_revision}
\end{equation}
Thus \(b_N^i\) denotes the drift appearing in the covariant generator, whereas \(\widetilde b_N^i\) denotes the coordinate drift appearing in the local It\^o stochastic differential equation. This distinction disappears in the flat Cartesian examples used below, but it is necessary in a curved field-space parametrization. We take the generator \(\mathcal L_N\) as the primary stochastic object.

Equations~\eqref{eq:general_fieldspace_sde_revision}--\eqref{eq:fieldspace_generator_validity_revision} are assumed only when the coarse-graining procedure yields an approximately Markovian effective description on the time scales of interest. The survival construction does not determine whether this approximation is valid; it determines the conditioned process once the generator and loss prescription have been specified.

The white-noise normalization also does not imply a fundamental field-space resolution equal to the root-mean-square (rms) displacement in one e-fold.  Within the Markovian continuum approximation,
\begin{equation}
 {
 \left\langle
 \Delta\phi^i\Delta\phi^j
 \right\rangle
 \simeq
 2D_N^{ij}\Delta N
 }
 \label{eq:short_time_diffusion_scaling_revision}
\end{equation}
for sufficiently short intervals \(\Delta N\) over which the local coefficients may be treated as constant.  Hence the quantity \(\sqrt{2D_\perp}\) is a convenient one-e-fold rms displacement scale, not by itself a lower physical spatial resolution of the stochastic process.

The physically relevant minimum time scale is instead set by the coarse-graining prescription and by the correlation time of the effective noise.  Smooth coarse-graining windows can produce noise with finite temporal correlations rather than ideal white noise~\cite{Mahbub:2022osb}.  Denoting the corresponding correlation interval schematically by
\(\Delta N_{\rm corr}\), one may associate, only when a local diffusive description remains adequate, the comparison scale
\begin{equation}
 {
 \delta\phi_{{\rm corr},\perp}
 \sim
 \sqrt{
 2D_\perp\Delta N_{\rm corr}
 }.
 }
 \label{eq:correlation_displacement_scale_revision}
\end{equation}
A microscopic transition whose field-space width is not negligible on the relevant coarse-grained time and field scales should not automatically be replaced by a sharp absorbing surface; a finite-width killing profile or an enlarged stochastic description is then more appropriate.  We therefore use the one-e-fold displacement below strictly as a comparison scale, unless an independent coarse-graining argument supplies a physical resolution scale.

For explicitly time-dependent coefficients, killing profiles, or controlled domains, the survival problem must retain both the current time and the prescribed final time. Let \(\Omega(N)\) denote the controlled domain at e-fold \(N\), and define the absolute first-exit time
\begin{equation}
 T_{\rm exit}
 \equiv
 \inf
 \left\{
 N'>N:
 \phi(N')\notin\Omega(N')
 \right\}.
 \label{eq:nonautonomous_exit_time_revision}
\end{equation}
For a nonnegative killing rate \(\kappa_{\rm kill}(\phi,N)\), the generalized finite-horizon survival weight is
\begin{equation}
 H(\phi,N;N_f)
 \equiv
 \mathbb E_{\phi,N}
 \left[
 \exp\!\left(
 -\int_N^{N_f}
 \kappa_{\rm kill}\!\left(
 \phi(N'),N'
 \right)
 \,\mathrm dN'
 \right)
 \mathbf{1}_{\{T_{\rm exit}>N_f\}}
 \right],
 \qquad
 N\leq N_f,
 \label{eq:nonautonomous_survival_revision}
\end{equation}
where the subscript \((\phi,N)\) indicates that the process starts at field-space point \(\phi\) at e-fold \(N\), and \(\mathbf{1}_{\{T_{\rm exit}>N_f\}}\) is the indicator that no hard boundary has been crossed before \(N_f\). In the absence of soft killing, this expression reduces to the ordinary hard-wall survival probability,
\begin{equation}
 \left.
 H(\phi,N;N_f)
 \right|_{\kappa_{\rm kill}=0}
 =
 \mathbb{P}_{\phi,N}
 \left[
 T_{\rm exit}>N_f
 \right].
 \label{eq:nonautonomous_hard_survival_limit}
\end{equation}

The Feynman--Kac formula~\cite{KaratzasShreve1991,Pinsky:1995} implies that the generalized survival weight obeys
\begin{equation}
 \left(
 \partial_N
 +
 \mathcal L_N
 -
 \kappa_{\rm kill}(\phi,N)
 \right)
 H(\phi,N;N_f)
 =
 0,
 \label{eq:nonautonomous_backward_revision}
\end{equation}
with absorbing and terminal conditions
\begin{equation}
 H(\phi,N;N_f)
 =
 0
 \quad
 \text{for}
 \quad
 \phi\in\partial\Omega(N),
 \label{eq:nonautonomous_absorbing_condition_revision}
\end{equation}
and
\begin{equation}
 H(\phi,N_f;N_f)
 =
 1
 \quad
 \text{for}
 \quad
 \phi\in\Omega(N_f).
 \label{eq:nonautonomous_terminal_condition_revision}
\end{equation}
The associated finite-horizon conditioned covariant drift is
\begin{equation}
 b_{\rm surv}^{i}(\phi,N;N_f)
 =
 b_N^i(\phi,N)
 +
 2D_N^{ij}(\phi,N)
 \nabla_j
 \ln H(\phi,N;N_f).
 \label{eq:nonautonomous_doob_drift_revision}
\end{equation}
This is the appropriate formulation for an explicitly time-dependent generator, a moving controlled domain, or a time-dependent soft-killing profile.

In particular, replacing a moving boundary by a static one is an additional approximation, not a consequence of the local wall law.  Its accuracy must be established by comparing the accumulated boundary motion over the relevant interval with the characteristic distances entering the first-passage problem.

If the generator, controlled domain, and killing profile are autonomous, time-translation invariance permits the reduction
\begin{equation}
 H(\phi,N;N_f)
 =
 h(\phi,\tau),
 \qquad
 \tau
 \equiv
 N_f-N.
 \label{eq:autonomous_remaining_horizon_reduction}
\end{equation}
The remaining-horizon survival function then satisfies
\begin{equation}
 \partial_\tau h
 =
 \left(
 \mathcal L
 -
 \kappa_{\rm kill}
 \right)
 h.
 \label{eq:autonomous_remaining_horizon_equation}
\end{equation}
For a static hard-loss domain without soft killing,
\(\kappa_{\rm kill}=0\), one may equivalently define
\begin{equation}
 T_{\rm exit}
 \equiv
 \inf
 \left\{
 \Delta N>0:
 \phi(N+\Delta N)\notin\Omega
 \right\},
 \label{eq:exit_time_validity_revision}
\end{equation}
and
\begin{equation}
 h(\phi,\tau)
 \equiv
 \mathbb{P}_{\phi}
 \left(
 T_{\rm exit}>\tau
 \right).
 \label{eq:survival_probability_validity_revision}
\end{equation}
The autonomous finite-horizon equations developed in the main construction below refer to this specialization. The more general function \(H(\phi,N;N_f)\) is required whenever explicit time dependence or a moving domain cannot be neglected.

Once the stochastic generator and the loss prescription have been justified, the corresponding backward survival problem and Doob transform follow independently of whether the generator originates from slow-roll inflation, a spectator sector, a phase-space reduction, or another controlled stochastic system.

For sufficiently light scalar modes evolving on a slow-roll attractor, the usual field-space specialization takes the form
\begin{equation}
 b_{N,\mathrm{SR}}^{i}
 \simeq
 -M_{\rm Pl}^{2}
 G^{ij}\nabla_j\ln V,
 \qquad
 D_{N,\mathrm{SR}}^{ij}
 \simeq
 \frac{H^2}{8\pi^2}\,
 G^{ij}.
 \label{eq:slowroll_specialization_revision}
\end{equation}
Here \(b_{N,\mathrm{SR}}^i\) is the covariant drift entering the generator, \(V(\phi)\) is the scalar potential, \(H\equiv\dot a/a\) is the Hubble parameter, and a dot denotes differentiation with respect to cosmological proper time. We use
\begin{equation}
 M_{\rm Pl}
 \equiv
 \left(
 8\pi G_{\rm N}
 \right)^{-1/2}
 \label{eq:reduced_planck_mass_revision}
\end{equation}
for the four-dimensional reduced Planck mass. The first relation in Eq.~\eqref{eq:slowroll_specialization_revision} follows from the slow-roll background equations, while the second is the standard light-field diffusion tensor in e-fold time. When reduced Planck units are used in the numerical examples below, we set \(M_{\rm Pl}=1\).

The diffusion relation in Eq.~\eqref{eq:slowroll_specialization_revision} is not universal for an arbitrary compactification modulus. A more general parametrization is
\begin{equation}
 D_N^{ij}
 =
 \frac{H^2}{8\pi^2}\,
 {\cal Q}^{ij},
 \label{eq:noise_matrix_revision}
\end{equation}
where \({\cal Q}^{ij}(\phi,N)\) is a symmetric positive-semidefinite field-space tensor encoding the mode-function normalization, effective masses, field mixing, choice of gauge-invariant perturbation variables, and coarse-graining prescription.

To characterize diffusion normal to a regular control surface, let \(n_i\) be the inward-pointing unit normal covector, normalized by
\begin{equation}
 G^{ij}n_i n_j
 =
 1.
 \label{eq:normal_normalization_validity_revision}
\end{equation}
The normal diffusion coefficient and projected noise factor are
\begin{equation}
 D_{N,\perp}
 \equiv
 n_iD_N^{ij}n_j,
 \qquad
 {\cal Q}_{\perp}
 \equiv
 n_i{\cal Q}^{ij}n_j,
 \qquad
 D_{N,\perp}
 =
 \frac{H^2}{8\pi^2}
 {\cal Q}_{\perp}.
 \label{eq:normal_noise_projection_revision}
\end{equation}
In the light, weakly mixed canonical limit, \({\cal Q}^{ij}\simeq G^{ij}\), and hence \({\cal Q}_{\perp}\simeq1\). A heavy or strongly stabilized normal mode may instead have \({\cal Q}_{\perp}\ll1\), corresponding to suppressed normal diffusion. If
\begin{equation}
 D_{N,\perp}
 =
 0,
 \label{eq:vanishing_normal_diffusion_revision}
\end{equation}
there is no direct local diffusive increment normal to the wall; normal drift or coupling to other variables may still reach it, but the nondegenerate hard-wall law \(\Delta b_\perp^{\rm Doob}\sim2D_{N,\perp}/s\) does not apply.

Likewise, if a tower or another sector becomes light compared with the Hubble scale, the assumption that the stochastic state consists only of the original moduli must be reassessed.  Integrating out such a sector can modify the background, drift, and noise precisely in the region in which a later cutoff surface would otherwise be probed.  A species/Hubble surface is therefore not automatically an admissible stopping surface for a reduced moduli-only stochastic process.

Outside the slow-roll regime, Eq.~\eqref{eq:slowroll_specialization_revision} must not be assumed. If an approximately Markovian phase-space description remains controlled, the stochastic state may instead be enlarged to
\begin{equation}
 X^A
 =
 \left(
 \phi^i,\pi_i
 \right),
 \qquad
 A=1,\ldots,2n_\phi,
 \label{eq:phase_space_coordinates_revision}
\end{equation}
where \(\pi_i\) denotes the momentum variable conjugate to \(\phi^i\) in the chosen phase-space formulation. In local phase-space coordinates, one may write
\begin{equation}
 \mathrm dX^A
 =
 \widetilde B^A(X,N)\,\mathrm dN
 +
 \Sigma^A{}_{a}(X,N)\,\mathrm dW_N^a,
 \qquad
 D^{AB}
 \equiv
 \frac12
 \sum_{a=1}^{n_{\rm noise}}
 \Sigma^A{}_{a}\Sigma^B{}_{a}.
 \label{eq:phase_space_sde_revision}
\end{equation}
The corresponding local backward generator is
\begin{equation}
 \mathcal L_{\rm ph}f
 =
 \widetilde B^A\partial_A f
 +
 D^{AB}\partial_A\partial_B f,
 \qquad
 \partial_A
 \equiv
 \frac{\partial}{\partial X^A}.
 \label{eq:phase_space_generator_revision}
\end{equation}
No curved phase-space covariantization is assumed in Eq.~\eqref{eq:phase_space_generator_revision}; if a covariant phase-space geometry is introduced, the coordinate and covariant drifts must again be distinguished in the same manner as in Eq.~\eqref{eq:ito_covariant_drift_relation_revision}.

For an autonomous phase-space generator and static controlled region
\(\Omega_{\rm ph}\), the finite-horizon survival probability is
\begin{equation}
 h_{\rm ph}(X,\tau)
 \equiv
 \mathbb P_X
 \left(
 T_{\rm exit}^{\rm ph}>\tau
 \right),
 \label{eq:phase_space_survival_revision}
\end{equation}
where \(T_{\rm exit}^{\rm ph}\) is the first time at which \(X(N)\) leaves \(\Omega_{\rm ph}\). The corresponding Doob-conditioned phase-space drift is
\begin{equation}
 \widetilde B_{\rm cond}^{A}
 =
 \widetilde B^A
 +
 \Delta B_{\rm Doob}^{A},
 \qquad
 \Delta B_{\rm Doob}^{A}
 =
 2D^{AB}
 \partial_B\ln h_{\rm ph}.
 \label{eq:phase_space_doob_revision}
\end{equation}
For an explicitly time-dependent phase-space generator or moving phase-space domain, \(h_{\rm ph}(X,\tau)\) must be replaced by the corresponding two-time function
\begin{equation}
 H_{\rm ph}(X,N;N_f),
 \label{eq:nonautonomous_phase_space_survival_revision}
\end{equation}
which obeys the phase-space analogue of Eq.~\eqref{eq:nonautonomous_backward_revision}.

Phase-space formulations and explicit beyond-slow-roll constructions show that stochastic inflation is not identical to its slow-roll field-space reduction. At the same time, known limitations of the standard stochastic-\(\delta N\) treatment show that the noise prescription, the choice of stochastic variables, and their regime of validity must be checked rather than assumed~\cite{Grain:2017dqa,Pattison:2019hef,Cruces:2018cvq}. The present paper therefore uses the survival map conditionally on a valid generator and employs
Eq.~\eqref{eq:slowroll_specialization_revision} only in calculations explicitly identified as using the slow-roll, light-field normalization.

This distinction is especially relevant in asymptotic regions of string compactification moduli space. Let \(d_{\rm st}>2\) denote the number of noncompact spacetime dimensions in the asymptotic effective theory, and let \(M_{{\rm Pl},d_{\rm st}}\) denote the corresponding reduced \(d_{\rm st}\)-dimensional Planck scale. With the field-space metric
normalized in the associated \(d_{\rm st}\)-dimensional Planck units, define the dimensionless potential-slope parameter
\begin{equation}
 \gamma_{d_{\rm st}}
 \equiv
 M_{{\rm Pl},d_{\rm st}}
 \frac{
 |\nabla V|
 }{
 V
 },
 \qquad
 |\nabla V|
 \equiv
 \sqrt{
 G^{ij}
 \nabla_iV
 \nabla_jV
 }.
 \label{eq:asymptotic_slope_definition_revision}
\end{equation}
In asymptotic regimes where the bound
\begin{equation}
 \gamma_{d_{\rm st}}
 \gtrsim
 \frac{2}{
 \sqrt{d_{\rm st}-2}
 }
 \label{eq:asymptotic_slope_context_revision}
\end{equation}
is applicable, slow-roll inflation driven by that same runaway direction is unavailable as a spatially flat late-time attractor \cite{Rudelius:2022gbz,Andriot:2023wvg}. This statement restricts the admissible stochastic generator; it does not invalidate the abstract survival construction after a different controlled generator has been supplied.

For example, the fluctuating direction may be a separately justified light spectator or entropic modulus during inflation driven by another field, the operational control surface may occur at finite field distance before the strict asymptotic regime is reached, or the appropriate stochastic state may require the phase-space description in Eqs.~\eqref{eq:phase_space_sde_revision}--\eqref{eq:phase_space_doob_revision}. No claim is made that every asymptotic modulus is light, that a Markovian field-space process describes it, or that it automatically receives the canonical fluctuation amplitude \(H/(2\pi)\). That amplitude is specific to the light canonical limit \({\cal Q}_{\perp}\simeq1\), for which
\begin{equation}
 D_{N,\perp}
 \simeq
 \frac{H^2}{8\pi^2}.
 \label{eq:canonical_normal_diffusion_revision}
\end{equation}

The hierarchy of assumptions is therefore explicit: compactification physics supplies candidate control surfaces and the spectrum that determines which degrees of freedom become relevant; the cosmological construction supplies a background and stochastic generator for a specified set of retained variables; and the survival calculation is applied only on the overlap of these two domains of validity.

\section{String Boundary Data}
\label{sec:string_towers}

We now translate string and compactification control data into the survival inputs used by the stochastic problem. The output of this section is not yet a conditioned drift, but the operational data from which that drift will be
computed. Hard loss of EFT control is encoded by boundary functions \(F_A(\phi)=0\), with \(F_A>0\) on the prescribed controlled side; gradual degradation is encoded by nonnegative killing rates \(\kappa_A(\phi)\); and finite-duration restrictions enter through a remaining horizon \(\tau\). Together with the independently specified stochastic generator \((b^i,D^{ij})\), these data determine the survival probability.

A compactification scale defines a candidate loss surface before it defines an admissible first-passage boundary for a particular reduced stochastic EFT.  The latter additionally requires that the retained stochastic variables and their generator remain valid up to the proposed surface.  We therefore distinguish below between the microscopic construction of candidate cutoff surfaces and the ordering of those surfaces relative to thresholds at which additional degrees of freedom must enter the stochastic description.

Tower spectra and species counting produce cutoff/Hubble functions~\cite{Ooguri:2006in,Grimm:2018ohb,Lee:2019wij,Dvali:2007hz,Dvali:2007wp,Dvali:2010vm}; several towers contributing to the same gravitational cutoff can define one combined species wall~\cite{Castellano:2021mmx,Bedroya:2024uva,Caron-Huot:2024lbf}; weak-coupling limits produce magnetic-cutoff ratios~\cite{Arkani-Hamed:2006emk,Heidenreich:2015nta,Saraswat:2016eaz}; and TCC-type conditions provide finite-duration inputs~\cite{Bedroya:2019snp,Bedroya:2019tba,Brahma:2019vpl,Brandenberger:2021pzy}. De Sitter-type criteria~\cite{Obied:2018sgi,Agrawal:2018own,Murayama:2018lie,Ooguri:2018wrx,Garg:2018reu,Andriot:2018mav} can be used as diagnostic comparisons with the conditioned flow. They may also be promoted to potential-based survival data when an independent operational prescription specifies that their violation represents loss or degradation of the chosen description.

\subsection{Survival channels and cutoff/Hubble walls}
\label{sec:survival_channels_cutoff_walls}

The most direct hard-loss input compares a local ultraviolet control scale with the local Hubble scale. Let \(A\) label a specified loss channel, and let \(\Lambda_A(\phi)>0\) denote the corresponding field-dependent cutoff or compactification scale. Depending on the channel, \(\Lambda_A\) may be a species scale, a string scale, a Kaluza--Klein scale, a weak-coupling cutoff, or another operational scale beyond which the chosen low-energy description should not be extrapolated. The local Hubble parameter is denoted by \(H(\phi)>0\). We first define the nominal cutoff/Hubble function
\begin{equation}
 F_{A/H}(\phi)
 \equiv
 \ln\!\left(
 \frac{\Lambda_A(\phi)}{H(\phi)}
 \right).
 \label{eq:general_qg_hubble_function}
\end{equation}
The nominal controlled domain and its limiting surface are
\begin{equation}
 \Omega_A^{\rm nom}
 =
 \left\{
 \phi:
 F_{A/H}(\phi)>0
 \right\},
 \qquad
 \partial\Omega_A^{\rm nom}
 =
 \left\{
 \phi:
 F_{A/H}(\phi)=0
 \right\},
 \label{eq:nominal_channel_domain}
\end{equation}
or, equivalently,
\begin{equation}
 \Lambda_A(\phi)>H(\phi)
 \quad\text{in}\quad
 \Omega_A^{\rm nom},
 \qquad
 \Lambda_A(\phi)=H(\phi)
 \quad\text{on}\quad
 \partial\Omega_A^{\rm nom}.
 \label{eq:nominal_scale_interpretation}
\end{equation}

The equality \(\Lambda_A=H\) is only a nominal threshold. The retained EFT may already be marginal when the two scales become comparable. To avoid extrapolating the stochastic description all the way to that limiting equality, we introduce a dimensionless control factor
\begin{equation}
 q_{\rm ctrl}\geq1.
 \label{eq:control_factor_definition}
\end{equation}
The corresponding conservative cutoff/Hubble function is
\begin{equation}
 F_{A/H}^{(q_{\rm ctrl})}(\phi)
 \equiv
 \ln\!\left(
 \frac{\Lambda_A(\phi)}
 {q_{\rm ctrl}H(\phi)}
 \right)
 =
 F_{A/H}(\phi)-\ln q_{\rm ctrl}.
 \label{eq:safety_wall_revision}
\end{equation}
Its controlled domain and operational wall are
\begin{align}
 \Omega_A^{(q_{\rm ctrl})}
 &=
 \left\{
 \phi:
 F_{A/H}^{(q_{\rm ctrl})}(\phi)>0
 \right\},
 \label{eq:controlled_domain_qctrl}
 \\
 \partial\Omega_A^{(q_{\rm ctrl})}
 &=
 \left\{
 \phi:
 F_{A/H}^{(q_{\rm ctrl})}(\phi)=0
 \right\}.
 \label{eq:controlled_boundary_qctrl}
\end{align}
Equivalently,
\begin{equation}
 \Lambda_A(\phi)>q_{\rm ctrl}H(\phi)
 \quad\text{inside}\quad
 \Omega_A^{(q_{\rm ctrl})},
 \qquad
 \Lambda_A(\phi)=q_{\rm ctrl}H(\phi)
 \quad\text{on}\quad
 \partial\Omega_A^{(q_{\rm ctrl})}.
 \label{eq:qctrl_scale_interpretation}
\end{equation}
The choice \(q_{\rm ctrl}=1\) reproduces the nominal prescription,
\begin{equation}
 F_{A/H}^{(q_{\rm ctrl}=1)}
 =
 F_{A/H},
 \qquad
 \Omega_A^{(q_{\rm ctrl}=1)}
 =
 \Omega_A^{\rm nom}.
 \label{eq:nominal_qctrl_relation}
\end{equation}
For \(q_{\rm ctrl}>1\), the operational wall is placed on the more conservative side of field space, where the ultraviolet scale still exceeds the Hubble scale by the prescribed factor.

The parameter \(q_{\rm ctrl}\) belongs to the loss prescription, not to the microscopic compactification model.  Consequently, a robustness test under variation of \(q_{\rm ctrl}\) must hold the physical quantities that define \(\Lambda_A\), the cosmological background, and the initial state fixed.  Re-adjusting those quantities so that the initial distance from the wall remains fixed would instead define a different physical model.  We implement this distinction explicitly in the spectator calculation below.

The surface \(\partial\Omega_A^{(q_{\rm ctrl})}\) is an operational boundary of the chosen stochastic EFT, not a microscopic singularity and not, by itself, a boundary between quantum-gravity-consistent and inconsistent theories. Imposing an absorbing condition there means that histories are removed from the retained ensemble once they leave the prescribed control domain.

This hard-wall idealization is appropriate only when the transition can be treated as sharp on the physical field and time scales resolved by the chosen coarse-grained stochastic description.  The one-e-fold rms displacement \(\sqrt{2D_\perp}\) is not itself such a fundamental resolution limit, as discussed in \cref{sec:stochastic_validity_revision}.

If the degradation of EFT control occurs over a resolvable field-space interval, the corresponding channel should instead be represented by a nonnegative killing profile
\begin{equation}
 \kappa_A(\phi)\geq0,
\end{equation}
whose magnitude specifies the local probability-removal rate in the stochastic time variable used by the generator.

There is a second, logically independent requirement.  A candidate wall may be used with a reduced stochastic generator only if that generator remains valid up to the wall.  In the notation of Eq.~\eqref{eq:stochastic_validity_exit_time_revision}, the relevant stopping prescription must therefore account for
\begin{equation}
 {
 T_{\rm stop}
 =
 \min
 \left(
 T_A,
 T_{\rm valid}
 \right),
 }
 \label{eq:channel_validity_stop_revision}
\end{equation}
unless the degrees of freedom responsible for the earlier validity threshold have explicitly been added to the stochastic state.  This ordering issue is particularly important for species-scale surfaces.

To display the wall geometry explicitly, consider a local trajectory parameterized by an increasing dimensionless canonical-distance coordinate \(d\). If \(\ell\) denotes proper length along that trajectory, we define
\begin{equation}
 d
 \equiv
 \frac{\ell-\ell_{\rm ref}}{M_{\rm Pl}},
 \label{eq:dimensionless_trajectory_distance}
\end{equation}
where \(\ell_{\rm ref}\) is the proper length at an arbitrary reference point and \(M_{\rm Pl}\) is the four-dimensional reduced Planck mass. The local logarithmic rates of the cutoff and the Hubble scale are
\begin{equation}
 \lambda_A(d)
 \equiv
 -\frac{\mathrm d\ln\Lambda_A}{\mathrm dd},
 \qquad
 \beta(d)
 \equiv
 -\frac{\mathrm d\ln H}{\mathrm dd}.
 \label{eq:local_logarithmic_rates}
\end{equation}
Over a sufficiently small interval, these rates may be approximated as constants. Choosing \(d=0\) at the reference point then gives
\begin{equation}
 \Lambda_A(d)
 =
 \Lambda_{A,0}\mathrm e^{-\lambda_A d},
 \qquad
 H(d)
 =
 H_0\mathrm e^{-\beta d},
 \label{eq:qg_and_hubble_exponential_profiles_section3}
\end{equation}
where \(\Lambda_{A,0}>0\) and \(H_0>0\) are the corresponding values at
\(d=0\). Defining
\begin{equation}
 F_{A,0}
 \equiv
 \ln\!\left(
 \frac{\Lambda_{A,0}}{H_0}
 \right),
 \qquad
 \Delta\lambda_A
 \equiv
 \lambda_A-\beta,
 \label{eq:channel_intercept_and_rate}
\end{equation}
the nominal and conservative control functions become
\begin{align}
 F_{A/H}(d)
 &=
 F_{A,0}
 -
 \Delta\lambda_A d,
 \label{eq:qg_hubble_linear_section3}
 \\
 F_{A/H}^{(q_{\rm ctrl})}(d)
 &=
 F_{A,0}
 -
 \ln q_{\rm ctrl}
 -
 \Delta\lambda_A d.
 \label{eq:qctrl_linear_channel}
\end{align}

A finite wall is encountered along the chosen increasing-\(d\) direction only if the cutoff decreases relative to the Hubble scale,
\begin{equation}
 \Delta\lambda_A
 =
 \lambda_A-\beta
 >
 0.
 \label{eq:relative_cutoff_decay_condition}
\end{equation}
Provided also that the reference point lies inside the conservative controlled domain,
\begin{equation}
 F_{A,0}
 >
 \ln q_{\rm ctrl},
 \label{eq:initial_control_condition_qctrl}
\end{equation}
the operational wall is located at
\begin{equation}
 d_{A,b}(q_{\rm ctrl})
 =
 \frac{
 F_{A,0}-\ln q_{\rm ctrl}
 }{
 \Delta\lambda_A
 }.
 \label{eq:safety_wall_location_revision}
\end{equation}
The nominal wall location is
\begin{equation}
 d_{A,b}^{\rm nom}
 \equiv
 \frac{F_{A,0}}{\Delta\lambda_A},
 \label{eq:qg_hubble_wall_position}
\end{equation}
and the conservative wall can therefore be written as
\begin{equation}
 d_{A,b}(q_{\rm ctrl})
 =
 d_{A,b}^{\rm nom}
 -
 \frac{\ln q_{\rm ctrl}}{\Delta\lambda_A}.
 \label{eq:qctrl_wall_shift}
\end{equation}
Increasing \(q_{\rm ctrl}\) moves the operational wall toward the reference point without changing the logarithmic slope,
\begin{equation}
 \frac{\mathrm dF_{A/H}^{(q_{\rm ctrl})}}{\mathrm dd}
 =
 -\Delta\lambda_A.
 \label{eq:qctrl_slope_invariance}
\end{equation}

Consequently, \(q_{\rm ctrl}\) changes the wall location but not the local logarithmic slope of the channel.  This does \emph{not} imply that a finite-horizon survival probability is insensitive to \(q_{\rm ctrl}\): first-passage probabilities can depend strongly on the absolute initial distance from the wall.  A physically meaningful robustness test therefore varies \(q_{\rm ctrl}\) while keeping the underlying compactification parameters, stochastic generator, background, and initial state fixed.

If \(\Delta\lambda_A=0\), the ratio \(\Lambda_A/H\) is constant along the local trajectory, so an initially controlled trajectory neither approaches nor recedes from the wall within this approximation. If \(\Delta\lambda_A<0\), the ratio \(\Lambda_A/H\) increases with \(d\), and the chosen increasing-\(d\) direction moves away from the loss surface. Finally, if \(F_{A,0}\leq\ln q_{\rm ctrl}\), the reference point \(d=0\) is already on or outside the conservative controlled domain, and no positive wall distance exists in that direction.

The following subsections specify how tower spectra, species counting, weak-coupling limits, and direct string and Kaluza--Klein thresholds determine the channel-dependent quantities \(\Lambda_A\), \(F_{A,0}\), and \(\lambda_A\), as well as when a hard surface should be replaced by a soft killing profile.

\subsection{Species and combined tower cutoffs}

A tower of states whose characteristic mass decreases along a canonically normalized trajectory provides a standard cutoff channel. For a single approximately exponential tower, let
\begin{equation}
 M_{\rm tower}(d)
 =
 M_0\mathrm e^{-\alpha d},
 \qquad
 \alpha>0,
 \label{eq:tower_mass_general}
\end{equation}
where \(M_0\) is the tower mass at \(d=0\) and \(\alpha\) is its dimensionless logarithmic mass rate. If the number of tower states below a cutoff
\(\Lambda\) grows as
\begin{equation}
 \mathcal N(\Lambda,d)
 \sim
 \left(
 \frac{\Lambda}{M_{\rm tower}(d)}
 \right)^p,
 \qquad
 p>0,
 \label{eq:tower_count_general}
\end{equation}
then the four-dimensional species estimate
\begin{equation}
 \Lambda_{\rm sp}
 \sim
 M_{\rm Pl}
 \left[
 \mathcal N(\Lambda_{\rm sp},d)
 \right]^{-1/2}
 \label{eq:species_bound_input}
\end{equation}
implies
\begin{equation}
 \Lambda_{\rm sp}^{p+2}
 \sim
 M_{\rm Pl}^{2}M_{\rm tower}^{p}.
\end{equation}
In reduced Planck units, the species scale therefore takes the local form
\begin{equation}
 \Lambda_{\rm sp}(d)
 =
 \Lambda_{{\rm sp},0}
 \mathrm e^{-\lambda_{\rm sp}d},
 \qquad
 \lambda_{\rm sp}
 =
 \frac{p\alpha}{p+2},
 \label{eq:species_exponent_general}
\end{equation}
up to order-one and convention-dependent factors~\cite{Ooguri:2006in,Dvali:2007hz,Dvali:2007wp,Dvali:2008jb,Brustein:2009ex,Dvali:2010vm,Grimm:2018ohb,Heidenreich:2018kpg,Lee:2019wij,Calderon-Infante:2023ler,vandeHeisteeg:2023ubh,
Scalisi:2024jhq}. The tower mass rate \(\alpha\) controls the descent of the characteristic tower spacing, whereas the species rate \(\lambda_{\rm sp}\) controls the descent of the collective gravitational cutoff. Defining
\begin{equation}
 F_{{\rm sp},0}
 \equiv
 \ln\!\left(
 \frac{\Lambda_{{\rm sp},0}}{H_0}
 \right),
\end{equation}
the corresponding candidate species boundary is
\begin{equation}
 F_{\rm sp/H}(d)
 \equiv
 \ln\!\left(
 \frac{\Lambda_{\rm sp}(d)}{H(d)}
 \right)
 =
 F_{{\rm sp},0}
 -
 \left(
 \lambda_{\rm sp}-\beta
 \right)d.
 \label{eq:species_hubble_boundary}
\end{equation}

The tower spectrum enters the candidate species surface through \(F_{\rm sp/H}\), but the tower mass remains independently relevant to the validity of the reduced stochastic EFT.  In particular, the tower can become
light compared with \(H\) before the collective species scale approaches \(H\).

For the single-tower counting law \(\mathcal N\sim(\Lambda/M_{\rm tower})^p\), imposing the candidate species surface
\begin{equation}
 {
 \Lambda_{\rm sp}
 =
 q_{\rm ctrl}H
 }
 \label{eq:species_candidate_surface_revision}
\end{equation}
in the species relation gives
\begin{equation}
 {
 \frac{M_{\rm tower}}{H}
 \sim
 q_{\rm ctrl}^{(p+2)/p}
 \left(
 \frac{H}{M_{\rm Pl}}
 \right)^{2/p}.
 }
 \label{eq:tower_mass_at_species_wall_revision}
\end{equation}
For \(p=1\),
\begin{equation}
 {
 \left.
 \frac{M_{\rm tower}}{H}
 \right|_{\Lambda_{\rm sp}=q_{\rm ctrl}H}
 \sim
 q_{\rm ctrl}^{3}
 \left(
 \frac{H}{M_{\rm Pl}}
 \right)^2.
 }
 \label{eq:p1_tower_mass_at_species_wall_revision}
\end{equation}
Thus for an inflationary scale \(H/M_{\rm Pl}\ll1\) and moderate \(q_{\rm ctrl}\), many tower states can already lie below \(H\) by the time the formal species/Hubble surface is reached.

In the reduced stochastic examples in this paper, the individual KK/tower modes are not retained as stochastic degrees of freedom. If those states are not retained in the stochastic state, the same few-field generator cannot in general be extrapolated to that species surface: the additional light sector can affect the background, dissipation, mixing, and noise.  In such a reduced description the first direct tower/Hubble threshold,
\begin{equation}
 {
 M_{\rm tower}
 =
 q_{\rm tower}H,
 }
 \label{eq:direct_tower_validity_threshold_revision}
\end{equation}
with an appropriately specified control factor \(q_{\rm tower}\), provides an earlier validity boundary.  The species surface remains useful compactification data, but it is a physical first-passage boundary of the same reduced stochastic process only if the tower sector has been included and its effects on the generator are controlled.

This distinction is central to the interpretation below: a candidate species wall characterizes the collective gravitational cutoff, whereas a direct KK or string threshold can determine where a moduli-only stochastic EFT must already be terminated.

For several towers, the required input is the \emph{full density of states} below the candidate cutoff,
\begin{equation}
 {
 \mathcal N_{\rm tot}(\Lambda,\phi)
 =
 \#\left\{
 \text{states with }
 m_{\boldsymbol n}(\phi)<\Lambda
 \right\},
 }
 \label{eq:full_species_count_revision}
\end{equation}
and the combined four-dimensional species scale is defined implicitly by
\begin{equation}
 {
 \Lambda_{\rm sp}^{2}
 \mathcal N_{\rm tot}
 \left(
 \Lambda_{\rm sp},\phi
 \right)
 \sim
 M_{\rm Pl}^{2}.
 }
 \label{eq:full_combined_species_equation_revision}
\end{equation}
The detailed spectrum therefore matters parametrically, not merely through order-one normalization factors~\cite{Etheredge:2024tok}.

An additive approximation,
\begin{equation}
 {
 \mathcal N_{\rm tot}
 \simeq
 \sum_A
 \left(
 \frac{\Lambda}{M_A}
 \right)^{p_A},
 }
 \label{eq:additive_species_restricted_revision}
\end{equation}
is appropriate only when the relevant states decompose into distinct sectors whose leading state counts can be added without a parametrically important population carrying simultaneous quantum numbers under several tower directions.  Under this restricted assumption, defining
\begin{equation}
 {
 \mathcal N_A
 \equiv
 \left(
 \frac{\Lambda_{\rm sp}}{M_A}
 \right)^{p_A},
 \qquad
 w_A
 \equiv
 \frac{\mathcal N_A}
 {\sum_B\mathcal N_B},
 \qquad
 \overline p
 \equiv
 \sum_Aw_Ap_A,
 }
 \label{eq:species_weights_restricted_revision}
\end{equation}
and
\begin{equation}
 {
 \mu_A(d)
 \equiv
 -\frac{\mathrm d}{\mathrm dd}
 \ln M_A\bigl(\phi(d)\bigr),
 }
 \label{eq:multi_tower_mass_rates_restricted_revision}
\end{equation}
one recovers
\begin{equation}
 {
 \lambda_{\rm eff}(d)
 \equiv
 -\frac{\mathrm d\ln\Lambda_{\rm sp}}{\mathrm dd}
 =
 \frac{
 \sum_Aw_Ap_A\mu_A
 }{
 2+\overline p
 }.
 }
 \label{eq:combined_species_effective_rate_restricted_revision}
\end{equation}
Equation~\eqref{eq:combined_species_effective_rate_restricted_revision} should therefore not be applied to a multidimensional KK lattice by simply adding its one-dimensional axis towers.

For example, consider two compact circles with KK spacings \(M_1\) and \(M_2\).  The spectrum contains jointly charged states
\begin{equation}
 {
 m_{n_1,n_2}^{2}
 =
 n_1^{2}M_1^{2}
 +
 n_2^{2}M_2^{2}.
 }
 \label{eq:two_circle_joint_kk_spectrum_revision}
\end{equation}
For a cutoff above both spacings, lattice-point counting gives parametrically
\begin{equation}
 {
 \mathcal N_{\rm KK}^{(2)}
 \sim
 \frac{\Lambda^{2}}{M_1M_2},
 }
 \label{eq:two_circle_joint_state_count_revision}
\end{equation}
and hence
\begin{equation}
 {
 \Lambda_{\rm sp}^{4}
 \sim
 M_{\rm Pl}^{2}M_1M_2.
 }
 \label{eq:two_circle_species_scale_revision}
\end{equation}
For equal spacings \(M_1=M_2=M\), this gives \(\Lambda_{\rm sp}^{4}\sim M_{\rm Pl}^{2}M^{2}\), rather than the \(\Lambda_{\rm sp}^{3}\sim M_{\rm Pl}^{2}M\) scaling obtained by treating the two axis towers as independent additive one-dimensional spectra.  The jointly charged states therefore change both the location and the gradient of the species surface.

More generally, if the leading state count in a region of moduli space has the monomial form
\begin{equation}
 {
 \mathcal N_{\rm tot}
 \sim
 C
 \Lambda^{p_{\rm eff}}
 \prod_A
 M_A^{-r_A},
 \qquad
 \sum_A r_A=p_{\rm eff},
 }
 \label{eq:general_joint_species_count_revision}
\end{equation}
then
\begin{equation}
 {
 \Lambda_{\rm sp}^{p_{\rm eff}+2}
 \sim
 \frac{M_{\rm Pl}^{2}}{C}
 \prod_A M_A^{r_A},
 }
 \label{eq:general_joint_species_scale_revision}
\end{equation}
and along a canonical trajectory
\begin{equation}
 {
 -\frac{\mathrm d\ln\Lambda_{\rm sp}}{\mathrm dd}
 =
 \frac{
 \sum_A r_A\mu_A
+
 \mathrm d\ln C/\mathrm dd
 }{
 p_{\rm eff}+2
 }.
 }
 \label{eq:general_joint_species_gradient_revision}
\end{equation}
Here \(C=C(\phi)>0\), and \(p_{\rm eff}\) and \(r_A\) are fixed within the counting regime being differentiated. When the prefactor \(C\) is locally constant, the last term vanishes.  This form includes both a single tower and a jointly charged KK lattice as special cases and makes explicit that the spectrum, rather than a chosen decomposition into ``towers,'' determines the species-wall geometry.

The cutoff/Hubble function and field-space gradient are in all cases
\begin{equation}
 F_{\rm sp/H}
 =
 \ln\!\left(
 \frac{\Lambda_{\rm sp}}{H}
 \right),
 \qquad
 \nabla_iF_{\rm sp/H}
 =
 \nabla_i\ln\Lambda_{\rm sp}
 -
 \nabla_i\ln H.
 \label{eq:combined_species_boundary_gradient}
\end{equation}

A combined species surface is distinct from a genuine multi-boundary problem, in which different microscopic loss mechanisms supply several independent functions \(F_A\).  Joint quantum numbers within one spectrum modify the state count defining a \emph{single} species scale; they should not be represented as several independent absorbing walls.  The restricted additive formula and the independent-boundary case are summarized in \cref{app:boundary_webs}.

\subsection[Circle and large-volume compactification data]{Circle and large-volume compactification}
\label{subsec:circle_lv_benchmarks}

We use two simple compactification limits to fix the canonical distance, tower-mass rate, and state-counting exponent entering the species cutoff. The quantities in this subsection are
four-dimensional Einstein-frame scalings in reduced Planck units. Order-one factors, threshold corrections, and anisotropies can shift their normalizations, but do not alter the local wall law once the corresponding loss surface has been specified~\cite{Klaewer:2016kiy,Grimm:2018ohb,vanBeest:2021lhn,Etheredge:2024tok}. The exponent \(p\) below is the state-counting exponent introduced in \cref{eq:tower_count_general}; it specifies how the number of tower states below a cutoff grows with \(\Lambda/M_{\rm tower}\).

\emph{\(5D\to4D\) circle decompactification.}
Let \(R\) denote the dimensionless Einstein-frame circle radius, normalized to its value at the reference point after the Weyl rescaling required to keep the four-dimensional reduced Planck mass fixed. The dimensionless canonical distance along the radius direction is
\begin{equation}
 d
 =
 \sqrt{\frac32}\ln R.
 \label{eq:kk_canonical_distance}
\end{equation}
The KK tower mass in four-dimensional Planck units scales as
\begin{equation}
 \frac{m_{\rm KK}}{M_{\rm Pl}}
 \sim
 R^{-3/2}
 =
 \exp\!\left[
 -\sqrt{\frac32}\,d
 \right],
 \qquad
 \alpha_{\rm KK}
 =
 \sqrt{\frac32}.
 \label{eq:kk_tower_scaling}
\end{equation}
For a single one-dimensional KK tower, the state-counting exponent is \(p=1\). Equation~\eqref{eq:species_exponent_general} then gives
\begin{equation}
 \lambda_{\rm sp}^{\rm KK}
 =
 \frac{\alpha_{\rm KK}}{3}
 =
 \frac{1}{\sqrt6},
 \qquad
 \frac{\Lambda_{\rm sp}^{\rm KK}}{M_{\rm Pl}}
 \sim
 \exp\!\left(
 -\frac{d}{\sqrt6}
 \right)
 \sim
 R^{-1/2}.
 \label{eq:kk_species_exponent}
\end{equation}
Defining
\begin{equation}
 F_{{\rm sp},0}^{\rm KK}
 \equiv
 \ln\!\left(
 \frac{\Lambda_{{\rm sp},0}^{\rm KK}}{H_0}
 \right),
\end{equation}
the associated candidate species/Hubble input is
\begin{equation}
 F_{\rm sp/H}^{\rm KK}(d)
 =
 F_{{\rm sp},0}^{\rm KK}
 -
 \left(
 \frac{1}{\sqrt6}-\beta
 \right)d,
 \label{eq:kk_species_hubble_boundary}
\end{equation}
where \(\beta\) is the local Hubble rate defined in \cref{eq:local_logarithmic_rates}.

For this \(p=1\) example, the direct KK threshold and the collective species threshold are parametrically different.  A stochastic EFT retaining only the modulus but not the KK tower should normally be terminated when the KK states become dynamically relevant, rather than being extrapolated automatically to the later species/Hubble equality.

\emph{Large-volume Calabi--Yau decompactification.}
For an isotropic large-volume direction, let \(\mathcal V\) denote the dimensionless Einstein-frame internal volume variable used to parameterize the canonical trajectory. We adopt the convention
\begin{equation}
 d
 =
 \sqrt{\frac23}\ln\mathcal V.
 \label{eq:lv_canonical_distance}
\end{equation}
The light KK tower then scales as
\begin{equation}
 \frac{m_{\rm KK}}{M_{\rm Pl}}
 \sim
 \mathcal V^{-2/3}
 =
 \exp\!\left[
 -\sqrt{\frac23}\,d
 \right],
 \qquad
 \alpha_{\rm LV}
 =
 \sqrt{\frac23}.
 \label{eq:lv_tower_scaling}
\end{equation}

For an isotropic six-dimensional KK lattice, the jointly charged lattice states are already included in the leading six-dimensional density of states, giving \(p=6\).  This is a full lattice count rather than the sum of six independent one-dimensional axis towers.

Hence
\begin{equation}
 \lambda_{\rm sp}^{\rm LV}
 =
 \frac{6\alpha_{\rm LV}}{8}
 =
 \sqrt{\frac38},
 \qquad
 \frac{\Lambda_{\rm sp}^{\rm LV}}{M_{\rm Pl}}
 \sim
 \mathcal V^{-1/2}.
 \label{eq:lv_species_exponent}
\end{equation}
This has the same leading large-volume dependence as the usual parametric string scale,
\begin{equation}
 \frac{M_s}{M_{\rm Pl}}
 \sim
 \mathcal V^{-1/2},
 \label{eq:lv_string_scale}
\end{equation}
up to order-one and convention-dependent factors~\cite{Palti:2019pca,vanBeest:2021lhn}. Defining
\begin{equation}
 F_{{\rm sp},0}^{\rm LV}
 \equiv
 \ln\!\left(
 \frac{\Lambda_{{\rm sp},0}^{\rm LV}}{H_0}
 \right),
\end{equation}
the corresponding species/Hubble input is
\begin{equation}
 F_{\rm sp/H}^{\rm LV}(d)
 =
 F_{{\rm sp},0}^{\rm LV}
 -
 \left(
 \sqrt{\frac38}-\beta
 \right)d.
 \label{eq:lv_species_hubble_boundary}
\end{equation}

The rates \(\lambda_{\rm sp}^{\rm KK}=1/\sqrt6\simeq0.408\) and \(\lambda_{\rm sp}^{\rm LV}=\sqrt{3/8}\simeq0.612\) show that, for equal initial cutoff/Hubble ratios and equal \(\beta\), the isotropic large-volume species cutoff decreases faster than the circle species cutoff. In the large-volume limit, the species scale and the direct string scale share the same leading volume dependence. Equality of these logarithmic rates does not imply equality of their microscopic origins or of their order-one threshold normalizations.

\subsection{Direct string and KK threshold data}
\label{subsec:direct_string_kk_data}

Species counting gives the collective nominal condition \(\Lambda_{\rm sp}/H=1\). There are also direct operational thresholds at which the Hubble scale reaches the string scale or the KK scale,
\begin{equation}
 \frac{M_s}{H}=1,
 \qquad
 \frac{M_{\rm KK}}{H}=1.
\end{equation}
Here \(M_{\rm KK}\) denotes the direct compactification threshold, whereas \(\Lambda_{\rm sp}\) denotes the gravitational species cutoff determined by the collective tower count.

These thresholds play different physical roles.  If KK or string states are not retained as stochastic degrees of freedom, the direct threshold can mark the end of validity of the reduced moduli-only process, even when the formal collective species scale remains above \(H\).  The species surface should then not be interpreted as an independently accessible later wall of that same reduced process.

The example below locates the direct string and KK walls in a string-frame volume coordinate \(\mathcal V_s\), converts them into proper distances, and supplies the local boundary forms used in \cref{sec:swampland_layers}.

The string-frame variable \(\mathcal V_s\) used below is the dimensionless internal volume measured in string units. It should not be identified without qualification with the Einstein-frame volume variable \(\mathcal V\) used in \cref{subsec:circle_lv_benchmarks}. In a weakly coupled isotropic string-frame normalization, with \(M_s^2=\alpha'^{-1}\), the four-dimensional Planck scale obeys
\begin{equation}
 M_{\rm Pl}^{2}
 =
 C_{\rm Pl}
 \frac{\mathcal V_s}{g_s^{2}}
 M_s^{2},
 \qquad
 \frac{M_s}{M_{\rm Pl}}
 =
 \frac{g_s}
 {\sqrt{C_{\rm Pl}\mathcal V_s}},
 \qquad
 \frac{M_{\rm KK}}{M_{\rm Pl}}
 \sim
 \frac{g_s}
 {\sqrt{C_{\rm Pl}}\mathcal V_s^{2/3}}.
 \label{eq:direct_string_kk_scales}
\end{equation}
Here \(g_s>0\) is the string coupling and \(C_{\rm Pl}>0\) collects convention-dependent factors in the relation between the string-frame volume, the string scale, and the four-dimensional reduced Planck mass~\cite{Giddings:2001yu,Grana:2005jc,Douglas:2006es,Denef:2007pq,vanBeest:2021lhn,Bedroya:2024uva,Caron-Huot:2024lbf}. The direct nominal hard-wall functions are
\begin{equation}
 F_{s/H}
 \equiv
 \ln\!\left(
 \frac{M_s}{H}
 \right),
 \qquad
 F_{\rm KK/H}
 \equiv
 \ln\!\left(
 \frac{M_{\rm KK}}{H}
 \right).
 \label{eq:string_kk_hubble_boundaries}
\end{equation}

For the numerical example, we take the Hubble scale to be an approximately constant inflationary value \(H_*\). Using the leading slow-roll scalar-amplitude relation
\begin{equation}
 \frac{H_*^{2}}{M_{\rm Pl}^{2}}
 =
 \frac{\pi^{2}}{2}A_s r,
 \label{eq:inflationary_scale_as_r}
\end{equation}
where \(A_s\) is the dimensionless scalar power-spectrum amplitude and \(r\) is the tensor-to-scalar ratio evaluated at the same pivot scale, the controlled-side conditions \(H_*<M_s\) and \(H_*<M_{\rm KK}\) give
\begin{equation}
 \mathcal V_s
 <
 \frac{2g_s^{2}}
 {\pi^{2}C_{\rm Pl}A_s r},
 \qquad
 \mathcal V_s
 <
 \left(
 \frac{2g_s^{2}}
 {\pi^{2}C_{\rm Pl}A_s r}
 \right)^{3/4},
 \label{eq:string_and_kk_cutoff_volume_bounds}
\end{equation}
respectively. Parametric scale separation corresponds to satisfying these inequalities strongly rather than merely approaching saturation.

For a representative weakly coupled choice, take
\begin{equation}
 A_s
 =
 2.1\times10^{-9},
 \qquad
 r
 =
 0.003,
 \qquad
 g_s
 =
 0.1,
 \qquad
 C_{\rm Pl}
 =
 \frac{1}{2\pi}.
\end{equation}
Here \(A_s\) is fixed by the CMB scalar-amplitude normalization~\cite{Planck:2018jri}. The value \(r=0.003\) is a low-tensor inflationary reference value, while \(g_s=0.1\) keeps the compactification in a perturbative regime. The value of \(C_{\rm Pl}\) fixes a convenient normalization convention. These choices determine only the numerical wall locations; the logarithmic rates and the leading local survival law are independent of this normalization. One finds
\begin{equation}
 \frac{H_*}{M_{\rm Pl}}
 \simeq
 5.6\times10^{-6},
 \qquad
 \mathcal V_{b,\rm nom}^{(s)}
 \simeq
 2.0\times10^{9},
 \qquad
 \mathcal V_{b,\rm nom}^{({\rm KK})}
 \simeq
 9.5\times10^{6}.
 \label{eq:direct_cutoff_benchmark_numbers}
\end{equation}
These are the nominal wall locations obtained from \(M_s=H_*\) and \(M_{\rm KK}=H_*\), respectively.

For the conservative surfaces \(M_s/H_*=q_{\rm ctrl}\) and \(M_{\rm KK}/H_*=q_{\rm ctrl}\), the same physical parameters give
\begin{equation}
 \mathcal V_b^{(s)}(q_{\rm ctrl})
 =
 q_{\rm ctrl}^{-2}
 \mathcal V_{b,\rm nom}^{(s)},
 \qquad
 \mathcal V_b^{({\rm KK})}(q_{\rm ctrl})
 =
 q_{\rm ctrl}^{-3/2}
 \mathcal V_{b,\rm nom}^{({\rm KK})}.
 \label{eq:regulated_volume_walls_revision}
\end{equation}
Both walls are thereby moved toward smaller volume, where the cutoff/Hubble hierarchy is safer. Their locations would exchange order only when
\begin{equation}
 q_{\rm ctrl}
 =
 q_{\rm cross}
 \equiv
 \left[
 \frac{
 \mathcal V_{b,\rm nom}^{(s)}
 }{
 \mathcal V_{b,\rm nom}^{({\rm KK})}
 }
 \right]^2
 \simeq
 4.4\times10^{4}.
 \label{eq:string_kk_wall_crossing_qctrl}
\end{equation}
Their ordering is therefore unchanged for conservative control factors of order unity or ten in this example.

The stability of this ordering should not be confused with stability of a first-passage probability.  A survival probability for an initial condition placed very close to one of these surfaces can be highly sensitive to a much smaller change of \(q_{\rm ctrl}\).  Such sensitivity measures dependence on the chosen loss prescription and is examined below with the physical model and initial condition held fixed.

The example should be interpreted conditionally. It combines compactification data that determine the wall geometry with a prescribed inflationary stochastic normalization; it does not derive the modulus potential and its noise matrix from a fully stabilized compactification. A simple consistent interpretation is that \(H_*\) is generated predominantly by a separate adiabatic direction, while the canonical volume direction is a light, weakly mixed spectator or entropic mode. Let \(m_{\perp,\rm eff}\) denote the effective mass of the fluctuation normal to the control surface, let \(\Omega_{\rm turn}\) denote the covariant turn rate of the inflationary background trajectory in field space, and let \({\cal Q}_{\perp}\) be the projected noise factor defined in \cref{eq:normal_noise_projection_revision}. The standard canonical normal-diffusion coefficient requires, at least schematically,
\begin{equation}
 \frac{m_{\perp,\rm eff}^{2}}{H_*^{2}}
 \ll
 1,
 \qquad
 \frac{|\Omega_{\rm turn}|}{H_*}
 \ll
 1,
 \qquad
 {\cal Q}_{\perp}
 \simeq
 1.
 \label{eq:benchmark_lightness_revision}
\end{equation}
If the modulus is heavy or strongly stabilized, \({\cal Q}_{\perp}\), and hence the normal diffusion coefficient \(D_{N,\perp}\), is suppressed.

This reduces the rms stochastic displacement over a specified interval and the magnitude of the conditioned wall response at fixed proper distance. The wall locations derived in this subsection are therefore examples of operational control data, not evidence that the asymptotic volume direction itself supports slow-roll inflation. Likewise, the direct threshold should be understood as a stopping prescription for the retained variables, rather than as a claim that the full theory becomes singular when \(M_{\rm KK}\) or \(M_s\) approaches \(H\).  If the corresponding tower is instead retained dynamically, the stochastic state and its generator must be enlarged across that threshold.

Only the hierarchy and logarithmic rates are used below; the numerical wall locations shift with normalization, \(g_s\), threshold conventions, and anisotropy. The values in \cref{eq:direct_cutoff_benchmark_numbers} specify wall locations, not initial conditions. The corresponding hypersurfaces are treated as absorbing boundaries of the retained ensemble. The controlled side is \(F_A>0\). For a specified direct channel \(A\), this is equivalently \(\mathcal V_s<\mathcal V_b^{(A)}(q_{\rm ctrl})\), and the boundary layer is approached from positive inward distance \(s_A>0\). The associated stochastic scales are computed in \cref{sec:compactification_layer_benchmark}.

For slowly varying \(g_s\), define the dimensionless canonical volume coordinate by
\begin{equation}
 d
 =
 \sqrt{\frac23}\ln\mathcal V_s.
 \label{eq:direct_volume_canonical_distance}
\end{equation}
For a specified value of \(q_{\rm ctrl}\), the inward proper distances to the direct KK and string walls are
\begin{equation}
 s_{\rm KK}
 =
 \sqrt{\frac23}
 \ln\!\left[
 \frac{
 \mathcal V_b^{({\rm KK})}(q_{\rm ctrl})
 }{
 \mathcal V_s
 }
 \right],
 \qquad
 s_s
 =
 \sqrt{\frac23}
 \ln\!\left[
 \frac{
 \mathcal V_b^{(s)}(q_{\rm ctrl})
 }{
 \mathcal V_s
 }
 \right].
 \label{eq:direct_kk_string_boundary_distances}
\end{equation}
The direct string wall and the isotropic large-volume species wall share the same leading volume exponent in this example, but remain distinct operational channels with potentially different normalizations and microscopic interpretations.

Indeed,
\begin{equation}
 \ln M_s
 =
 -\frac12\ln\mathcal V_s+\text{const.},
 \qquad
 \ln M_{\rm KK}
 =
 -\frac23\ln\mathcal V_s+\text{const.},
\end{equation}
and therefore, using Eq.~\eqref{eq:direct_volume_canonical_distance}, the direct threshold rates are
\begin{equation}
 \lambda_s
 =
 \sqrt{\frac38},
 \qquad
 \lambda_{\rm KK}
 =
 \sqrt{\frac23}.
 \label{eq:direct_string_kk_rates}
\end{equation}
At fixed \(H_*\), the conservative cutoff functions can be written as
\begin{align}
 F_{\rm KK/H}^{(q_{\rm ctrl})}
 &=
 \ln\!\left(
 \frac{M_{\rm KK}}{q_{\rm ctrl}H_*}
 \right)
 =
 \frac23
 \ln\!\left[
 \frac{
 \mathcal V_b^{({\rm KK})}(q_{\rm ctrl})
 }{
 \mathcal V_s
 }
 \right]
 =
 \lambda_{\rm KK}s_{\rm KK},
 \label{eq:direct_kk_qctrl_boundary_form}
 \\
 F_{s/H}^{(q_{\rm ctrl})}
 &=
 \ln\!\left(
 \frac{M_s}{q_{\rm ctrl}H_*}
 \right)
 =
 \frac12
 \ln\!\left[
 \frac{
 \mathcal V_b^{(s)}(q_{\rm ctrl})
 }{
 \mathcal V_s
 }
 \right]
 =
 \lambda_s s_s.
 \label{eq:direct_string_qctrl_boundary_form}
\end{align}
Thus the boundary function \(F_A\) and the inward proper distance \(s_A\) encode the same leading local boundary data, while the survival response itself is computed in \cref{sec:swampland_layers}.

The physical ordering is therefore explicit.  Compactification data first determine the spectrum and candidate scales \(M_{\rm KK}\), \(M_s\), and \(\Lambda_{\rm sp}\).  The chosen stochastic EFT then determines which of these surfaces is actually reachable before its retained degrees of freedom cease to provide a controlled description. Only a surface lying within that common domain is promoted to an absorbing first-passage boundary of the reduced process.

\section[Stochastic Survival near EFT-Loss Boundaries]{Stochastic Survival near EFT-Loss Boundaries}
\label{sec:swampland_layers}

We now apply the survival map to the boundary data constructed above. For a regular hard-wall input, compactification physics fixes the loss function \(F_A\), while the stochastic problem determines the survival probability \(h_A\), the survival action \(\mathcal S_A=-\ln h_A\), and the Doob-transformed response \(\Delta b_{\rm Doob}\). Soft killing, explicitly time-dependent controlled domains, and other nonautonomous finite-horizon variants are discussed separately; here we focus on the local hard-wall layer, its finite-horizon normal benchmark, and the diagnostics induced by conditioning on survival.

\subsection[Local response map and one-e-fold stochastic scale]{Local response map and one-e-fold stochastic scale}
\label{subsec:local_response_resolution}

Let a hard EFT-loss channel \(A\) be described by
\begin{equation}
 F_A(\phi)>0 \quad\text{on the controlled side}, \qquad F_A(\phi)=0 \quad\text{on the loss surface}.
 \label{eq:hard_channel_definition_section4}
\end{equation}
The field-space norm of its gradient is
\begin{equation}
 |\nabla F_A| \equiv \sqrt{ G^{ij}\nabla_iF_A\nabla_jF_A }.
 \label{eq:channel_gradient_norm_section4}
\end{equation}
At a regular point of the wall,
\begin{equation}
 \left. |\nabla F_A| \right|_{\partial\Omega_A} \neq0,
\end{equation}
and the inward-pointing unit normal covector is
\begin{equation}
 n_i^{(A)} \equiv \left. \frac{\nabla_iF_A}{|\nabla F_A|} \right|_{\partial\Omega_A}.
 \label{eq:channel_normal_definition}
\end{equation}
The orientation is inward because \(F_A\) increases into the controlled domain. The corresponding inward proper distance satisfies
\begin{equation}
 s_A = \frac{ F_A }{ |\nabla F_A|_{\partial\Omega_A} } + \mathcal O(F_A^2).
 \label{eq:master_proper_distance}
\end{equation}
Here and below, the subscript \(\partial\Omega_A\) means that the smooth prefactor is evaluated at the nearest regular point of the wall.

The normal diffusion coefficient is
\begin{equation}
 D_{\perp}^{(A)} \equiv n_i^{(A)}D^{ij}n_j^{(A)}.
 \label{eq:channel_normal_diffusion_section4}
\end{equation}
When no confusion can arise, we suppress the channel label and write \(D_\perp\equiv D_\perp^{(A)}\). In the driftless local normal benchmark, and to leading order in a region where the wall is locally flat and \(D_\perp\) may be treated as constant, the finite-horizon survival probability is
\begin{equation}
 h_A^{(0)}(\phi,\tau) \simeq \operatorname{erf}\!\left[ \frac{s_A}{2\sqrt{D_\perp\tau}} \right].
 \label{eq:master_survival_probability}
\end{equation}
The superscript \((0)\) denotes the driftless constant-diffusion normal benchmark, and \(\tau\geq0\) is the remaining conditioning horizon measured in the stochastic time variable used by the generator.

The same solution defines an operational control margin. To retain at least a fraction \(p\in(0,1)\) of histories until the remaining horizon \(\tau\), one needs
\begin{equation}
 h_A^{(0)}(s_A,\tau) \geq p
 \qquad\Longleftrightarrow\qquad
 s_A \geq s_{A,p}^{(0)}(\tau)
 \equiv
 2\sqrt{D_\perp\tau}\,\operatorname{erf}^{-1}(p).
 \label{eq:channel_survival_margin_revision}
\end{equation}
Using Eq.~\eqref{eq:master_proper_distance}, this becomes the leading local boundary-function condition
\begin{equation}
 F_A(\phi) \gtrsim
 2|\nabla F_A|_{\partial\Omega_A}
 \sqrt{D_\perp\tau}\,
 \operatorname{erf}^{-1}(p).
 \label{eq:channel_control_margin_revision}
\end{equation}
The symbol \(\gtrsim\) emphasizes that the conversion from \(F_A\) to \(s_A\) retains only the leading regular-wall term. Unlike the location \(F_A=0\), which is supplied by microscopic control data, Eq.~\eqref{eq:channel_control_margin_revision} is a stochastic finite-time margin: it states how far inside the controlled domain an initial condition must lie to achieve a chosen reliability level over a specified duration. A nonzero normal drift, a moving wall, or state-dependent diffusion replaces this closed form by the corresponding backward boundary-value problem.

In particular, the finite-horizon margin in Eq.~\eqref{eq:channel_control_margin_revision} assumes that the loss surface is stationary in the stochastic coordinates used to formulate the normal problem. If the microscopic control condition depends explicitly on the background evolution, the boundary motion must be included whenever its displacement is not negligible compared with the field-space scales explored by the stochastic process over the same horizon.

The compactification data enter the local benchmark through the proper distance \(s_A\), while the additional stochastic inputs are \(D_\perp\) and \(\tau\). The driftless benchmark survival action is
\begin{equation}
 \mathcal S_A^{(0)}(\phi,\tau) \equiv -\ln h_A^{(0)}(\phi,\tau).
 \label{eq:local_survival_action_section4}
\end{equation}
The Doob transform gives the normal conditioned response
\begin{equation}
 \Delta b_{{\rm Doob},\perp}^{(A)}
 =
 -2D_\perp\partial_{s_A}\mathcal S_A
 =
 2D_\perp\partial_{s_A}\ln h_A,
 \label{eq:local_doob_from_survival_probability}
\end{equation}
where the formula applies to the relevant survival probability, not only to the driftless benchmark. In the near-wall regime
\begin{equation}
 s_A \ll \sqrt{D_\perp\tau},
\end{equation}
the driftless finite-horizon solution, and more generally any regular absorbing-wall solution with nonzero normal diffusion, gives
\begin{equation}
 \Delta b_{{\rm Doob},\perp}^{(A)}
 \simeq
 \frac{2D_\perp}{s_A}.
 \label{eq:master_normal_response}
\end{equation}
Equivalently, in covariant boundary-function form,
\begin{equation}
 \Delta b_{\rm Doob}^{i}
 \simeq
 2D^{ij}\frac{\nabla_jF_A}{F_A}.
 \label{eq:master_doob_response}
\end{equation}
Thus, once \(F_A\) and \(D^{ij}\) are supplied, the leading local singular response is fixed. The microscopic data determine the location and orientation of the wall, while the diffusion tensor determines the direction \(D^{ij}\nabla_jF_A\) of the singular drift. Its inward normal projection is universally \(2D_\perp/s_A\) for a regular hard wall with nondegenerate normal diffusion.

Suppose that the unconditioned drift points toward the wall. Its inward normal component can then be written as
\begin{equation}
 b_{\perp}^{\rm cl}
 \equiv
 n_i^{(A)}b^i
 =
 -\mu,
 \qquad
 \mu>0.
 \label{eq:classical_boundary_speed_section4}
\end{equation}
The distance at which the near-wall Doob response matches the magnitude of this boundary-directed drift is
\begin{equation}
 s_{\rm Doob}
 \equiv
 \frac{2D_\perp}{\mu}.
 \label{eq:sdoob_definition_section4}
\end{equation}
It is useful to define the local competition ratio
\begin{equation}
 R_{\rm Doob}(s_A)
 \equiv
 \frac{\Delta b_{{\rm Doob},\perp}^{(A)}}{|b_\perp^{\rm cl}|}
 \simeq
 \frac{2D_\perp}{\mu s_A}
 =
 \frac{s_{\rm Doob}}{s_A}.
 \label{eq:doob_competition_ratio}
\end{equation}
The conditioned response is competitive with or larger than the unconditioned normal drift when
\begin{equation}
 R_{\rm Doob}\gtrsim1,
 \qquad\text{equivalently}\qquad
 s_A\lesssim s_{\rm Doob}.
 \label{eq:doob_competition_condition}
\end{equation}

The following normalization is a specialization of the general survival map, not a condition for the validity of the map itself. In stochastic slow roll, using e-fold time \(N=\ln a\), reduced Planck units \(M_{\rm Pl}=1\), and \(3H^2\simeq V\), the drift and diffusion tensors are
\begin{equation}
 b_{\rm SR}^{i}
 \simeq
 -G^{ij}\nabla_j\ln V,
 \qquad
 D_{\rm SR}^{ij}
 \simeq
 \frac{H^2}{8\pi^2}{\cal Q}^{ij}.
 \label{eq:slowroll_noise_general_revision}
\end{equation}
Here \({\cal Q}^{ij}\) is the noise tensor introduced in \cref{eq:noise_matrix_revision}; the light, weakly mixed canonical limit is \({\cal Q}^{ij}=G^{ij}\). Define
\begin{equation}
 \partial_\perp
 \equiv
 n^{i}_{(A)}\nabla_i,
 \qquad
 n^i_{(A)}
 \equiv
 G^{ij}n_j^{(A)},
\end{equation}
and
\begin{equation}
 {\cal Q}_\perp
 \equiv
 n_i^{(A)}{\cal Q}^{ij}n_j^{(A)}.
\end{equation}
Then
\begin{equation}
 D_\perp
 \simeq
 \frac{H^2}{8\pi^2}{\cal Q}_\perp,
 \qquad
 \mu
 \simeq
 |b_{{\rm SR},\perp}|
 \simeq
 |\partial_\perp\ln V|,
 \qquad
 \delta\phi_{q,\perp}
 \equiv
 \sqrt{2D_\perp}
 \simeq
 \frac{H}{2\pi}\sqrt{{\cal Q}_\perp}.
 \label{eq:normal_noise_general_revision}
\end{equation}
The quantity \(\delta\phi_{q,\perp}\) is the root-mean-square coarse-grained stochastic displacement in the normal direction per e-fold.

It is used below only as a convenient one-e-fold comparison scale; it is not a fundamental field-space resolution limit. Within the white-noise Markov approximation, the rms displacement generated by diffusion over an interval
\(\Delta N\) is
\begin{equation}
 \delta\phi_{\perp}(\Delta N)
 =
 \sqrt{2D_\perp\Delta N}
 =
 \delta\phi_{q,\perp}\sqrt{\Delta N}.
 \label{eq:short_time_stochastic_displacement}
\end{equation}
Accordingly, the continuum stochastic process itself remains defined at distances smaller than \(\delta\phi_{q,\perp}\). If the coarse-graining prescription produces a finite noise-correlation time \(\tau_c\), the corresponding stochastic displacement scale is parametrically
\begin{equation}
 \delta\phi_{\perp,c}
 \sim
 \sqrt{2D_\perp\tau_c}.
 \label{eq:correlation_time_field_scale}
\end{equation}
This estimate presupposes that a diffusive approximation remains adequate on the correlation interval; it is not an exact short-time colored-noise variance. The validity of a sharp Markovian absorbing-boundary description must then be assessed relative to the coarse-graining prescription and to the field-space width over which EFT control is actually lost. Smooth coarse-graining windows can, in particular, generate temporally correlated rather than strictly white noise~\cite{Mahbub:2022osb}. We therefore do not interpret the one-e-fold displacement as a microscopic cutoff on the stochastic description.

The competition distance is therefore
\begin{equation}
 s_{\rm Doob}
 =
 \frac{2D_\perp}{\mu}
 \simeq
 \frac{H^2{\cal Q}_\perp}
 {4\pi^2|\partial_\perp\ln V|}.
 \label{eq:sdoob_noise_general_revision}
\end{equation}
Similarly, a target positive normal drift increment \(X>0\) is reached, within the near-wall approximation, at
\begin{equation}
 s_X
 \equiv
 \frac{2D_\perp}{X}
 \simeq
 \frac{H^2{\cal Q}_\perp}{4\pi^2X}.
 \label{eq:sx_noise_general_revision}
\end{equation}
Relative to one normal stochastic kick,
\begin{equation}
 \frac{s_X}{\delta\phi_{q,\perp}}
 \simeq
 \frac{H\sqrt{{\cal Q}_\perp}}{2\pi X},
 \qquad
 \frac{s_{\rm Doob}}{\delta\phi_{q,\perp}}
 \simeq
 \frac{H\sqrt{{\cal Q}_\perp}}
 {2\pi|\partial_\perp\ln V|}.
 \label{eq:resolved_scales_noise_general_revision}
\end{equation}

Without assigning \(\delta\phi_{q,\perp}\) the status of a resolution limit, one may instead ask whether the region in which the conditioned response competes with the unconditioned drift extends to distances at least as large as the rms stochastic displacement accumulated over one e-fold. This occurs when
\begin{equation}
 s_{\rm Doob}
 \gtrsim
 \delta\phi_{q,\perp},
 \label{eq:one_efold_competition_comparison}
\end{equation}
or equivalently,
\begin{equation}
 |\partial_\perp\ln V|
 \lesssim
 \frac{H}{2\pi}\sqrt{{\cal Q}_\perp}.
 \label{eq:stochastic_threshold_noise_general_revision}
\end{equation}
This comparison concerns the relative sizes of the classical and one-e-fold stochastic displacements. It does not imply that distances \(s_A<\delta\phi_{q,\perp}\) are unresolved by the continuum stochastic process.

For \({\cal Q}_\perp=1\), these expressions reduce to the canonical formulas used in the numerical benchmark below. If the normal mode is massive or strongly mixed, the noise matrix must be computed rather than set to unity, and the corresponding stochastic displacement scales inherit the corresponding \({\cal Q}_\perp\) dependence.

Equation~\eqref{eq:stochastic_threshold_noise_general_revision} is the normal-direction diffusion-dominance criterion. It coincides with the familiar local stochastic-inflation or eternal-inflation criterion only when the normal direction is the relevant adiabatic inflationary direction. For a spectator or entropic modulus, it instead states that the stochastic displacement in that direction is at least as large as its classical displacement per e-fold. Explicitly,
\begin{equation}
 \delta\phi_{{\rm cl},\perp}
 \equiv
 |b_{{\rm SR},\perp}|
 \simeq
 |\partial_\perp\ln V|,
 \qquad
 \delta\phi_{q,\perp}
 \simeq
 \frac{H}{2\pi}\sqrt{{\cal Q}_\perp},
 \label{eq:classical_quantum_normal_steps}
\end{equation}
so that
\begin{equation}
 |\partial_\perp\ln V|
 \lesssim
 \frac{H}{2\pi}\sqrt{{\cal Q}_\perp}
 \qquad\Longleftrightarrow\qquad
 \delta\phi_{q,\perp}
 \gtrsim
 \delta\phi_{{\rm cl},\perp}.
 \label{eq:normal_diffusion_dominance}
\end{equation}
For the adiabatic inflationary direction, this reproduces the usual local stochastic threshold~\cite{Starobinsky:1986fx,Starobinsky:1994bd,Vennin:2015hra,Assadullahi:2016gkk}.

At a proper distance equal to the one-e-fold rms stochastic displacement,
\begin{equation}
 s_A
 \sim
 \delta\phi_{q,\perp}
 \qquad\Longrightarrow\qquad
 \left|\Delta b_{{\rm Doob},\perp}^{(A)}\right|
 \sim
 \delta\phi_{q,\perp}.
 \label{eq:kick_sized_doob_response}
\end{equation}
This is a comparison of two one-e-fold scales: at a wall distance equal to one rms stochastic displacement, the singular conditioned drift also produces an increment of that order. It does not identify \(\delta\phi_{q,\perp}\) with a lower field-space resolution. By contrast, an order-one target \(X=\mathcal O(1)\) lies at
\begin{equation}
 \frac{s_X}{\delta\phi_{q,\perp}}
 \sim
 \frac{H\sqrt{{\cal Q}_\perp}}{2\pi},
\end{equation}
which is sub-kick for semiclassical \(H\ll1\) and \({\cal Q}_\perp=\mathcal O(1)\). This separation shows that boundary distances comparable with ordinary one-e-fold stochastic displacements and the much thinner region associated with an order-one drift increment are parametrically distinct notions.

\subsection[Stochastic scales near direct KK/string walls]{Stochastic scales near direct KK/string walls}
\label{sec:compactification_layer_benchmark}

The direct threshold data in \cref{eq:string_kk_hubble_boundaries,eq:direct_kk_string_boundary_distances,eq:direct_kk_qctrl_boundary_form,eq:direct_string_qctrl_boundary_form} provide a concrete hard-wall input. For the numerical example we use reduced Planck units and adopt the light canonical-noise limit
\begin{equation}
 {\cal Q}_\perp=1.
 \label{eq:benchmark_canonical_noise_choice}
\end{equation}
For the values quoted in \cref{eq:direct_cutoff_benchmark_numbers}, the inflationary and coarse-grained stochastic scales are
\begin{equation}
 \frac{H_*}{M_{\rm Pl}}
 \simeq
 5.6\times10^{-6},
 \qquad
 D_\perp
 =
 \frac{H_*^2}{8\pi^2}
 \simeq
 3.9\times10^{-13},
 \qquad
 \delta\phi_q
 \equiv
 \sqrt{2D_\perp}
 =
 \frac{H_*}{2\pi}
 \simeq
 8.9\times10^{-7}.
 \label{eq:benchmark_stochastic_scales}
\end{equation}
The last two numerical expressions use \(M_{\rm Pl}=1\). The corresponding one-e-fold rms stochastic displacement is therefore extremely small in Planck units. It is natural to introduce the kick-normalized boundary distance
\begin{equation}
 u_A
 \equiv
 \frac{s_A}{\delta\phi_q}.
 \label{eq:compactification_kick_distance}
\end{equation}

We retain the control factor \(q_{\rm ctrl}\) introduced in \cref{sec:survival_channels_cutoff_walls}. For the direct KK channel,
\begin{equation}
 F_{{\rm KK}/H}^{(q_{\rm ctrl})}
 =
 \frac23
 \ln\!\left[
 \frac{\mathcal V_b^{({\rm KK})}(q_{\rm ctrl})}{\mathcal V_s}
 \right]
 =
 \lambda_{\rm KK}s_{\rm KK},
 \qquad
 \lambda_{\rm KK}
 =
 \sqrt{\frac23}.
 \label{eq:kk_boundary_to_distance}
\end{equation}
It follows that
\begin{equation}
 s_{\rm KK}
 =
 \frac{F_{{\rm KK}/H}^{(q_{\rm ctrl})}}{\lambda_{\rm KK}}
 =
 \sqrt{\frac23}
 \ln\!\left[
 \frac{\mathcal V_b^{({\rm KK})}(q_{\rm ctrl})}{\mathcal V_s}
 \right],
 \label{eq:kk_distance_from_volume_section4}
\end{equation}
and
\begin{equation}
 u_{\rm KK}(\mathcal V_s)
 =
 \frac{1}{\delta\phi_q}
 \sqrt{\frac23}
 \ln\!\left[
 \frac{\mathcal V_b^{({\rm KK})}(q_{\rm ctrl})}{\mathcal V_s}
 \right].
 \label{eq:ukk_from_volume}
\end{equation}
The variable \(u_A\) is therefore only a dimensionless normalization of the boundary distance by the one-e-fold rms stochastic displacement. It does not introduce a physical short-distance cutoff: values \(u_A<1\) remain part of the continuum white-noise first-passage problem whenever that Markovian description and the sharp-wall approximation remain applicable.

Using
\begin{equation}
 \delta\phi_q^2=2D_\perp,
\end{equation}
the driftless local half-line survival probability becomes
\begin{equation}
 h_A^{(0)}(\mathcal V_s,\tau)
 \simeq
 \operatorname{erf}\!\left[
 \frac{s_A(\mathcal V_s)}{2\sqrt{D_\perp\tau}}
 \right]
 =
 \operatorname{erf}\!\left[
 \frac{u_A(\mathcal V_s)}{\sqrt{2\tau}}
 \right],
 \qquad
 \mathcal S_A^{(0)}
 =
 -\ln h_A^{(0)}.
 \label{eq:compactification_halfline_survival}
\end{equation}
Equivalently, in kick units the normal diffusion coefficient is
\begin{equation}
 D_u
 \equiv
 \frac{D_\perp}{\delta\phi_q^2}
 =
 \frac12,
 \label{eq:kick_unit_diffusion}
\end{equation}
and the finite-horizon normal problem takes the scale-independent form
\begin{equation}
 h_A^{(0)}(u_A,\tau)
 =
 \operatorname{erf}\!\left(
 \frac{u_A}{\sqrt{2\tau}}
 \right).
 \label{eq:survival_in_kick_units}
\end{equation}
Here \(\tau\) is the remaining conditioning time measured in e-folds. The finite-horizon curves below describe a stationary auxiliary half-line process. The separate \(r=0.003\) normalization does not justify holding its KK/Hubble surface fixed for 60 or 200 e-folds; the quantitative wall-motion test in \cref{subsec:two_field_survival_benchmark} applies.

Fixed-survival contours obey
\begin{equation}
 u_{A,p}(\tau)
 =
 \sqrt{2\tau}\,
 \operatorname{erf}^{-1}(p),
 \qquad
 p\in(0,1),
 \label{eq:fixed_survival_contours}
\end{equation}
so a longer horizon requires a larger kick-normalized distance from the wall at fixed target survival probability.

For a representative near-wall point measured in one-e-fold stochastic units, take
\begin{equation}
 u_{\rm KK}=10.
\end{equation}
In the compactification coordinate this corresponds to
\begin{equation}
 \mathcal V_s
 =
 \mathcal V_b^{({\rm KK})}(q_{\rm ctrl})
 \exp\!\left[
 -\frac{u_{\rm KK}\delta\phi_q}{\sqrt{2/3}}
 \right].
 \label{eq:u10_volume_distance_exact}
\end{equation}
Using the benchmark value of \(\delta\phi_q\),
\begin{equation}
 \mathcal V_s
 \simeq
 \mathcal V_b^{({\rm KK})}(q_{\rm ctrl})
 \left(1-1.1\times10^{-5}\right),
 \qquad
 u_{\rm KK}=10.
 \label{eq:u10_volume_distance}
\end{equation}
Thus even a distance ten times the one-e-fold rms stochastic displacement corresponds to an extremely narrow interval in the logarithmic compactification coordinate. For a remaining horizon \(\tau=200\), the same point has
\begin{equation}
 h^{(0)}(u_0=10,\tau=200)
 =
 \operatorname{erf}\!\left(
 \frac{10}{\sqrt{400}}
 \right)
 =
 \operatorname{erf}\!\left(\frac12\right)
 \simeq
 0.52.
 \label{eq:benchmark_survival_u10_200}
\end{equation}

Differentiating Eq.~\eqref{eq:survival_in_kick_units} gives the finite-horizon Doob response in kick units,
\begin{equation}
 \frac{\Delta b_{{\rm Doob},\perp}^{(A)}}{\delta\phi_q}
 =
 \sqrt{\frac{2}{\pi\tau}}\,
 \frac{\exp[-u_A^2/(2\tau)]}
 {\operatorname{erf}[u_A/\sqrt{2\tau}]}.
 \label{eq:finite_horizon_doob_kick_units}
\end{equation}
For
\begin{equation}
 u_A\ll\sqrt{\tau},
\end{equation}
this reduces to
\begin{equation}
 \frac{\Delta b_{{\rm Doob},\perp}^{(A)}}{\delta\phi_q}
 \simeq
 \frac1{u_A}.
 \label{eq:finite_horizon_doob_kick_wall_limit}
\end{equation}
The value \(u_A=1\) simply denotes a wall distance equal to the rms stochastic displacement accumulated over one e-fold. It is not a resolution threshold of the white-noise process. The point \(u_A=10\) therefore represents a wall distance ten times this convenient comparison scale, whose survival probability nevertheless becomes nontrivial over a sufficiently long horizon. Whether a sharp absorbing wall remains a faithful physical description at smaller \(u_A\) depends instead on the coarse-graining correlation scale and on the microscopic width of the transition out of EFT control.

\begin{figure}[!htb]
 \centering
 \includegraphics[width=\textwidth]{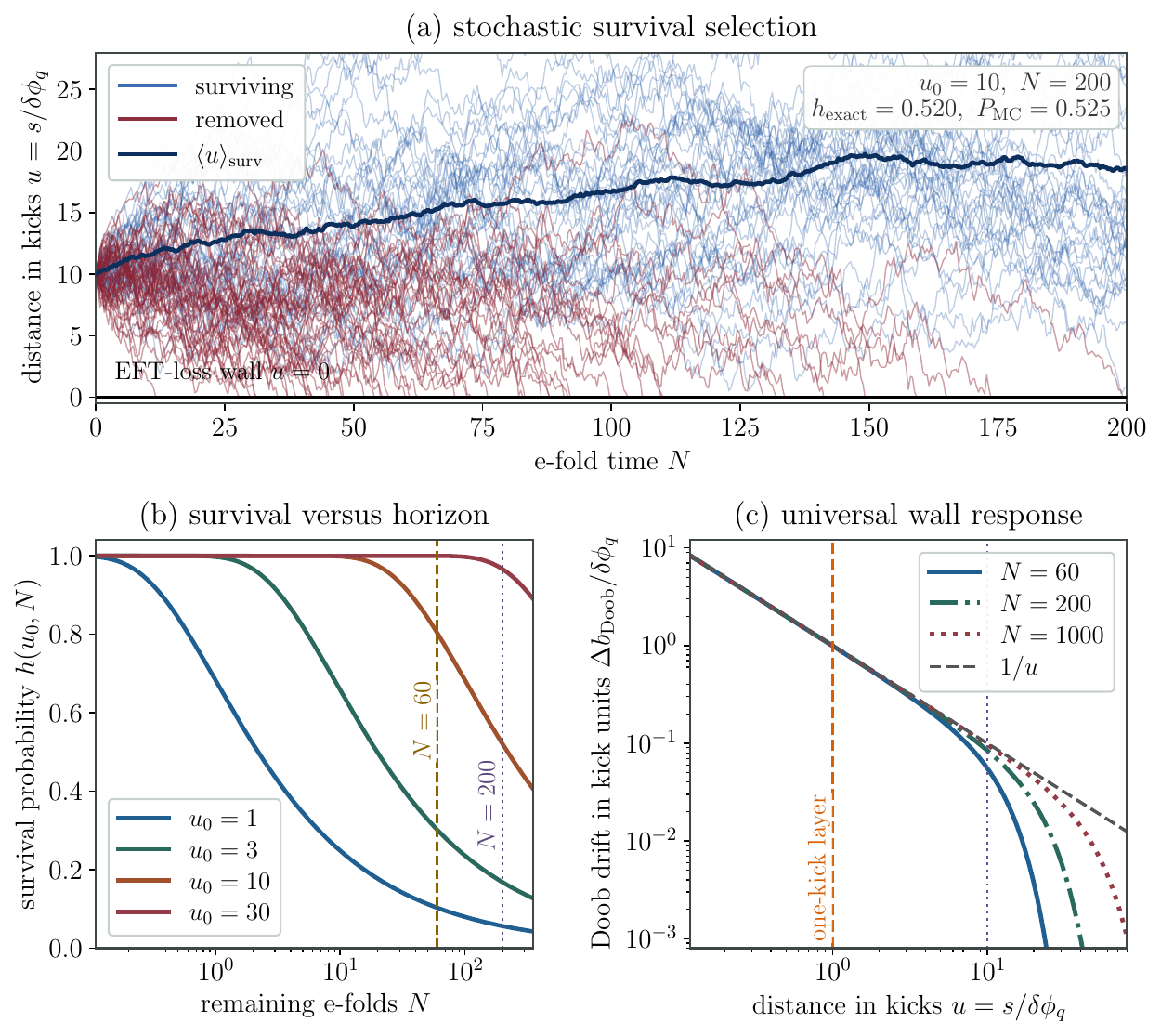}
 \caption{Finite-horizon survival and conditioned drift near a hard EFT-loss wall. Panel~(a) shows a killed half-line ensemble, panel~(b) the finite-horizon survival probability, and panel~(c) the conditioned normal drift compared with the universal near-wall law. Distances are expressed in units of the one-e-fold rms stochastic displacement; this normalization is a comparison scale rather than a physical resolution limit.} The example uses \((u_0,\tau)=(10,200)\) and the light canonical-noise limit \({\cal Q}_\perp=1\).
 \label{fig:localDoobResponseKK}
\end{figure}

\Cref{fig:localDoobResponseKK} illustrates this kick-normalized description. Panel~(a) visualizes the killed diffusion for the benchmark point \((u_0,\tau)=(10,200)\); panel~(b) shows the analytic survival probability
\[
 h^{(0)}(u_0,\tau)
 =
 \operatorname{erf}\!\left[
 \frac{u_0}{\sqrt{2\tau}}
 \right]
\]
for several initial distances; and panel~(c) compares the finite-horizon Doob response with the universal wall law \(1/u\). Agreement near the wall is the universal local contribution, while departure at larger \(u\) records finite-horizon survival data.

To keep track of the order-one normalizations implicit in the scaling relations, write
\begin{equation}
 M_s
 =
 C_s\,g_s\,\mathcal V_s^{-1/2}M_{\rm Pl},
 \qquad
 M_{\rm KK}
 =
 C_{\rm KK}\,g_s\,\mathcal V_s^{-2/3}M_{\rm Pl},
 \label{eq:direct_scale_normalization_constants}
\end{equation}
where \(C_s\) and \(C_{\rm KK}\) are positive convention- and geometry-dependent coefficients. For a common value of \(q_{\rm ctrl}\), the direct KK and string functions then satisfy
\begin{equation}
 F_{{\rm KK}/H}^{(q_{\rm ctrl})}
 -
 F_{s/H}^{(q_{\rm ctrl})}
 =
 -\frac16\ln\mathcal V_s
 +
 \ln\!\left(\frac{C_{\rm KK}}{C_s}\right).
 \label{eq:direct_kk_string_difference}
\end{equation}
The \(-\ln q_{\rm ctrl}\) shifts cancel between the two channels. In the equal-normalization convention used in the numerical benchmark, \(C_{\rm KK}=C_s\), and, more generally, parametrically at sufficiently large volume for order-one coefficients, the direct KK wall is encountered before the direct string wall. The precise finite-volume ordering can be shifted by threshold normalizations. Compactification data determine which microscopic threshold is encountered first and where the wall lies; stochastic data determine how the survival layer near that wall is resolved.

An order-one normal drift increment probes a parametrically thinner region. For \(X=\mathcal O(1)\),
\begin{equation}
 u_X
 \equiv
 \frac{s_X}{\delta\phi_q}
 =
 \frac{\delta\phi_q}{X}
 \sim
 10^{-6}.
 \label{eq:order_one_target_subkick}
\end{equation}
Thus the operational wall is fixed by \(F_A^{(q_{\rm ctrl})}=0\), while \(u_A=\mathcal O(1)\) merely identifies distances comparable with the one-e-fold stochastic displacement. Order-one gradient-like drift targets occur at parametrically smaller distances for semiclassical \(H\ll1\). Whether those distances remain within the physical domain of a Markovian sharp-wall description is a separate coarse-graining question.

\subsection[Light spectator with a field-dependent KK threshold]{Light spectator with a field-dependent KK threshold}
\label{subsec:two_field_survival_benchmark}

The compactification examples above identify candidate control surfaces and their local normal geometry, but they do not evolve an explicit inflationary path ensemble. We now close this gap with a minimal two-field model in which one scalar supports the inflationary background while a light spectator controls a field-dependent KK threshold. The construction is not intended as a fully stabilized compactification. It isolates a specific physical question: given a light stabilized spectator and a prescribed KK/Hubble loss surface, what fraction of stochastic histories retain that EFT description for a specified duration, when do the lost histories first reach the surface, and how is the surviving ensemble statistically biased? The resulting calculation realizes the chain
\begin{equation}
 \text{inflationary generator}
 \longrightarrow
 \text{KK control surface}
 \longrightarrow
 h
 \longrightarrow
 f_{\rm exit}
 \longrightarrow
 \text{surviving-ensemble bias}.
 \label{eq:two_field_operational_chain}
\end{equation}

We use reduced Planck units in the dynamical equations and consider two canonically normalized fields: an adiabatic field \(\phi\) and a spectator modulus \(\chi\), with potential
\begin{equation}
 V(\phi,\chi)
 =
 V_{\rm inf}(\phi)
 +
 \frac12m_\chi^2(\chi-\chi_0)^2.
 \label{eq:two_field_benchmark_potential}
\end{equation}
The inflationary energy density is assumed to be dominated by \(V_{\rm inf}\), and the spectator contribution is taken to remain subleading over the interval of interest. We therefore approximate
\begin{equation}
 H^2(\phi,\chi)
 \simeq
 \frac{V_{\rm inf}(\phi)}{3}
 \simeq
 H_*^2.
 \label{eq:two_field_quasi_constant_hubble}
\end{equation}

The stationary-wall calculation below is defined in the limit in which the evolution of \(H\) is negligible on the field-space scales relevant to first passage. This requirement is stronger than merely demanding a small fractional change in \(H\). If \(H=H(N)\), the KK wall is
\begin{equation}
 \chi_b(N)
 =
 \chi_0
 +
 \frac{1}{\alpha_\chi}
 \ln\!\left[
 \frac{M_{{\rm KK},0}}{q_{\rm ctrl}H(N)}
 \right],
 \label{eq:two_field_moving_wall}
\end{equation}
and therefore
\begin{equation}
 \frac{\mathrm d\chi_b}{\mathrm dN}
 =
 \frac{\epsilon_H}{\alpha_\chi},
 \qquad
 \epsilon_H
 \equiv
 -\frac{\mathrm d\ln H}{\mathrm dN}.
 \label{eq:two_field_wall_velocity}
\end{equation}
The inward distance \(s=\chi_b-\chi\) consequently acquires the additional drift \(+\epsilon_H/\alpha_\chi\). Over a finite interval,
\begin{equation}
 \Delta\chi_b(N)
 =
 \frac{1}{\alpha_\chi}
 \int_0^N\epsilon_H(N')\,\mathrm dN'.
 \label{eq:two_field_wall_displacement}
\end{equation}
A conservative condition for treating the wall as stationary over a horizon \(N\) is
\begin{equation}
 |\Delta\chi_b(N)|
 \ll
 \min\!\left[
 s_0,\,
 \sqrt{2D_\chi N}
 \right],
 \label{eq:two_field_stationary_wall_condition}
\end{equation}
where \(s_0=\chi_b(0)-\chi_{\rm ini}\). The numerical probabilities quoted below are therefore results of this explicitly stationary-wall model and should not be applied to a background that violates Eq.~\eqref{eq:two_field_stationary_wall_condition}.

For a light, weakly mixed spectator,
\begin{equation}
 \frac{m_\chi^2}{H_*^2}<1,
 \qquad
 {\cal Q}_\chi\simeq1,
 \label{eq:two_field_light_spectator_conditions}
\end{equation}
the coarse-grained slow-roll dynamics in e-fold time is
\begin{equation}
 \mathrm d\chi
 =
 -\nu_\chi(\chi-\chi_0)\mathrm dN
 +
 \sqrt{2D_\chi}\,\mathrm dW_N,
 \label{eq:two_field_chi_sde}
\end{equation}
where
\begin{equation}
 \nu_\chi
 \equiv
 \frac{m_\chi^2}{3H_*^2},
 \qquad
 D_\chi
 \equiv
 \frac{H_*^2}{8\pi^2}{\cal Q}_\chi,
 \label{eq:two_field_chi_parameters}
\end{equation}
and
\begin{equation}
 \left\langle\mathrm dW_N\right\rangle=0,
 \qquad
 \left\langle\mathrm dW_N^2\right\rangle=\mathrm dN.
 \label{eq:two_field_wiener_normalization}
\end{equation}
The deterministic drift restores \(\chi\) toward \(\chi_0\), while the stochastic term allows histories to explore the side of field space on which the KK threshold decreases.

We model the direct KK scale by
\begin{equation}
 M_{\rm KK}(\chi)
 =
 M_{{\rm KK},0}
 \exp\!\left[
 -\alpha_\chi(\chi-\chi_0)
 \right],
 \qquad
 \alpha_\chi>0.
 \label{eq:two_field_kk_profile}
\end{equation}
For a conservative control factor \(q_{\rm ctrl}\geq1\), the operational control function is
\begin{equation}
 F_{\rm KK}^{(q_{\rm ctrl})}(\phi,\chi)
 \equiv
 \ln\!\left[
 \frac{M_{\rm KK}(\chi)}
 {q_{\rm ctrl}H(\phi,\chi)}
 \right].
 \label{eq:two_field_control_function}
\end{equation}
The controlled domain is
\begin{equation}
 F_{\rm KK}^{(q_{\rm ctrl})}>0,
 \qquad
 M_{\rm KK}>q_{\rm ctrl}H,
 \label{eq:two_field_controlled_domain}
\end{equation}
and the selected loss surface is \(F_{\rm KK}^{(q_{\rm ctrl})}=0\). Under the stationary-\(H_*\) approximation,
\begin{equation}
 F_{\rm KK}^{(q_{\rm ctrl})}(\chi)
 \simeq
 \ln\!\left[
 \frac{M_{{\rm KK},0}}
 {q_{\rm ctrl}H_*}
 \right]
 -
 \alpha_\chi(\chi-\chi_0),
 \label{eq:two_field_control_function_linear}
\end{equation}
so the absorbing surface lies at
\begin{equation}
 \chi_b(q_{\rm ctrl})
 =
 \chi_0
 +
 \frac{1}{\alpha_\chi}
 \ln\!\left[
 \frac{M_{{\rm KK},0}}
 {q_{\rm ctrl}H_*}
 \right].
 \label{eq:two_field_wall_location}
\end{equation}
Changing the control factor moves the operational surface according to
\begin{equation}
 \chi_b(q_2)-\chi_b(q_1)
 =
 -\frac{1}{\alpha_\chi}
 \ln\!\left(\frac{q_2}{q_1}\right),
 \label{eq:two_field_qctrl_wall_shift}
\end{equation}
without changing the local logarithmic KK rate \(\alpha_\chi\). Thus \(q_{\rm ctrl}\) determines where stochastic evolution is terminated, whereas the local first-passage structure is controlled by the drift, diffusion, and remaining horizon near that selected surface.

Equation~\eqref{eq:two_field_qctrl_wall_shift} also makes clear that varying \(q_{\rm ctrl}\) while holding \(u_0\) and \(u_{\rm eq}\) fixed does \emph{not} test sensitivity to the loss prescription: doing so retunes \(M_{{\rm KK},0}\), \(\chi_{\rm ini}\), or both so as to preserve the same dimensionless distances. A prescription test must instead keep the underlying physical quantities
\begin{equation}
 H_*,
 \qquad
 M_{{\rm KK},0},
 \qquad
 m_\chi,
 \qquad
 \chi_0,
 \qquad
 \chi_{\rm ini}
 \label{eq:qctrl_fixed_physical_data}
\end{equation}
fixed while varying \(q_{\rm ctrl}\).

We introduce the inward proper distance
\begin{equation}
 s
 \equiv
 \chi_b(q_{\rm ctrl})-\chi.
 \label{eq:two_field_inward_distance}
\end{equation}
The controlled domain is \(s>0\), and the absorbing surface is \(s=0\). Since \(\mathrm ds=-\mathrm d\chi\), Eq.~\eqref{eq:two_field_chi_sde} becomes
\begin{equation}
 \mathrm ds
 =
 \nu_\chi(s_{\rm eq}-s)\mathrm dN
 +
 \sqrt{2D_\chi}\,\mathrm dW_N^{(s)},
 \label{eq:two_field_normal_sde}
\end{equation}
where
\begin{equation}
 s_{\rm eq}
 \equiv
 \chi_b(q_{\rm ctrl})-\chi_0
 =
 \frac{1}{\alpha_\chi}
 \ln\!\left[
 \frac{M_{{\rm KK},0}}
 {q_{\rm ctrl}H_*}
 \right],
 \label{eq:two_field_equilibrium_distance}
\end{equation}
and \(\mathrm dW_N^{(s)}=-\mathrm dW_N\) is again a normalized Wiener increment. The classical normal drift points toward \(s_{\rm eq}>0\), away from the absorbing wall. Loss therefore occurs through the stochastic tail that reaches \(s=0\) despite the restoring drift.

The one-e-fold rms stochastic displacement is
\begin{equation}
 \delta\chi_q
 \equiv
 \sqrt{2D_\chi}
 =
 \frac{H_*}{2\pi}\sqrt{{\cal Q}_\chi}.
 \label{eq:two_field_kick_size}
\end{equation}
We then measure the spectator distance directly in units of \(\delta\chi_q\) and define
\begin{equation}
 u
 \equiv
 \frac{s}{\delta\chi_q},
 \qquad
 u_{\rm eq}
 \equiv
 \frac{s_{\rm eq}}{\delta\chi_q},
 \qquad
 u_0
 \equiv
 \frac{\chi_b(q_{\rm ctrl})-\chi_{\rm ini}}
 {\delta\chi_q}.
 \label{eq:two_field_kick_coordinates}
\end{equation}

At fixed physical data, changing the loss prescription from \(q_1\) to \(q_2\) shifts both dimensionless distances by the same amount,
\begin{equation}
 u_{\rm eq}(q_2)-u_{\rm eq}(q_1)
 =
 u_0(q_2)-u_0(q_1)
 =
 -\frac{\ln(q_2/q_1)}
 {\alpha_\chi\delta\chi_q}.
 \label{eq:qctrl_fixed_model_u_shift}
\end{equation}
Consequently, the survival probability can be highly prescription-sensitive when the physical initial state has deliberately been placed within only a few one-e-fold stochastic displacement units of the selected surface. This sensitivity is information about the near-boundary construction rather than a prediction independent of the chosen definition of EFT loss.

The normal dynamics reduces to the dimensionless Ornstein--Uhlenbeck first-passage process
\begin{equation}
 \mathrm du
 =
 \nu_\chi(u_{\rm eq}-u)\mathrm dN
 +
 \mathrm dW_N^{(u)},
 \qquad
 u>0,
 \label{eq:two_field_ou_sde}
\end{equation}
with absorption at \(u=0\). Its backward generator is
\begin{equation}
 {\cal L}_u
 =
 \nu_\chi(u_{\rm eq}-u)\partial_u
 +
 \frac12\partial_u^2.
 \label{eq:two_field_ou_generator}
\end{equation}

For a remaining horizon \(\tau\), the survival probability \(h(u,\tau)\) satisfies
\begin{equation}
 \partial_\tau h
 =
 {\cal L}_u h
 =
 \nu_\chi(u_{\rm eq}-u)\partial_u h
 +
 \frac12\partial_u^2h,
 \label{eq:two_field_backward_equation}
\end{equation}
with
\begin{equation}
 h(0,\tau)=0,
 \qquad
 h(u,0)=1,
 \qquad
 h(u,\tau)\rightarrow1
 \quad\text{as}\quad
 u\rightarrow\infty
 \quad\text{for finite }\tau.
 \label{eq:two_field_backward_conditions}
\end{equation}
The probability of remaining in the controlled domain for \(N\) e-folds is
\begin{equation}
 P_{\rm surv}(N)
 =
 h(u_0,N),
 \label{eq:two_field_survival_probability}
\end{equation}
while the first-exit density and cumulative loss probability are
\begin{equation}
 f_{\rm exit}(u_0,N)
 =
 -\partial_Nh(u_0,N),
 \qquad
 P_{\rm loss}(u_0,N)
 =
 1-h(u_0,N)
 =
 \int_0^N
 f_{\rm exit}(u_0,N')\,\mathrm dN'.
 \label{eq:two_field_exit_observables}
\end{equation}
Because the one-dimensional Ornstein--Uhlenbeck process repeatedly explores the neighborhood of its equilibrium point, absorption at \(u=0\) occurs with unit probability over an arbitrarily long horizon, even though the classical drift points away from the wall. The long-horizon extension used below is therefore an asymptotic first-passage diagnostic, not an assumption that the observationally relevant inflationary phase lasts \(10^5\) e-folds.

Conditioning a trajectory on survival over a specified remaining horizon modifies its normal drift through the Doob transform:
\begin{equation}
 b_{\rm cond}^{u}(u,\tau)
 =
 \nu_\chi(u_{\rm eq}-u)
 +
 \Delta b_{\rm Doob}^{u}(u,\tau),
 \qquad
 \Delta b_{\rm Doob}^{u}
 =
 \partial_u\ln h.
 \label{eq:two_field_conditioned_drift}
\end{equation}
Here \(2D_u=1\) in kick-normalized units. Returning to proper-distance units gives
\begin{equation}
 \Delta b_{\rm Doob}^{s}
 =
 2D_\chi\partial_s\ln h.
 \label{eq:two_field_doob_proper_units}
\end{equation}
Regularity at a smooth absorbing surface implies
\begin{equation}
 h(u,\tau)
 =
 C(\tau)u
 +
 {\cal O}(u^2),
 \qquad
 u\rightarrow0^+,
 \label{eq:two_field_near_wall_h}
\end{equation}
and hence the universal wall law
\begin{equation}
 \Delta b_{\rm Doob}^{u}
 \simeq
 \frac1u,
 \qquad
 \Delta b_{\rm Doob}^{s}
 \simeq
 \frac{2D_\chi}{s}.
 \label{eq:two_field_near_wall_doob}
\end{equation}
The spectator model therefore embeds the local singular drift in a global first-passage problem with a nonzero stabilizing spectator drift.

The killed ensemble also develops a selection bias. We define the time-local surviving mean by
\begin{equation}
 \left\langle u(N)\right\rangle_{\rm surv}
 \equiv
 \mathbb E\!\left[
 u(N)\,\middle|\,T_{\rm exit}>N
 \right],
 \label{eq:two_field_time_local_surviving_mean}
\end{equation}
where \(T_{\rm exit}\) is the first hitting time of \(u=0\). The unconditioned Ornstein--Uhlenbeck mean is
\begin{equation}
 \left\langle u(N)\right\rangle_{\rm uncond}
 =
 u_{\rm eq}
 +
 (u_0-u_{\rm eq})\mathrm e^{-\nu_\chi N}.
 \label{eq:two_field_unconditioned_mean}
\end{equation}
The surviving-ensemble bias is
\begin{equation}
 \Delta\left\langle u(N)\right\rangle_{\rm surv}
 \equiv
 \left\langle u(N)\right\rangle_{\rm surv}
 -
 \left\langle u(N)\right\rangle_{\rm uncond}.
 \label{eq:two_field_survival_bias_u}
\end{equation}
Histories undergoing sufficiently large wall-directed fluctuations are removed, while histories remaining farther inside the controlled domain are overrepresented among the survivors. Consequently, \(\Delta\langle u\rangle_{\rm surv}>0\).

Since larger \(u\) corresponds to smaller \(\chi\),
\begin{equation}
 \Delta\left\langle\chi(N)\right\rangle_{\rm surv}
 =
 -\delta\chi_q\,
 \Delta\left\langle u(N)\right\rangle_{\rm surv}
 <0.
 \label{eq:two_field_survival_bias_chi}
\end{equation}
Moreover, within the stationary-\(H_*\) approximation,
\begin{equation}
 F_{\rm KK}^{(q_{\rm ctrl})}
 \simeq
 \alpha_\chi s
 =
 \alpha_\chi\delta\chi_q\,u,
 \label{eq:two_field_F_in_u}
\end{equation}
so that
\begin{equation}
 \Delta
 \left\langle
 F_{\rm KK}^{(q_{\rm ctrl})}(N)
 \right\rangle_{\rm surv}
 =
 \alpha_\chi\delta\chi_q\,
 \Delta\left\langle u(N)\right\rangle_{\rm surv}
 >0.
 \label{eq:two_field_survival_bias_F}
\end{equation}
Survival conditioning therefore selects histories with a larger KK/Hubble control margin, even though stochastic fluctuations reduce the survival probability of the original unconditioned ensemble.

\begin{figure}[!htb]
 \centering
 \includegraphics[width=\textwidth]
 {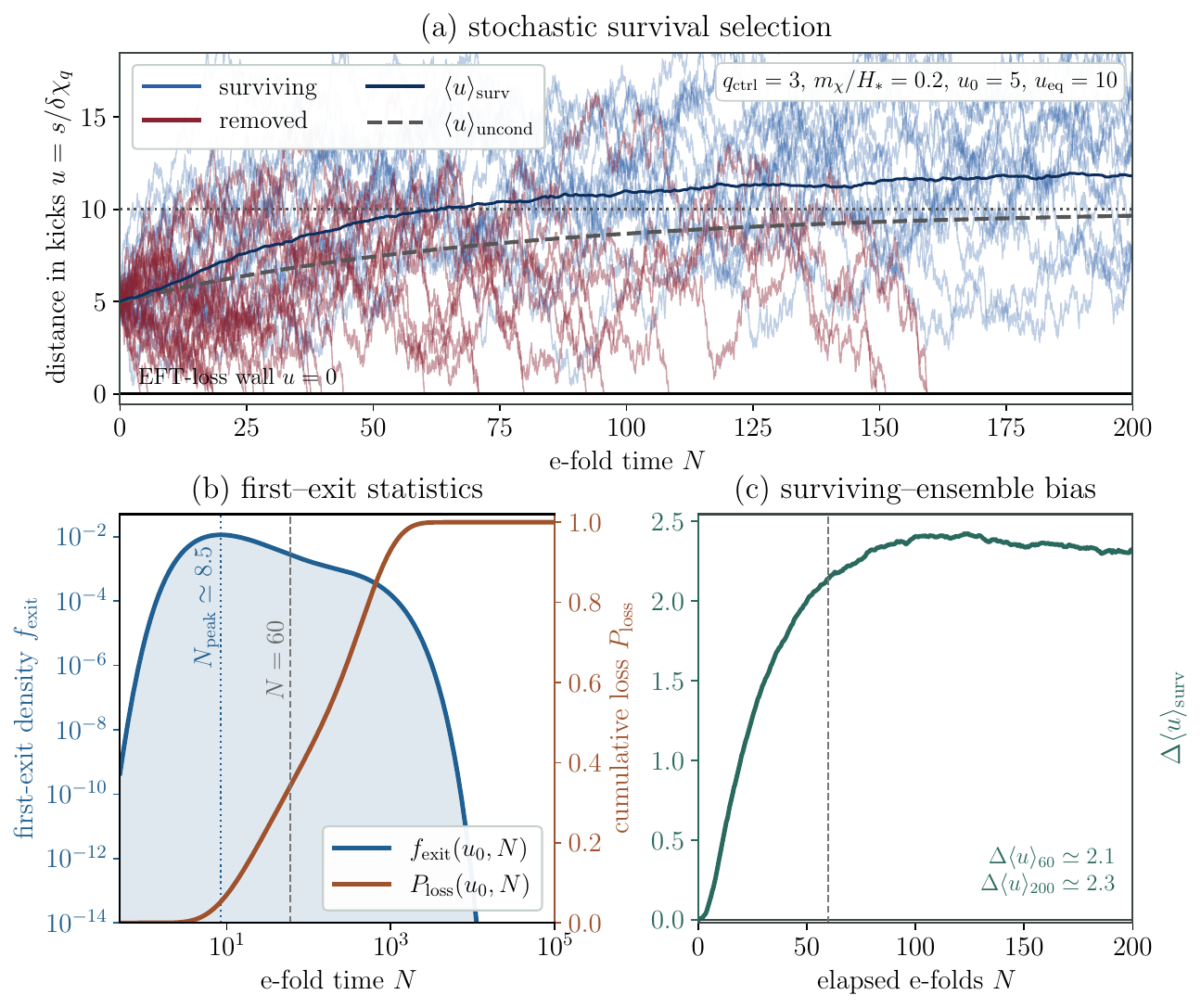}
 \caption{Stochastic survival of a light spectator with a restoring quadratic potential and a stationary KK/Hubble loss surface with \(q_{\rm ctrl}=3\), \(m_\chi/H_*=0.2\), \(u_0=5\), and \(u_{\rm eq}=10\). Panel~(a) shows histories absorbed at \(u=0\) and the surviving and unconditioned means. Panel~(b) shows the first-exit density and cumulative loss probability, with \(N_{\rm peak}\simeq8.5\);
 the extension to \(N=10^5\) is an asymptotic first-passage diagnostic. Panel~(c) shows the positive surviving-ensemble bias, with \(\Delta\langle u\rangle_{60}\simeq2.1\) and \(\Delta\langle u\rangle_{200}\simeq2.3\) for the displayed Monte Carlo realization. The quoted first-passage and conditional-mean results refer to the stationary-wall loss prescription specified in the text.}
 \label{fig:twoFieldSurvivalBenchmark}
\end{figure}

For the stationary-wall example shown in \cref{fig:twoFieldSurvivalBenchmark}, we choose
\begin{equation}
 \frac{H_*}{M_{\rm Pl}}
 =
 5.6\times10^{-6},
 \qquad
 {\cal Q}_\chi=1,
 \qquad
 q_{\rm ctrl}=3,
 \qquad
 \alpha_\chi=\sqrt{\frac23},
 \label{eq:two_field_numerical_parameters_a}
\end{equation}
together with
\begin{equation}
 \frac{m_\chi}{H_*}=0.2,
 \qquad
 u_0=5,
 \qquad
 u_{\rm eq}=10,
 \qquad
 N_*=60,
 \qquad
 N_{\max}=200.
 \label{eq:two_field_numerical_parameters_b}
\end{equation}
These values imply
\begin{equation}
 \nu_\chi
 =
 \frac{m_\chi^2}{3H_*^2}
 =
 \frac{0.2^2}{3}
 \simeq
 1.33\times10^{-2},
 \label{eq:two_field_numerical_nu}
\end{equation}
and
\begin{equation}
 D_\chi
 =
 \frac{H_*^2}{8\pi^2}
 \simeq
 3.97\times10^{-13}M_{\rm Pl}^2,
 \qquad
 \delta\chi_q
 =
 \frac{H_*}{2\pi}
 \simeq
 8.91\times10^{-7}M_{\rm Pl}.
 \label{eq:two_field_numerical_diffusion}
\end{equation}
The equilibrium point is separated from the selected wall by ten one-e-fold rms stochastic displacement units, while the ensemble begins five such units from the wall. The initial classical normal drift is therefore
\begin{equation}
 b_{\rm cl}^{u}(u_0)
 =
 \nu_\chi(u_{\rm eq}-u_0)
 \simeq
 6.67\times10^{-2}
 \label{eq:two_field_initial_normal_drift}
\end{equation}
per e-fold. This stationary-wall parameter choice retains a stabilizing drift while deliberately placing the system in a narrow region where diffusion, absorption, and relaxation compete. It is therefore designed to expose first-passage selection rather than to represent a generic, parametrically well-separated compactification.

The stationary-background requirement can now be quantified for these same parameters. For orientation, the value \(r=0.003\) used earlier to set an inflationary normalization in the direct-threshold example would, for a canonical slow-roll adiabatic background, correspond to \(\epsilon_H\simeq r/16=1.875\times10^{-4}\). If that value were imposed here, Eq.~\eqref{eq:two_field_wall_velocity} would give
\begin{equation}
 \frac{1}{\delta\chi_q}
 \frac{\mathrm d\chi_b}{\mathrm dN}
 =
 \frac{\epsilon_H}
 {\alpha_\chi\delta\chi_q}
 \simeq
 2.58\times10^2,
 \label{eq:two_field_wall_velocity_kick_units}
\end{equation}
so the wall would move by roughly \(258\) one-e-fold stochastic displacement units per e-fold. Over \(60\) e-folds, approximately constant \(\epsilon_H\) would give
\begin{equation}
 \Delta\chi_b(60)
 \simeq
 \frac{60\epsilon_H}{\alpha_\chi}
 \simeq
 1.38\times10^{-2}M_{\rm Pl},
 \label{eq:two_field_r003_wall_motion}
\end{equation}
whereas the initial wall distance and driftless diffusion width are
\begin{equation}
 s_0
 =
 u_0\delta\chi_q
 \simeq
 4.46\times10^{-6}M_{\rm Pl},
 \qquad
 \sqrt{2D_\chi(60)}
 =
 \delta\chi_q\sqrt{60}
 \simeq
 6.90\times10^{-6}M_{\rm Pl}.
 \label{eq:two_field_stationary_comparison_scales}
\end{equation}
Such a background is therefore far outside the stationary-wall regime. For approximately constant \(\epsilon_H\), comparison with the diffusion width alone requires
\begin{equation}
 |\epsilon_H|
 \ll
 \alpha_\chi
 \sqrt{\frac{2D_\chi}{60}}
 \simeq
 9.4\times10^{-8},
 \label{eq:two_field_epsilon_diffusion_bound}
\end{equation}
while the requirement that the wall move much less than the initial distance \(s_0\) is slightly stronger,
\begin{equation}
 |\epsilon_H|
 \ll
 \frac{\alpha_\chi s_0}{60}
 =
 \frac{\alpha_\chi u_0\delta\chi_q}{60}
 \simeq
 6.1\times10^{-8}.
 \label{eq:two_field_epsilon_initial_margin_bound}
\end{equation}
The two-field calculation below should accordingly be read as an explicitly much flatter quasi-de Sitter limit, independent of the \(r=0.003\) normalization used in the preceding direct-threshold example. A background
with appreciable \(H(N)\) evolution requires the nonautonomous moving-wall problem rather than the stationary Ornstein--Uhlenbeck calculation quoted here.

For the displayed extension to \(N_{\max}=200\), the same initial-margin test strengthens to \(|\epsilon_H|\ll1.82\times10^{-8}\). All quoted stationary results are defined in the fixed-\(H_*\) model; these inequalities state when an evolving background could approximate it.

The prescribed value of \(u_{\rm eq}\) fixes the local hierarchy at \(\chi=\chi_0\):
\begin{equation}
 \frac{M_{{\rm KK},0}}
 {q_{\rm ctrl}H_*}
 =
 \exp\!\left(
 \alpha_\chi\delta\chi_q u_{\rm eq}
 \right).
 \label{eq:two_field_local_hierarchy_choice}
\end{equation}
For the quoted parameters,
\begin{equation}
 \ln\!\left(
 \frac{M_{{\rm KK},0}}
 {q_{\rm ctrl}H_*}
 \right)
 =
 \alpha_\chi\delta\chi_q u_{\rm eq}
 \simeq
 7.28\times10^{-6},
 \label{eq:two_field_small_equilibrium_margin}
\end{equation}
while the logarithmic control margin at the initial point is
\begin{equation}
 F_{\rm KK}^{(q_{\rm ctrl})}(\chi_{\rm ini})
 =
 \alpha_\chi\delta\chi_q u_0
 \simeq
 3.64\times10^{-6}.
 \label{eq:two_field_small_initial_margin}
\end{equation}
The example is therefore intentionally extremely close to the prescribed KK/Hubble surface. At fixed physical data, even a fractional change \(q_{\rm ctrl}\rightarrow q_{\rm ctrl}(1+10^{-5})\) shifts the wall by
\begin{equation}
 \frac{|\Delta\chi_b|}{\delta\chi_q}
 =
 \frac{\ln(1+10^{-5})}
 {\alpha_\chi\delta\chi_q}
 \simeq
 13.7,
 \label{eq:two_field_qctrl_tiny_shift}
\end{equation}
one-e-fold stochastic displacement units and would place the same initial state outside the newly defined controlled domain. The numerical survival probabilities below are consequently conditional on the stated loss prescription; they are not prescription-insensitive predictions of the compactification data alone.

A direct fixed-model test illustrates this dependence. Writing \(\eta=q_{\rm ctrl}/3-1\), the physical parameters and initial field value are held fixed and both \(u_0\) and \(u_{\rm eq}\) are shifted by Eq.~\eqref{eq:qctrl_fixed_model_u_shift}. Solving the same backward problem gives the following values (rounded to three decimals):
\begin{center}
\begin{tabular}{r r r r r}
\toprule
\(\eta\) & \(u_0\) & \(u_{\rm eq}\) & \(P_{\rm surv}(60)\) & \(P_{\rm surv}(200)\)\\
\midrule
\(-2\times10^{-6}\) & 7.748 & 12.748 & 0.867 & 0.706\\
\(-1\times10^{-6}\) & 6.374 & 11.374 & 0.780 & 0.582\\
\(0\) & 5.000 & 10.000 & 0.658 & 0.440\\
\(+1\times10^{-6}\) & 3.626 & 8.626 & 0.501 & 0.294\\
\(+2\times10^{-6}\) & 2.252 & 7.252 & 0.316 & 0.159\\
\bottomrule
\end{tabular}
\end{center}
For \(\eta=10^{-5}\), the initial point is outside the new domain, so survival is zero from the start. The regular hard-wall pole law and the interpretation as statistical conditioning are insensitive to these shifts.
The finite-time probabilities, exit-time distribution, and magnitude of the surviving-ensemble bias are not; substantial loss in this example depends on its deliberately selected near-wall regime.

More generally, at fixed physical data the relevant regimes are transparent: if the initial margin satisfies \(s_0\gg\delta\chi_q\), loss over a short horizon is strongly suppressed unless drift or accumulated diffusion brings
the process close to the wall; if \(s_0\) is comparable with the stochastic displacement scales explored over the chosen horizon, first-passage probabilities become sensitive to the precise loss prescription; and if \(s_0<0\), the initial state already lies outside the domain selected by that prescription.

If the transition out of EFT control has a finite microscopic width, the appropriate comparison is obtained by keeping the same underlying physical model fixed and replacing the sharp absorbing condition by a killing profile \(\kappa_{\rm KK}(\chi)\) spanning that transition region. The hard-wall calculation presented here corresponds to the limiting prescription in which this transition is treated as sharp. Its universal near-wall \(2D_\chi/s\) behavior is therefore a statement about that specified hard-loss problem, whereas finite-horizon survival probabilities depend on the location and width assigned to the physical loss channel.

To generate the killed stochastic ensemble, we use the exact finite-step Ornstein--Uhlenbeck endpoint update
\begin{equation}
 u_{n+1}
 =
 u_{\rm eq}
 +
 (u_n-u_{\rm eq})\mathrm e^{-\nu_\chi\Delta N}
 +
 \sigma_{\Delta N}\,\xi_n,
 \qquad
 \xi_n\sim{\cal N}(0,1),
 \label{eq:two_field_exact_ou_step}
\end{equation}
with
\begin{equation}
 \sigma_{\Delta N}^2
 =
 \frac{1-\mathrm e^{-2\nu_\chi\Delta N}}
 {2\nu_\chi}.
 \label{eq:two_field_exact_ou_variance}
\end{equation}
A history is removed if its endpoint lies at \(u\leq0\). When both endpoints remain positive, unresolved within-step crossings are included through the short-step Brownian-bridge correction
\begin{equation}
 p_{\rm cross}
 \simeq
 \exp\!\left[
 -\frac{2u_nu_{n+1}}{\sigma_{\Delta N}^2}
 \right].
 \label{eq:two_field_bridge_crossing_probability}
\end{equation}
Individual histories are classified according to whether they survive the full interval \(0\leq N\leq N_{\max}\). By contrast, the conditioned mean \(\langle u(N)\rangle_{\rm surv}\) is evaluated from all histories that are still alive at the corresponding time \(N\), in accordance with Eq.~\eqref{eq:two_field_time_local_surviving_mean}. This distinction keeps the visual classification of complete histories separate from the time-local conditioning used to define the ensemble bias.

The backward problem \eqref{eq:two_field_backward_equation} is solved independently with an implicit finite-difference scheme. The original numerical grid uses \(u_{\max}=40\), \(\Delta u=1/30\), \(\Delta N=0.05\), and \(h(u_{\max},N)=1\). An independent Crank--Nicolson check at \(u_{\max}=60\), \(\Delta u=0.025\), \(\Delta N=0.01\) yields \(h(5,60)=0.657794\) and \(h(5,200)=0.440348\). Halving both grid spacings changes these probabilities by less than \(4\times10^{-7}\); increasing \(u_{\max}\) to 80 has negligible effect at these horizons. The fixed-model threshold test above uses the latter scheme. The original first-exit peak at \(N=8.55\) shifts to \(N\simeq8.48\) under refinement; it should be quoted only as approximately 8.5 e-folds. The short-horizon backward calculation used in Figure~\ref{fig:twoFieldSurvivalBenchmark} employs the original implicit method refined to \(u_{\max}=60\), \(\Delta u=0.025\), and \(\Delta N=0.005\), with the Monte Carlo parameters unchanged. For the long-horizon asymptotic test, we use
\begin{equation}
 N_{\rm long}=10^5,
 \qquad
 \Delta N_{\rm long}=0.25,
 \qquad
 u_{\max}=50,
 \label{eq:two_field_long_horizon_numerics}
\end{equation}
together with the truncated-domain boundary conditions
\begin{equation}
 h(0,N)=0,
 \qquad
 \partial_u h(u_{\max},N)=0.
 \label{eq:two_field_truncated_long_boundary_conditions}
\end{equation}
The long-horizon spatial grid has 1201 points. To retain the early-time peak resolution, its implicit evolution uses \(\Delta N=0.005\) through \(N=200\), followed by \(\Delta N_{\rm long}=0.25\), continuing the same solution state. The far-interior Neumann condition avoids the artificial late-time plateau that would arise from imposing \(h(u_{\max},N)=1\) at a finite numerical boundary. The first-exit density is differentiated on the piecewise-uniform implicit time grid and only then interpolated onto the logarithmic display grid.

\Cref{fig:twoFieldSurvivalBenchmark} summarizes the resulting stochastic and backward calculations. Panel~(a) displays the killed path ensemble together with the time-local surviving mean and the exact unconditioned Ornstein--Uhlenbeck mean. Their separation is the direct ensemble-level signature of survival selection. Panel~(b) shows the first-exit density and cumulative loss probability, including the early first-exit peak and the asymptotic long-horizon tail. Panel~(c) shows the corresponding positive displacement of the surviving ensemble relative to the unconditioned mean. Taken together, the panels connect the stochastic generator, the KK control surface, the global first-passage statistics, and the bias of the histories that retain EFT control. For the stated stationary-wall prescription the backward solution gives \(P_{\rm surv}(60)\simeq0.658\) and \(P_{\rm surv}(200)\simeq0.440\), while the first-exit density peaks at \(N_{\rm peak}\simeq8.5\). These numbers quantify the survival of this specified near-boundary stochastic EFT model; they are not claimed as generic probabilities for an evolving inflationary compactification.

\subsection{Exponential cutoff/Hubble channels}
\label{subsec:exponential_cutoff_channels}

Consider any exponential channel introduced in \cref{sec:survival_channels_cutoff_walls}. Its conservative cutoff/Hubble function is
\begin{equation}
 F_{A/H}^{(q_{\rm ctrl})}(d)
 =
 F_{A,0}
 -
 \ln q_{\rm ctrl}
 -
 \Delta\lambda_A d,
 \qquad
 \Delta\lambda_A
 \equiv
 \lambda_A-\beta.
 \label{eq:qg_common_linear_form_section4}
\end{equation}
A finite wall is approached along increasing \(d\) when
\begin{equation}
 \Delta\lambda_A>0,
 \qquad
 F_{A,0}>\ln q_{\rm ctrl}.
\end{equation}
Its location is
\begin{equation}
 d_{A,b}(q_{\rm ctrl})
 =
 \frac{F_{A,0}-\ln q_{\rm ctrl}}
 {\Delta\lambda_A}.
 \label{eq:qg_wall_location}
\end{equation}
On the controlled side, define the inward proper-distance coordinate
\begin{equation}
 s_A
 \equiv
 d_{A,b}(q_{\rm ctrl})-d.
 \label{eq:qg_distance_to_wall}
\end{equation}
The control function then takes the exact local linear form
\begin{equation}
 F_{A/H}^{(q_{\rm ctrl})}
 =
 \Delta\lambda_A s_A.
 \label{eq:exponential_channel_distance_form}
\end{equation}

The coordinate \(d\) increases toward the wall, whereas \(s_A\) increases inward, so
\begin{equation}
 \partial_d=-\partial_{s_A}.
\end{equation}
For a one-dimensional diffusion coefficient \(D^{dd}\), the boundary-function law gives
\begin{equation}
 \Delta b_{\rm Doob}^{d}
 \simeq
 2D^{dd}
 \frac{\partial_dF_{A/H}^{(q_{\rm ctrl})}}
 {F_{A/H}^{(q_{\rm ctrl})}}
 =
 -\frac{2D^{dd}}
 {d_{A,b}(q_{\rm ctrl})-d}.
 \label{eq:qg_common_doob}
\end{equation}
Equivalently, in the inward coordinate,
\begin{equation}
 \Delta b_{\rm Doob}^{s_A}
 \simeq
 \frac{2D_\perp}{s_A}.
 \label{eq:qg_common_doob_inward}
\end{equation}
The coordinate sign is conventional; the invariant statement is that the conditioned response points inward, away from the absorbing wall.

In the slow-roll light-mode normalization,
\begin{equation}
 D_\perp
 \simeq
 \frac{H^2{\cal Q}_\perp}{8\pi^2},
\end{equation}
and therefore
\begin{equation}
 \Delta b_{\rm Doob}^{d}
 \simeq
 -\frac{H_b^2{\cal Q}_{\perp,b}}
 {4\pi^2[d_{A,b}(q_{\rm ctrl})-d]},
 \qquad
 H_b
 \equiv
 H\!\left(d_{A,b}(q_{\rm ctrl})\right),
 \label{eq:qg_common_doob_slowroll}
\end{equation}
up to regular corrections from the variation of \(H\), \({\cal Q}_\perp\), and the wall geometry across the local layer. The canonical result follows by setting \({\cal Q}_{\perp,b}=1\).

Several microscopic inputs fit this exponential family:
\begin{equation}
 \lambda_A
 =
 \begin{cases}
 \dfrac{p\alpha}{p+2},
 &
 \text{tower/species channel},
 \\[7pt]
 \lambda_s
 \ \text{or}\
 \lambda_{\rm KK},
 &
 \text{direct string/KK threshold},
 \\[7pt]
 \alpha_g,
 &
 \text{magnetic weak-coupling cutoff},
 \\[7pt]
 \dfrac{p\delta a}{p+2},
 &
 \text{gravitino-controlled tower}.
 \end{cases}
 \label{eq:qg_lambda_cases}
\end{equation}
The magnetic weak-gravity entry uses
\begin{equation}
 \Lambda_{\rm WGC}
 \sim
 gM_{\rm Pl},
 \qquad
 g(d)
 =
 g_0\mathrm e^{-\alpha_gd},
\end{equation}
with \(\alpha_g>0\)~\cite{Arkani-Hamed:2006emk,Harlow:2015lma,Heidenreich:2015nta,Heidenreich:2018kpg,Saraswat:2016eaz}.
The gravitino-controlled entry assumes
\begin{equation}
 m_{3/2}
 \sim
 \mathrm e^{-ad},
 \qquad
 M_{\rm tower}
 \sim
 m_{3/2}^{\delta},
\end{equation}
with \(a,\delta>0\)~\cite{Castellano:2021yye,Cribiori:2021gbf}. These microscopic choices alter the control function, wall location, and regular rate data, but not the local \(2D_\perp/s_A\) wall law.

Several towers contributing to one gravitational species count determine the single combined function \(F_{\rm sp/H}\) discussed around \cref{eq:inverse_full_species_definition_revision}. Independent multi-boundary problems instead require distinct microscopic loss mechanisms and distinct functions \(F_A\).

In either case, a candidate cutoff/Hubble zero can be used as an absorbing surface for the retained stochastic theory only if that theory remains self-consistent up to the surface. If an additional tower becomes dynamically relevant before the formal species or cutoff wall is reached, the first such loss of control terminates the reduced stochastic description unless the newly light degrees of freedom have been included explicitly.

\subsection[Potential-gradient and Hessian comparisons]{Potential-gradient and Hessian comparisons}
\label{subsec:normal_hessian_diagnostics}

Cutoff/Hubble surfaces and potential inequalities have different microscopic status. A string, KK, or species threshold diagnoses the breakdown of a specified low-energy truncation, whereas a gradient or Hessian Swampland criterion is a conjectural inequality whose violation does not by itself define an absorbing stochastic boundary. The functions \(F_\nabla\) and \(F_{\rm Hess}\) introduced below may therefore be used as hard survival data only when an independent operational prescription declares their violation to terminate the retained ensemble. Otherwise, they serve as diagnostics of the conditioned flow or as inputs to a model-dependent soft-killing profile. The mathematical boundary-normal equivalence discussed below does not identify these distinct microscopic interpretations.~\footnote{Throughout this subsection, field-space distances and covariant derivatives appearing in the potential criteria are expressed in four-dimensional reduced Planck units. Equivalently, the canonically normalized field coordinates are treated as dimensionless after setting \(M_{\rm Pl}=1\). With explicit dimensions restored, the gradient and Hessian quantities below correspond respectively to \(M_{\rm Pl}|\nabla V|/V\) and \(M_{\rm Pl}^{2}\lambda_{\min}(\nabla_i\nabla_jV/V)\).}

The de Sitter gradient criterion is conventionally written as
\begin{equation}
 |\nabla\ln V|
 \gtrsim
 c,
 \qquad
 c=\mathcal O(1),
 \label{eq:ds_gradient_condition}
\end{equation}
while the refined criterion takes the alternative form
\begin{equation}
 |\nabla\ln V|
 \gtrsim
 c
 \qquad\text{or}\qquad
 \lambda_{\min}\!\left(
 \frac{\nabla_i\nabla_jV}{V}
 \right)
 \lesssim
 -c',
 \qquad
 c'=\mathcal O(1).
 \label{eq:refined_ds_condition}
\end{equation}
Here \(|\nabla\ln V|=\sqrt{G^{ij}\nabla_i\ln V\nabla_j\ln V}\), and the minimum Hessian eigenvalue is evaluated in an orthonormal field-space frame~\cite{Obied:2018sgi,Agrawal:2018own,Murayama:2018lie,Ooguri:2018wrx,Garg:2018reu,Andriot:2018mav}.

Subject to the operational qualification above, the gradient branch can be represented by
\begin{equation}
 F_\nabla(\phi)
 \equiv
 \ln\!\left(
 \frac{|\nabla\ln V|}{c}
 \right),
 \qquad
 F_\nabla>0
 \quad\Longleftrightarrow\quad
 |\nabla\ln V|>c.
 \label{eq:ds_survival_function}
\end{equation}
If histories are conditioned to remain in the region \(F_\nabla>0\), the local response near a regular hard-wall component is
\begin{equation}
 \Delta b_{\rm Doob}^{i}
 \simeq
 2D^{ij}
 \frac{\nabla_jF_\nabla}{F_\nabla}.
 \label{eq:ds_survival_doob_response}
\end{equation}
The conjectural potential criterion supplies the boundary data only after the additional operational prescription has been chosen; the survival problem then supplies the conditioned statistical response.

The refined criterion is naturally represented by two alternative channels,
\begin{equation}
 F_\nabla(\phi)
 \equiv
 \ln\!\left(
 \frac{|\nabla\ln V|}{c}
 \right),
 \qquad
 F_{\rm Hess}(\phi)
 \equiv
 \ln\!\left[
 \frac{-\lambda_{\min}(\nabla_i\nabla_jV/V)}
 {c'}
 \right].
 \label{eq:refined_ds_two_channels}
\end{equation}
The second function is defined only in the tachyonic region
\begin{equation}
 \lambda_{\min}\!\left(
 \frac{\nabla_i\nabla_jV}{V}
 \right)
 <0.
\end{equation}
The corresponding operational domain is the union
\begin{equation}
 \mathcal M_{\rm rdS}
 =
 \left\{
 \phi:F_\nabla(\phi)>0
 \right\}
 \cup
 \left\{
 \phi:F_{\rm Hess}(\phi)>0
 \right\}.
 \label{eq:refined_ds_union_domain}
\end{equation}
Each regular component of the boundary of this union may be treated by the local wall law. A surface \(F_\nabla=0\) is an absorbing component only where the Hessian branch does not already retain the point inside the union, and conversely for \(F_{\rm Hess}=0\).

It is useful to compare such potential-based loss functions with the cutoff/Hubble channels above. Along a canonically normalized trajectory \(d\), an exponential cutoff/Hubble channel has the local form
\begin{equation}
 F_{A/H}^{(q_{\rm ctrl})}(d)
 =
 \gamma_A
 [d_{A,b}(q_{\rm ctrl})-d],
 \qquad
 \gamma_A
 \equiv
 \Delta\lambda_A>0.
 \label{eq:qg_normal_form_for_diagnostics}
\end{equation}
A potential-based loss function belongs to the same leading boundary-normal Doob class whenever it has a regular simple zero,
\begin{equation}
 F_{\rm pot}(d)
 =
 a(d_b-d)
 +
 \mathcal O[(d_b-d)^2],
 \qquad
 a>0.
 \label{eq:general_doob_class_matching}
\end{equation}
The coefficient \(a\) does not affect the leading logarithmic derivative \(\partial_d\ln F_{\rm pot}\). Hence equality of the leading Doob class does not require \(a=\gamma_A\). To construct a representative that matches the cutoff/Hubble function itself, rather than only its logarithmic boundary-normal class, one may make the stronger choice
\begin{equation}
 a=\gamma_A.
 \label{eq:exact_representative_matching_choice}
\end{equation}

For the gradient branch, this exact representative matching is
\begin{equation}
 F_\nabla(d)
 =
 \ln\!\left(
 \frac{|\partial_d\ln V|}{c}
 \right)
 =
 \gamma(d_b-d)
 +
 \mathcal O[(d_b-d)^2],
 \qquad
 \gamma>0.
 \label{eq:gradient_branch_normal_matching}
\end{equation}
At leading order, an exact linear representative satisfies
\begin{equation}
 |\partial_d\ln V|
 =
 c\,\mathrm e^{\gamma(d_b-d)}.
 \label{eq:gradient_branch_slope_profile}
\end{equation}
Choosing a sign
\begin{equation}
 \partial_d\ln V
 =
 \sigma c\,\mathrm e^{\gamma(d_b-d)},
 \qquad
 \sigma=\pm1,
\end{equation}
gives
\begin{equation}
 V(d)
 =
 V_0
 \exp\!\left[
 -\sigma\frac{c}{\gamma}
 \mathrm e^{\gamma(d_b-d)}
 \right].
 \label{eq:gradient_matched_potential}
\end{equation}
If \(d\) is denoted by a canonical scalar \(\varphi\), this becomes
\begin{equation}
 V(\varphi)
 =
 V_0
 \exp\!\left[
 -\sigma\frac{c}{\gamma}
 \mathrm e^{\gamma(\varphi_b-\varphi)}
 \right],
 \qquad
 \left|
 \frac{V_{,\varphi}}{V}
 \right|
 =
 c\,\mathrm e^{\gamma(\varphi_b-\varphi)}.
 \label{eq:gradient_matched_canonical_potential}
\end{equation}
The logarithmic slope reaches the chosen threshold at \(\varphi=\varphi_b\) and varies exponentially away from the wall.

In string or supergravity variables it can be useful to define the exponential coordinate
\begin{equation}
 \rho
 \equiv
 \mathrm e^{\gamma(d_b-d)}.
 \label{eq:gradient_matched_exponential_modulus}
\end{equation}
The same representative then reads
\begin{equation}
 V(\rho)
 =
 V_0
 \exp\!\left[
 -\sigma\frac{c}{\gamma}\rho
 \right].
 \label{eq:gradient_matched_modulus_potential}
\end{equation}
This is a local normal-form representative of the operational boundary data, not a unique microscopic model.

To display its regular near-wall behavior, define
\begin{equation}
 x\equiv d_b-d.
\end{equation}
Then
\begin{equation}
 V(d)
 =
 V_b
 \exp\!\left[
 -\sigma cx
 -\sigma\frac{c\gamma}{2}x^2
 +\mathcal O(x^3)
 \right],
 \qquad
 V_b
 \equiv
 V_0\mathrm e^{-\sigma c/\gamma},
 \label{eq:gradient_matched_near_wall_exponential}
\end{equation}
and hence
\begin{equation}
 V(d)
 =
 V_b
 \left[
 1-\sigma cx
 +\frac12(c^2-\sigma c\gamma)x^2
 +\mathcal O(x^3)
 \right].
 \label{eq:gradient_matched_near_wall_taylor}
\end{equation}
The potential is regular at the wall, while its logarithmic slope reaches the selected threshold.

If the canonical distance is logarithmic in a modulus, the same representative takes a familiar asymptotic form. Let
\begin{equation}
 d
 =
 \alpha\ln\mathcal V
 +
 \text{const.},
 \qquad
 \alpha>0.
 \label{eq:logarithmic_volume_distance_diagnostic}
\end{equation}
Then
\begin{equation}
 \mathrm e^{\gamma(d_b-d)}
 =
 A\mathcal V^{-p},
 \qquad
 p
 \equiv
 \alpha\gamma,
 \qquad
 A>0.
 \label{eq:gradient_matched_volume_coordinate}
\end{equation}
The representative becomes
\begin{equation}
 V(\mathcal V)
 =
 V_0
 \exp\!\left[
 -\sigma B\mathcal V^{-p}
 \right],
 \qquad
 B>0.
 \label{eq:gradient_matched_volume_potential}
\end{equation}
At large volume,
\begin{equation}
 V(\mathcal V)
 =
 V_0
 \left[
 1-\sigma B\mathcal V^{-p}
 +\frac{B^2}{2}\mathcal V^{-2p}
 +\mathcal O(\mathcal V^{-3p})
 \right].
 \label{eq:gradient_matched_volume_expansion}
\end{equation}
Since \(\mathcal V^{-p}\sim\mathrm e^{-\gamma d}\), this is a plateau-like asymptotic expansion with exponentially suppressed corrections in the canonical distance. In the large-volume convention of \cref{eq:lv_canonical_distance},
\begin{equation}
 \alpha
 =
 \sqrt{\frac23},
 \qquad
 p
 =
 \gamma\sqrt{\frac23}.
 \label{eq:diagnostic_volume_power}
\end{equation}
The resemblance to Starobinsky- or \(\alpha\)-attractor-type asymptotic expansions is structural rather than microscopic: the survival construction fixes only the operational boundary-normal class, while coefficients and microscopic origins remain model-dependent.

For comparison, consider a pure exponential Hubble profile with
\(\beta>0\),
\begin{equation}
 H(d)
 =
 H_0\mathrm e^{-\beta d},
 \qquad
 V(d)
 \simeq
 3H^2(d)
\end{equation}
in reduced Planck units. It gives
\begin{equation}
 |\partial_d\ln V|
 =
 2\beta,
 \qquad
 F_\nabla
 =
 \ln\!\left(
 \frac{2\beta}{c}
 \right).
 \label{eq:exponential_potential_ds_constant}
\end{equation}
The gradient criterion is then a global pass/fail condition along the trajectory rather than a finite local wall. A finite gradient-type survival wall requires a field-dependent logarithmic slope. The same pure exponential profile has
\begin{equation}
 \frac{V''}{V}
 =
 4\beta^2
 >0,
\end{equation}
so its one-field tachyonic Hessian branch is inactive. If the trajectory must also approach a finite cutoff/Hubble wall, then
\begin{equation}
 \lambda_A-\beta>0.
\end{equation}
Combining this with the gradient condition \(2\beta\gtrsim c\) gives
\begin{equation}
 \lambda_A
 >
 \beta
 \gtrsim
 \frac{c}{2}.
 \label{eq:exponential_ds_cutoff_compatibility}
\end{equation}
This is not a universal bound. It is only the trajectory-level compatibility condition for this simple exponential profile to satisfy the gradient diagnostic while approaching a finite cutoff/Hubble wall.

The Hessian branch analogously constrains second-derivative data. In a single canonical direction,
\begin{equation}
 F_{\rm Hess}(d)
 =
 \ln\!\left[
 \frac{-V''(d)/V(d)}{c'}
 \right]
 \label{eq:hessian_branch_one_field}
\end{equation}
inside the tachyonic region. Exact matching to the same linear representative requires
\begin{equation}
 -\frac{V''(d)}{V(d)}
 =
 c'\mathrm e^{\gamma(d_b-d)},
 \label{eq:hessian_branch_matching}
\end{equation}
or equivalently
\begin{equation}
 V''(d)
 +
 c'\mathrm e^{\gamma(d_b-d)}V(d)
 =
 0.
 \label{eq:hessian_matched_potential_equation}
\end{equation}
This is the Hessian-branch analogue of \cref{eq:gradient_branch_slope_profile}. It determines a local representative whose tachyonic curvature reaches the chosen refined de Sitter threshold at the wall.

The two exact matchings should not generally be imposed simultaneously, because the refined criterion is an alternative rather than a conjunction. If the gradient representative is imposed exactly, then
\begin{equation}
 -\frac{V''}{V}
 =
 \sigma\gamma c\,
 \mathrm e^{\gamma(d_b-d)}
 -
 c^2\mathrm e^{2\gamma(d_b-d)}.
 \label{eq:gradient_matching_implied_hessian}
\end{equation}
This cannot equal \(c'\mathrm e^{\gamma(d_b-d)}\) with constant \(c'\) throughout a finite interval except in degenerate limits. At the wall,
\begin{equation}
 c'_{\rm wall}
 =
 \sigma\gamma c-c^2,
\end{equation}
so a tachyonic local matching requires \(\sigma\gamma c-c^2>0\). The two branches can therefore be tuned to agree at leading order at a point, but not generically over a finite region.

It remains useful to distinguish microscopic potential curvature from curvature of the conditioned statistical flow. For a general regular hard wall \(F=0\),
\begin{equation}
 \mathcal S_{\rm surv}
 =
 -\ln h
 \simeq
 -\ln F
 +
 \mathcal O(1).
 \label{eq:survival_action_as_logF}
\end{equation}
Consequently,
\begin{equation}
 \nabla_i\mathcal S_{\rm surv}
 \simeq
 -\frac{\nabla_iF}{F},
 \qquad
 \nabla_i\nabla_j\mathcal S_{\rm surv}
 \simeq
 \frac{\nabla_iF\nabla_jF}{F^2}
 -
 \frac{\nabla_i\nabla_jF}{F}.
 \label{eq:survival_hessian_general_F}
\end{equation}
In boundary-normal coordinates,
\begin{equation}
 \mathcal S_{\rm surv}
 \sim
 -\ln s,
 \qquad
 \partial_s\mathcal S_{\rm surv}
 \sim
 -\frac1s,
 \qquad
 \partial_s^2\mathcal S_{\rm surv}
 \sim
 \frac1{s^2}.
 \label{eq:survival_hessian_compact}
\end{equation}
The corresponding conditioned normal flow obeys
\begin{equation}
 \Delta b_\perp^{\rm Doob}
 \simeq
 \frac{2D_\perp}{s},
 \qquad
 \partial_s\Delta b_\perp^{\rm Doob}
 \simeq
 -\frac{2D_\perp}{s^2}.
 \label{eq:linearized_doob_compact}
\end{equation}
This motivates the statistical curvature of the conditioned flow,
\begin{equation}
 \mathcal C_{\rm Doob}(s)
 \equiv
 \left|
 \partial_s\Delta b_\perp^{\rm Doob}
 \right|.
 \label{eq:statistical_curvature_diagnostic}
\end{equation}
Using
\begin{equation}
 \delta\phi_{q,\perp}^2
 =
 2D_\perp,
 \qquad
 u
 \equiv
 \frac{s}{\delta\phi_{q,\perp}},
\end{equation}
one obtains
\begin{equation}
 \mathcal C_{\rm Doob}(s)
 \simeq
 \frac{2D_\perp}{s^2}
 =
 \frac1{u^2}.
 \label{eq:statistical_curvature_in_kick_units}
\end{equation}
Thus, when the distance is expressed relative to the one-e-fold rms stochastic displacement,
\begin{equation}
 u\sim1
 \qquad\Longrightarrow\qquad
 \mathcal C_{\rm Doob}
 \sim
 \mathcal O(1).
 \label{eq:statistical_curvature_order_one}
\end{equation}
This is a dimensionless statement tied to the chosen one-e-fold normalization: it says that the Jacobian of the conditioned drift becomes order unity at a distance comparable with the rms displacement accumulated over one e-fold. It does not identify \(u=1\) with a physical resolution limit, nor does it define a universal microscopic curvature threshold. It must not be identified with the refined de Sitter Hessian constant \(c'\), because \(\mathcal C_{\rm Doob}\) is constructed from the Jacobian of the Doob drift rather than from the Hessian of the microscopic scalar potential.

If the operational loss function is chosen to be \(F_{\rm Hess}\), the microscopic input is the refined de Sitter Hessian branch. If instead \(F\) is a cutoff/Hubble ratio, the same order-one \(\mathcal C_{\rm Doob}\) is a statistical boundary-layer effect. In both cases, the local singular response is controlled by the vanishing of the chosen operational loss function, not by the appearance of a new microscopic scalar potential.

This motivates the inverse question addressed next: whether a conditioned statistical drift determines a scalar survival boundary and, when it does, which Doob-equivalence class of operational loss surfaces it reconstructs.
\section{Boundary-Normal Universality and Inverse Reconstruction}
\label{sec:inverse_reconstruction}

The preceding sections used the survival map in the forward direction: compactification or Swampland data specify a loss channel, and the survival problem computes the conditioned drift. The same structure can be read backward. Given a compatible statistical drift, one may ask whether it comes
from a scalar operational loss surface, and if so which local boundary class it reconstructs. The inverse map does not identify a unique microscopic mechanism; it reconstructs the operational survival data seen by the conditioned ensemble. For a regular hard wall, finite-horizon effects and global survival data do not change the leading \(1/s_A\) behavior of the inward normal projection. The direction of the full singular vector remains sensitive to the diffusion tensor, while soft killing constitutes a distinct survival prescription and need not produce a hard-wall singularity at all.

There are therefore two logically different inverse problems.  The first asks whether a measured conditioned response is generated by a scalar
survival function \(h\).  The second, stronger question asks whether that scalar survival function has the regular linear zero required by a hard
absorbing boundary.  Only the latter reconstructs a boundary-normal first-passage class.

\subsection{Doob-equivalent boundary data}

For a regular hard-wall channel \(A\), the forward local map gives
\begin{equation}
 \Delta b^i_{\rm Doob}
 \simeq
 2D^{ij}\nabla_j\ln F_A
 +
 \mathcal O(1).
 \label{eq:inverse_forward_map}
\end{equation}
Only the singular logarithmic behavior is universal. Two regular positive functions are locally Doob-equivalent if they vanish on the same hypersurface and have the same linear zero,
\begin{equation}
 F_1
 =
 aF_2+\mathcal O(F_2^2),
 \qquad
 a>0 .
 \label{eq:doob_equivalence_form_condition}
\end{equation}
Equivalently,
\begin{equation}
 \nabla_i\ln F_1
 =
 \nabla_i\ln F_2+\mathcal O(1).
 \label{eq:doob_equivalence_log_condition}
\end{equation}
The induced drifts therefore differ only by regular terms and share the same singular contribution. This is an equivalence of survival-conditioned
responses, not of microscopic origins: different cutoff normalizations, potential diagnostics, or compactification thresholds may define the same
leading absorbing surface while retaining different subleading and global data.

In particular, multiplication of a regular boundary function by a smooth strictly positive factor,
\begin{equation}
 \widetilde F_A(\phi)
 =
 \mathrm e^{g(\phi)}F_A(\phi),
 \label{eq:doob_positive_rescaling_revision}
\end{equation}
changes
\begin{equation}
 \nabla_i\ln \widetilde F_A
 =
 \nabla_i\ln F_A+\nabla_i g
\end{equation}
only by a regular term.  Smooth positive rescaling leaves the domain, and hence the entire survival problem for a fixed generator, unchanged. Changing a physical threshold instead moves the zero set: it preserves the local pole law at the new regular wall but generally changes finite-distance survival probabilities.

\subsection{Inverse reconstruction and obstruction}

Suppose that a local conditioned statistical drift \(\Delta b_{\rm stat}^{i}\) is known and that the diffusion tensor is invertible on the relevant stochastic subspace. Define the lowered one-form
\begin{equation}
 \omega_i
 \equiv
 \frac12
 \left(D^{-1}\right)_{ij}
 \Delta b_{\rm stat}^{j}.
 \label{eq:inverse_survival_one_form}
\end{equation}
If the observed response is the full Doob correction generated by a scalar survival function \(h\), then at each fixed time or remaining horizon
\begin{equation}
 \omega_i
 =
 \nabla_i\ln h.
 \label{eq:inverse_full_doob_exactness}
\end{equation}
Thus \(\omega\) must be locally exact. In a simply connected patch, the corresponding local integrability condition is
\begin{equation}
 \mathcal C_{ij}^{\rm inv}
 \equiv
 \nabla_i\omega_j-\nabla_j\omega_i
 =
 0.
 \label{eq:inverse_survival_obstruction}
\end{equation}
When this condition holds, the scalar survival weight can be reconstructed locally as
\begin{equation}
 h(\phi)
 =
 h_\star
 \exp\!\left(
 \int_{\phi_\star}^{\phi}
 \omega_i\,\mathrm d\phi^i
 \right),
 \qquad
 h_\star>0,
 \label{eq:inverse_survival_weight_reconstruction}
\end{equation}
with path-independent integral in the patch. On a multiply connected domain, local closure is not sufficient for a global reconstruction: the periods of \(\omega\) around all noncontractible closed curves must also vanish.

For an explicitly time-dependent or finite-horizon problem, this reconstruction is performed separately on each fixed-\(N\) or fixed-\(\tau\) slice.  The normalization \(h_\star\) may then depend on that time label. Spatial exactness alone does not determine its time evolution.

Moreover, Eq.~\eqref{eq:inverse_survival_weight_reconstruction} reconstructs a positive scalar weight, but local exactness by itself is not sufficient to show that the weight is a physical survival probability.  For a specified stochastic generator and loss prescription, the reconstructed function must also satisfy the appropriate backward equation, terminal data, boundary or
killing conditions, and probability range.  Schematically,
\begin{equation}
 0\leq h\leq1,
 \qquad
 \partial_\tau h
 =
 \left(
 {\cal L}-\kappa_{\rm kill}
 \right)h
 \label{eq:inverse_survival_pde_requirement_revision}
\end{equation}
in the autonomous remaining-horizon representation, with the corresponding nonautonomous equation when the domain or generator depends explicitly on time.  Exactness is therefore a necessary local structural test of a scalar Doob representation, not by itself a complete survival model.

This exactness requirement is a test of a scalar Doob representation for the assumed diffusion tensor and stochastic state variables. Multiple absorbing components, soft killing, explicit time dependence, and finite-horizon conditioning do not by themselves violate it; each still produces a scalar survival function and hence an exact \(\omega=\mathrm d\ln h\) on every fixed time slice. A statistically significant nonzero curl can instead indicate a misspecified diffusion tensor, estimation error, degenerate diffusion, hidden or projected stochastic variables, a non-Markovian reduced state, or a response that is not a pure scalar Doob correction.

If the diffusion tensor is degenerate, \(D^{-1}\) in Eq.~\eqref{eq:inverse_survival_one_form} should not be interpreted as an ordinary inverse on the full field space.  The reconstruction can then be posed only on the stochastic subspace on which diffusion has nonzero rank, or with an appropriate reduced description.  Apparent nonintegrability produced by inverting directions that carry no stochastic noise is not a physical obstruction to a scalar Doob transform.

A simple local example illustrates the obstruction. In coordinates
\((s,y)\),
\begin{equation}
 \omega
 =
 \dd\ln s
 \qquad\Longrightarrow\qquad
 \dd\omega=0,
 \qquad
 h\propto s,
 \label{eq:inverse_exact_example}
\end{equation}
whereas
\begin{equation}
 \omega
 =
 \dd\ln s+\epsilon y\,\dd s
 \qquad\Longrightarrow\qquad
 \dd\omega
 =
 \epsilon\,\dd y\wedge\dd s
 \neq0.
 \label{eq:inverse_obstruction_example}
\end{equation}
The second response cannot be represented as a scalar Doob correction for the assumed \(D^{ij}\) and state space.

A weaker inverse question concerns only the leading hard-wall singularity. If a regular absorbing surface is represented by \(F_{\rm eff}=0\), then
\begin{equation}
 \omega_i
 =
 \nabla_i\ln F_{\rm eff}
 +
 \mathcal O(1)
 \label{eq:inverse_log_gradient_condition}
\end{equation}
near the wall. Only the singular part is then fixed by the hard-wall geometry; the regular remainder need not be identified with the same local boundary function.

Consequently, the full one-form \(\omega\) need not coincide with \(\dd\ln F_{\rm eff}\) away from the wall for two responses to belong to the same leading hard-wall class.  What is universal in this weaker reconstruction is the coefficient and orientation of the singular \(1/s_A\) contribution.

Reconstructing a scalar logarithmic weight is therefore not yet sufficient to establish a hard-wall interpretation. Such an interpretation additionally requires a regular zero set and linear proper-distance behavior,
\begin{equation}
 h
 \propto
 s
 \qquad
 \text{as}
 \qquad
 s\rightarrow0^+.
 \label{eq:inverse_hard_wall_requirement}
\end{equation}
An integrable scalar weight without such a zero may instead describe soft killing, finite-horizon reweighting, or another nonsingular conditioning prescription.

Conversely, a higher-order or otherwise nonregular zero would define a different local asymptotic class and should not be identified with the regular absorbing-wall law derived in this work.

\subsection{Boundary-normal universality}
\label{subsec:boundary_normal_universality}

The repeated one-dimensional result \(F_{\rm eff}\propto d_b-d\) is a coordinate expression of a more general normal form. Let \(F_A=0\) be a regular hard boundary, with
\begin{equation}
 F_A>0,
 \qquad
 \Sigma_A:\ F_A=0,
 \qquad
 \nabla_iF_A\neq0
 \quad
 \text{on }
 \Sigma_A .
 \label{eq:boundary_normal_regular_wall}
\end{equation}
Let \(s_A\) be the inward proper distance to \(\Sigma_A\), and let \(\sigma^a\) denote coordinates along the boundary. Taylor expansion in boundary-normal coordinates gives
\begin{equation}
 F_A(\phi)
 =
 \lambda_A(\sigma)s_A
 +
 \mathcal O(s_A^2),
 \qquad
 \lambda_A(\sigma)
 =
 |\nabla F_A|_{\Sigma_A}
 >
 0 .
 \label{eq:boundary_normal_expansion}
\end{equation}
Therefore
\begin{equation}
 \nabla_i\ln F_A
 =
 \frac{\nabla_i s_A}{s_A}
 +
 \nabla_i\ln\lambda_A
 +
 \mathcal O(1),
 \label{eq:boundary_normal_log_gradient}
\end{equation}
and the singular conditioned drift is
\begin{equation}
 \Delta b_{\rm Doob}^i
 \simeq
 2D^{ij}
 \frac{\nabla_j s_A}{s_A}.
 \label{eq:boundary_normal_drift}
\end{equation}
The positive normalization \(\lambda_A\), threshold conventions, and smooth rescalings of \(F_A\) affect only regular terms. Conversely, a compatible drift with this singular normal form reconstructs
\begin{equation}
 F_{\rm eff}
 \propto
 s_A
 \label{eq:boundary_normal_inverse}
\end{equation}
up to Doob-equivalent regular redefinitions.

This is the local universality statement. For a single regular hard survival channel with nonzero normal diffusion, the inward normal component of the leading finite-horizon Doob drift is universally \(2D_\perp/s_A\). The full singular vector is directed along \(D^{ij}\nabla_j s_A\); anisotropic or mixed diffusion can therefore generate singular components that are not aligned with the metric normal. Global survival, horizon dependence away from the wall, boundary curvature, tangential data, multi-boundary corners, and soft killing determine the regular terms and the global conditioned process.

This universality is strictly local.  It does not imply that a stationary flat-wall model reproduces a finite-horizon survival probability for a moving boundary, nor that two Doob-equivalent local zeros have identical global first-passage statistics.  Boundary motion, finite transition width, global domain geometry, and the full drift and diffusion fields remain essential for quantitative survival probabilities away from the asymptotic wall layer.

\subsection{Cutoff/Hubble and potential diagnostics as realizations}

A broad one-dimensional cutoff/Hubble channel has the local form
\begin{equation}
 F_A(d)
 =
 F_{0,A}-\gamma_A d,
 \qquad
 \gamma_A>0,
 \qquad
 d_b^{(A)}
 =
 \frac{F_{0,A}}{\gamma_A}.
 \label{eq:inverse_cutoff_hubble_normal_form}
\end{equation}
On the controlled side \(d<d_b^{(A)}\),
\begin{equation}
 F_A(d)
 =
 \gamma_A
 \bigl(
 d_b^{(A)}-d
 \bigr),
 \qquad
 s_A
 =
 d_b^{(A)}-d .
 \label{eq:inverse_cutoff_hubble_wall_form}
\end{equation}
The forward drift in the \(d\)-coordinate is then
\begin{equation}
 \Delta b_A^d
 \simeq
 -\frac{2D}{d_b^{(A)}-d},
 \label{eq:inverse_cutoff_hubble_drift}
\end{equation}
so for constant local \(D\),
\begin{equation}
 \omega_d
 =
 \frac{1}{2D}\Delta b_A^d
 =
 -\frac{1}{d_b^{(A)}-d},
 \qquad
 F_{\rm eff}^{(A)}
 \propto
 d_b^{(A)}-d
 \propto
 F_A .
 \label{eq:inverse_cutoff_hubble_reconstruction}
\end{equation}
This is the one-dimensional realization of the boundary-normal result above.

When \(H\) varies along the trajectory, the relevant rates are relative rates. For the channels used in the paper,
\begin{equation}
\begin{array}{c|c|c}
\text{channel}
&
F_A(d)
&
\gamma_A
\\ \hline
\text{species/Hubble}
&
F_{0,{\rm sp}}-(\lambda_{\rm sp}-\beta)d
&
\lambda_{\rm sp}-\beta
\\[2pt]
\text{direct string/Hubble}
&
F_{0,s}-(\lambda_s-\beta)d
&
\lambda_s-\beta
\\[2pt]
\text{direct KK/Hubble}
&
F_{0,{\rm KK}}-(\lambda_{\rm KK}-\beta)d
&
\lambda_{\rm KK}-\beta
\\[2pt]
\text{magnetic WGC/Hubble}
&
F_{0,{\rm WGC}}-(\alpha_g-\beta)d
&
\alpha_g-\beta
\end{array}
\label{eq:inverse_channel_table}
\end{equation}
with \(\gamma_A>0\) when the wall is reached at finite increasing \(d\).

For the species row, \(\lambda_{\rm sp}\) denotes the local logarithmic rate of the \emph{combined} species scale obtained from the actual density of states.  For a jointly charged multidimensional lattice, it must therefore be computed from the full spectrum, rather than by summing one-dimensional axis towers.  In the monomial counting regime of
Eq.~\eqref{eq:general_joint_species_count_revision}, for locally constant
\(C\),
\begin{equation}
 \lambda_{\rm sp}
 =
 -\frac{\mathrm d\ln\Lambda_{\rm sp}}{\mathrm dd}
 =
 \frac{\sum_A r_A\mu_A}{p_{\rm eff}+2}.
 \label{eq:inverse_joint_species_rate_revision}
\end{equation}

The fixed-\(H_*\) compactification example corresponds to \(\beta=0\). In the volume coordinate used there,
\begin{equation}
 d
 =
 \sqrt{\frac23}\ln\mathcal V_s,
 \qquad
 F_{{\rm KK}/H}^{(q_{\rm ctrl})}
 =
 \lambda_{\rm KK}s_{\rm KK},
 \qquad
 F_{s/H}^{(q_{\rm ctrl})}
 =
 \lambda_s s_s,
 \label{eq:inverse_direct_volume_normal_form}
\end{equation}
so the inverse map reconstructs the same direct KK and string loss surfaces, up to their positive rates, without erasing their distinct compactification origins.

Potential-based diagnostics fit the same logic once they are written as operational loss functions. For example, the gradient de Sitter diagnostic uses
\begin{equation}
 F_{\nabla}(\phi)
 =
 \ln\frac{|\nabla\ln V|}{c},
 \label{eq:inverse_gradient_ds_boundary}
\end{equation}
and the Hessian branch of the refined diagnostic uses
\begin{equation}
 F_{\rm Hess}(\phi)
 =
 \ln\frac{
 -\lambda_{\min}(\nabla_i\nabla_jV/V)
 }{c'},
 \label{eq:inverse_hessian_ds_boundary}
\end{equation}
in the tachyonic region. If either function has a regular zero surface, the inverse map reconstructs its boundary-normal class in exactly the same sense as for a cutoff/Hubble wall.

Thus, \emph{after} an independent operational prescription has promoted the potential criterion to survival data, a cutoff/Hubble threshold, a gradient de Sitter loss function, and a refined-Hessian loss function can belong to the same leading boundary-normal Doob class when their regular zeros share the same normal form. Their microscopic meanings remain different.

For a combined species cutoff, the reconstructed scalar is the combined operational function
\begin{equation}
 F_{\rm sp/H}
 =
 \ln\frac{\Lambda_{\rm sp}}{H},
 \label{eq:inverse_combined_species_boundary}
\end{equation}
not the mass of any individual tower.

Here \(\Lambda_{\rm sp}\) is understood to be determined from the full species count,
\begin{equation}
 \Lambda_{\rm sp}^{2}
 \mathcal N_{\rm tot}
 \bigl(
 \Lambda_{\rm sp},\phi
 \bigr)
 \sim
 M_{\rm Pl}^{2},
 \label{eq:inverse_full_species_definition_revision}
\end{equation}
so jointly charged states are already part of the scalar wall being reconstructed.

If
\begin{equation}
 \Delta b_{\rm stat}^i
 \simeq
 2D^{ij}\nabla_j\ln F_{\rm sp/H},
 \label{eq:inverse_combined_species_drift}
\end{equation}
then the obstruction vanishes at leading singular order and the inverse map returns the combined scalar wall. A nonexact lowered full response instead fails the scalar Doob test for the assumed generator; adding more absorbing channels alone does not remove that obstruction.

The reconstruction of a formal species surface does not, however, establish that this surface is an admissible stopping boundary of a reduced few-field stochastic EFT.  If a direct KK, string, or other light-state threshold is encountered first and invalidates the retained generator, the inverse reconstruction identifies compactification boundary data lying outside the domain of that reduced stochastic model.  Operationally, the relevant first loss of control remains the earliest admissible stopping surface unless the additional degrees of freedom have been incorporated into the stochastic state.

The forward and inverse maps therefore identify the same object from opposite directions: not a microscopic Swampland mechanism itself, but the operational boundary-normal class through which that mechanism acts on the stochastic ensemble. Microscopic distinctions re-enter through wall locations, rates, tangential dependence, regular terms, anisotropic diffusion, global survival data, and independent spectral information.

The inverse hierarchy is therefore
\begin{equation}
 \Delta b_{\rm stat}
 \;\longrightarrow\;
 \omega
 \;\longrightarrow\;
 \text{scalar Doob test}
 \;\longrightarrow\;
 h
 \;\longrightarrow\;
 \text{hard-wall test}
 \;\longrightarrow\;
 F_{\rm eff}\propto s,
 \label{eq:inverse_reconstruction_hierarchy_revision}
\end{equation}
supplemented by the independent requirement that the reconstructed stopping surface lie within the validity domain of the stochastic variables used to infer it.  This separates what can be learned from conditioned statistical flow alone from the microscopic information that must still be supplied by the underlying EFT or compactification.

\section{Discussion and Conclusions}
\label{sec:discussion}

We have developed a finite-horizon stochastic framework linking quantum-gravity-motivated control data to cosmological path ensembles. The terminology is essential. In the title and throughout this paper, ``Swampland boundaries'' refers to operational surfaces at which a prescribed effective-field-theory description loses control according to Swampland or compactification data. Crossing such a surface does not by itself prove that the underlying theory belongs to the Swampland. It means only that the chosen EFT, field content, or parametric expansion should no longer be extrapolated past that surface.

A further qualification is equally important. A candidate surface obtained from compactification or quantum-gravity data becomes an admissible stochastic first-passage boundary only if the stochastic description being used remains valid up to that surface. If additional degrees of freedom become dynamically relevant earlier, the reduced stochastic state must be enlarged or the evolution must be terminated at the earlier loss of validity.

The framework separates two logically distinct inputs. The loss prescription specifies what counts as remaining controlled through hard surfaces \(F_A=0\), soft killing profiles \(\kappa_A\), or finite-horizon completion conditions. The stochastic generator,
\begin{equation}
 {\cal L} = b^i\nabla_i + D^{ij}\nabla_i\nabla_j,
 \label{eq:conclusion_generator}
\end{equation}
specifies how the retained variables evolve before loss. These inputs are independent: microscopic control data do not by themselves determine the stochastic generator, while a stochastic generator does not determine which microscopic event should count as loss of EFT control. Once both inputs are given, they determine the finite-horizon survival probability \(h\), the survival action
\begin{equation}
 \mathcal S_{\rm surv} = -\ln h,
 \label{eq:conclusion_survival_action}
\end{equation}
and the Doob correction
\begin{equation}
 \Delta b_{\rm Doob}^{i}
 =
 2D^{ij}\nabla_j\ln h
 =
 -2D^{ij}\nabla_j\mathcal S_{\rm surv}.
 \label{eq:conclusion_doob_revision}
\end{equation}
This correction is the statistical drift of the ensemble conditioned on future survival. It is not a new microscopic interaction and does not alter the unconditioned equations of motion.

For a regular absorbing boundary, let \(s\) denote inward proper distance and let \(D_\perp\) be the normal diffusion coefficient. Whenever the normal diffusion is nondegenerate, regularity of the backward problem implies \(h\propto s\) near the wall and therefore
\begin{equation}
 \Delta b_{\perp}^{\rm Doob}
 \simeq
 \frac{2D_\perp}{s}.
 \label{eq:conclusion_normal_wall_law}
\end{equation}
If the same surface is represented locally by a regular function \(F=0\), this becomes
\begin{equation}
 \Delta b_{\rm Doob}^{i}
 \simeq
 2D^{ij}\frac{\nabla_jF}{F}.
 \label{eq:conclusion_wall_revision}
\end{equation}
The microscopic construction fixes the wall location, its normal rate, tangential structure, diffusion tensor, and global boundary-value problem. The leading inward normal projection depends only on the local proper distance and the normal diffusion coefficient, \(n_i\Delta b_{\rm Doob}^{i}\simeq2D_\perp/s\). The direction of the full singular drift vector is \(D^{ij}n_j\) and can therefore retain anisotropic or mixed-diffusion information.

This universality is local. It does not imply that a static planar hard-wall problem gives the correct finite-horizon survival probability for a moving or finite-width boundary. Boundary motion, global geometry, soft loss, the full drift and diffusion fields, and the duration of the conditioning problem remain part of any quantitative finite-time prediction.

The inverse construction reaches the same structure from the opposite direction. Given a conditioned response \(\Delta b^i\), define the lowered one-form
\begin{equation}
 \omega_i
 \equiv
 \frac12\left(D^{-1}\right)_{ij}\Delta b^j.
 \label{eq:conclusion_inverse_one_form}
\end{equation}
A single scalar survival function exists locally only when \(\omega_i\) is exact,
\begin{equation}
 \omega_i = \nabla_i\ln h.
 \label{eq:conclusion_inverse_exactness}
\end{equation}
A nonvanishing curl obstructs a scalar Doob representation for the assumed diffusion tensor and stochastic state variables. It can indicate estimation error, a misspecified \(D^{ij}\), degenerate diffusion, hidden or projected variables, a non-Markovian reduced description, or a response that is not a pure scalar Doob correction. Multiple absorbing components, soft killing, explicit time dependence, and finite-horizon conditioning still generate an exact form \(\omega=\mathrm d\ln h\) on each fixed time slice and are not, by themselves, sources of curl. On a multiply connected domain, global reconstruction additionally requires vanishing periods around noncontractible closed curves. A hard-wall interpretation further requires a regular zero set with linear proper-distance behavior.

Exactness is therefore a necessary local structural condition, not a complete survival model. The reconstructed scalar must additionally satisfy the appropriate backward equation, terminal and loss data, and \(0\leq h\leq1\). A regular hard-wall interpretation further requires \(h\propto s\) at the reconstructed boundary. Thus the inverse chain distinguishes a scalar Doob representation from a physical survival probability, and both from the stronger reconstruction of a regular absorbing surface.

The finite-horizon formulation yields more than the local Doob response. It returns the controlled-history fraction \(h\) and the total loss fraction \(1-h\). For a purely hard-wall loss problem without soft killing, the total loss occurs through first exit, and therefore
\begin{equation}
 f_{\rm exit} = -\partial_\tau h,
 \label{eq:conclusion_exit_density}
\end{equation}
with exit hazard
\begin{equation}
 \Gamma_{\rm exit}
 =
 \frac{f_{\rm exit}}{h}
 =
 \partial_\tau\mathcal S_{\rm surv}
 =
 -\partial_\tau\ln h.
 \label{eq:conclusion_survival_hazard}
\end{equation}

When a nonzero soft-killing profile is present, the same derivatives instead define the total loss density and total loss hazard,
\begin{equation}
 f_{\rm loss}
 =
 -\partial_\tau h,
 \qquad
 \Gamma_{\rm loss}
 =
 \frac{f_{\rm loss}}{h}
 =
 -\partial_\tau\ln h.
 \label{eq:conclusion_total_loss_observables}
\end{equation}
In that case, \(f_{\rm loss}\) contains both hard-boundary first-exit and bulk-killing contributions; the boundary first-exit density must be distinguished from the soft-killing contribution when the two mechanisms are analyzed separately.

For the driftless normal half-line model, the minimum initial proper distance required to retain a fraction \(p\) of histories for a duration \(\tau\) is
\begin{equation}
 s_p(\tau)
 =
 2\sqrt{D_\perp\tau}\,
 \operatorname{erf}^{-1}(p).
 \label{eq:conclusion_margin_revision}
\end{equation}
In local boundary-function language,
\begin{equation}
 F_A
 \gtrsim
 2|\nabla F_A|
 \sqrt{D_\perp\tau}\,
 \operatorname{erf}^{-1}(p).
 \label{eq:conclusion_function_margin_revision}
\end{equation}
This converts a microscopic control surface into a finite-time stochastic margin at a prescribed confidence level. The formula is local and specific to the static driftless hard-wall approximation, but the underlying principle is general: EFT control is not determined only by the instantaneous distance from a wall, but also by the diffusion strength and the duration for which control must be maintained.

The explicit two-field example illustrates this point beyond the driftless normal approximation. A light spectator modulus was coupled to an exponentially decreasing KK scale, while its classical drift restored it away from the selected loss surface. After measuring distance from the wall in units of the one-e-fold rms stochastic displacement, the problem reduced, under the stationary quasi-de Sitter assumptions of \cref{subsec:two_field_survival_benchmark}, to a killed Ornstein--Uhlenbeck process,
\begin{equation}
 \mathrm du
 =
 \nu_\chi\left(u_{\rm eq}-u\right)\mathrm dN
 +
 \mathrm dW_N,
 \qquad
 u>0,
 \label{eq:conclusion_ou_benchmark}
\end{equation}
with absorption at \(u=0\). The stationary example used
\begin{equation}
 q_{\rm ctrl}=3,
 \qquad
 \frac{m_\chi}{H_*}=0.2,
 \qquad
 u_0=5,
 \qquad
 u_{\rm eq}=10,
 \label{eq:conclusion_benchmark_parameters}
\end{equation}
so the classical drift pointed toward the controlled interior while stochastic fluctuations still produced a nonzero first-passage probability. The backward solution and the independently simulated killed ensemble showed the same qualitative structure: an early first-exit peak, cumulative loss that approaches unity over sufficiently long horizons, and a positive surviving-ensemble displacement
\begin{equation}
 \Delta\langle u(N)\rangle_{\rm surv}
 =
 \mathbb E\!\left[
 u(N)\,\middle|\,T_{\rm exit}>N
 \right]
 -
 \mathbb E[u(N)]
 >
 0.
 \label{eq:conclusion_survival_bias}
\end{equation}
The surviving histories are therefore statistically biased away from the loss surface and toward a larger KK/Hubble control margin. This does not mean that stochasticity improves the survival probability of the original ensemble. Stochastic fluctuations generate the first-exit events; the bias arises because wall-directed histories are removed from the ensemble being conditioned on survival.

The numerical probabilities quoted for this example are conditional on the stationary-wall construction. For the underlying control function
\begin{equation}
 F_{\rm KK}
 =
 \ln\!\left[
 \frac{M_{\rm KK}(\chi)}
 {q_{\rm ctrl}H(N)}
 \right],
 \label{eq:conclusion_moving_kk_function_revision}
\end{equation}
a varying Hubble scale moves the KK boundary. With \(\epsilon_H=-\mathrm d\ln H/\mathrm dN\), its displacement is
\begin{equation}
 \Delta\chi_b(N)
 =
 \frac{1}{\alpha_\chi}
 \int_0^N
 \epsilon_H(N')\,\mathrm dN'.
 \label{eq:conclusion_moving_kk_displacement_revision}
\end{equation}
The stationary probabilities therefore apply only when this accumulated motion is negligible relative to the boundary-layer distances relevant to the calculation. Generic slow-roll evolution need not satisfy that condition; in that case the correct problem is the nonautonomous moving-wall backward problem.

This example also clarifies the distinction between local and global statements. The near-wall law \(\Delta b_{\rm Doob}^{u}\simeq1/u\) is universal, but the complete survival probability, exit-time distribution, and surviving-ensemble bias depend on the global drift, the initial condition, the equilibrium location, and the chosen horizon. A local protected zero or candidate boundary is therefore not by itself a complete stochastic survival problem. It becomes one only after a domain, a generator, an absorbing or killing prescription, and a time horizon have been specified.

The stochastic interpretation is conditional on the validity of that generator. The general survival equation and Doob transform do not assume slow roll. The commonly used diffusion tensor
\begin{equation}
 D^{ij}
 \simeq
 \frac{H^2}{8\pi^2}G^{ij}
 \label{eq:conclusion_light_field_diffusion}
\end{equation}
is a specialization appropriate to sufficiently light, weakly mixed, coarse-grained fields. Outside that regime, the noise matrix must be derived from the relevant fluctuation modes, and a phase-space or otherwise enlarged stochastic system may be required. Evidence for steep potentials in strict asymptotic string regimes therefore prevents an automatic application of a slow-roll field-space generator along many runaway directions. It does not, however, exclude a finite-horizon survival problem built from a separately justified spectator sector, a finite-distance control surface, or a valid phase-space generator. The two-field construction presented here should accordingly be interpreted as a stationary quasi-de Sitter spectator model designed to isolate first-passage selection near a prescribed KK-control surface, not as a complete stabilized compactification model.

Likewise, the quantity
\begin{equation}
 \delta\phi_{q,\perp}
 =
 \sqrt{2D_\perp}
 \label{eq:conclusion_one_efold_displacement_revision}
\end{equation}
used throughout the examples is the rms stochastic displacement accumulated over one e-fold in the Markovian white-noise normalization. It is not a fundamental spatial resolution limit. More generally,
\begin{equation}
 \sqrt{\left\langle(\Delta\phi_\perp)^2\right\rangle}
 \simeq
 \sqrt{2D_\perp\Delta N},
 \label{eq:conclusion_time_dependent_displacement_revision}
\end{equation}
within that approximation. If the coarse-graining procedure produces a finite correlation time or colored noise, the white-noise description itself must be reassessed on shorter time scales. Consequently, ``one-kick'' comparisons in this paper diagnose proximity to a one-e-fold stochastic displacement scale rather than a microscopic resolution threshold.

A separate qualification concerns the operational placement of the wall. The nominal equality
\begin{equation}
 \Lambda_{\rm QG}=H
 \label{eq:conclusion_nominal_wall}
\end{equation}
may lie at the edge of the EFT regime one wishes to retain. We therefore introduced a conservative control factor \(q_{\rm ctrl}\geq1\) and placed the absorbing surface at
\begin{equation}
 \Lambda_{\rm QG}=q_{\rm ctrl}H.
 \label{eq:conclusion_control_wall}
\end{equation}
Increasing \(q_{\rm ctrl}\) moves the operational boundary toward the safer side of the hierarchy without changing the local logarithmic rate or the universal normal wall law. It should be regarded as part of the chosen loss prescription, not as a new microscopic scale. In particular, finite-horizon first-passage probabilities can be highly sensitive to \(q_{\rm ctrl}\) when the physical initial condition lies within only a few stochastic displacement scales of the selected surface. A meaningful sensitivity test must therefore hold the physical model and initial condition fixed while varying \(q_{\rm ctrl}\); holding the kick-normalized distances \(u_0\) or \(u_{\rm eq}\) fixed would instead retune the physical hierarchy as the wall is moved.

This distinction is especially important for the near-boundary two-field example. Its chosen values \(u_0=5\) and \(u_{\rm eq}=10\) deliberately place the system in a narrow stochastic first-passage layer. The resulting loss probabilities demonstrate how the survival framework operates once a loss prescription has been specified; they should not be interpreted as a robust compactification prediction insensitive to the microscopic placement of the EFT endpoint.

Species-scale boundaries require an additional ordering check. The species scale must be computed from the actual density of states, including jointly charged lattice states when present, rather than from an additive count of one-dimensional towers when that approximation is not justified. More importantly, a formal species/Hubble equality need not be the first failure of the reduced stochastic EFT. If a KK or other tower threshold is encountered earlier and new states become relevant to the cosmological dynamics, the few-field generator cannot consistently be evolved unchanged to a later species wall. The physically operational stopping event is then the earliest loss of validity among the retained control channels, unless the newly light sector has been incorporated explicitly into an enlarged stochastic description.

Potential-based gradient and Hessian criteria require a different treatment. Unless an independent physical prescription makes their violation absorbing, they do not automatically define EFT-loss surfaces. They can instead enter as comparison diagnostics, candidate walls, or soft-loss inputs. This distinction prevents a conjectural inequality from being promoted directly
to a stochastic absorbing boundary without specifying the physical mechanism by which histories are actually removed.

These qualifications make explicit the domain in which the survival construction has physical content.  Compactification or quantum-gravity input may identify candidate loss surfaces, but a candidate becomes an admissible stopping boundary only if the retained stochastic EFT remains valid up to that surface.  The survival probability, loss statistics, and conditioned dynamics are therefore defined only after this consistency requirement is imposed.  In particular, a protected zero, a formal cutoff equality, or a conjectural potential criterion does not by itself constitute a physical stochastic stopping event.

The present work establishes a general local and finite-horizon baseline rather than a model-specific global prediction from a fully stabilized compactification. Once the control data and a valid generator are supplied, the general framework computes
\begin{equation}
 \left\{
 h,\;
 1-h,\;
 f_{\rm loss},\;
 \Gamma_{\rm loss},\;
 \Delta b_{\rm Doob}^{i},\;
 \Delta\langle X\rangle_{\rm surv}
 \right\},
 \label{eq:conclusion_computable_outputs}
\end{equation}
where \(f_{\rm loss}=f_{\rm exit}\) and \(\Gamma_{\rm loss}=\Gamma_{\rm exit}\) for a purely hard-wall problem without soft killing. It also supplies an inverse test of whether a measured conditioned response is compatible with a single scalar survival function. What remains model-dependent is the derivation of the full drift and diffusion tensors, the microscopic placement and softness of the control surface, and the global solution in a complete multifield compactification.

The conceptual contribution is therefore not the use of absorbing boundaries or Doob conditioning by itself. Rather, it is the construction of a dictionary that takes specified compactification- or quantum-gravity-motivated EFT-control data, tests whether the associated loss surface is admissible for the retained stochastic description, and turns the resulting stopping prescription into finite-horizon first-loss probabilities and conditioned cosmological histories.

The natural extensions are therefore concrete. One may derive the stochastic generator and control surfaces within the same stabilized multifield compactification, include explicitly time-dependent Hubble profiles and moving controlled domains through the two-time backward problem, replace nominal hard thresholds by microscopic soft-killing rates, incorporate several nonseparable loss channels, or work directly in phase space when slow roll fails. One may also study how survival conditioning propagates into inflationary observables, reheating histories, primordial-black-hole statistics, and other rare-event sectors. These developments would add model-specific predictive content while preserving the central result established here: once a quantum-gravity-motivated control prescription and a self-consistent stochastic domain have been specified, their intersection defines a first-passage problem whose finite-time survival probabilities, first-loss statistics, and conditioned path ensemble can be computed systematically.

\section*{Acknowledgments}
The work of O.G. was supported in part by the Istanbul Technical University Research Fund under grant No.~2025-47239. O.G. also acknowledges stimulating discussions at Lotus \& Swamplandia 2026 and is particularly grateful to Antonio Padilla for brief but inspiring comments.
\appendix

\section{Technical Variants}
\label{app:technical_variants}

This appendix collects the technical variants needed by the main text without changing the central hard-wall survival-action construction.

\subsection{Combined species algebra and multi-boundary limits}
\label{app:boundary_webs}

Several towers that belong to one gravitational species count define one combined cutoff, but the relevant input is the density of \emph{states}, not in general a sum of independent one-dimensional tower counts.  Let
\begin{equation}
 {\cal N}_{\rm tot}(\Lambda,\phi)
 \equiv
 \#\{
 \text{states with }m_{\boldsymbol n}(\phi)<\Lambda
 \}.
 \label{eq:appendix_full_species_count}
\end{equation}
The species scale is determined implicitly by
\begin{equation}
 \Lambda_{\rm sp}^{2}\,
 {\cal N}_{\rm tot}
 \bigl(
 \Lambda_{\rm sp},\phi
 \bigr)
 \sim
 M_{\rm Pl}^{2}.
 \label{eq:appendix_species_implicit_equation}
\end{equation}
This formulation automatically includes states carrying several tower quantum numbers simultaneously.

For example, two KK directions with characteristic masses \(M_1\) and \(M_2\) have
\begin{equation}
 m_{n_1,n_2}^{2}
 =
 n_1^{2}M_1^{2}
 +
 n_2^{2}M_2^{2}.
 \label{eq:appendix_two_circle_spectrum}
\end{equation}
When the cutoff lies above both spacings, the asymptotic lattice count is
\begin{equation}
 {\cal N}_{\rm tot}(\Lambda)
 \sim
 C_2\,
 \frac{\Lambda^{2}}{M_1M_2},
 \label{eq:appendix_two_circle_count}
\end{equation}
with an order-one geometric coefficient \(C_2\).  Hence
\begin{equation}
 \Lambda_{\rm sp}^{4}
 \sim
 \frac{M_{\rm Pl}^{2}M_1M_2}{C_2},
 \label{eq:appendix_two_circle_species}
\end{equation}
rather than the cubic scaling that would follow from adding two independent one-dimensional axis towers.

More generally, suppose that in a given asymptotic regime the joint density of states takes the monomial form
\begin{equation}
 {\cal N}_{\rm tot}(\Lambda,\phi)
 \simeq
 C(\phi)\,
 \Lambda^{p_{\rm eff}}
 \prod_A
 M_A(\phi)^{-r_A},
 \qquad
 \sum_A r_A=p_{\rm eff}.
 \label{eq:appendix_joint_species_count_revision}
\end{equation}
Here the exponents are fixed within the counting regime; a change of active lattice directions requires a separate local count. Then
\begin{equation}
 \Lambda_{\rm sp}^{p_{\rm eff}+2}
 \simeq
 \frac{M_{\rm Pl}^{2}}{C(\phi)}
 \prod_A M_A(\phi)^{r_A}.
 \label{eq:appendix_joint_species_scale}
\end{equation}
Along a canonically normalized trajectory \(d\), define
\begin{equation}
 \mu_A
 \equiv
 -\frac{\mathrm d\ln M_A}{\mathrm dd}.
 \label{eq:appendix_tower_log_rates}
\end{equation}
Differentiating Eq.~\eqref{eq:appendix_joint_species_scale} gives
\begin{equation}
 \lambda_{\rm sp}
 \equiv
 -\frac{\mathrm d\ln\Lambda_{\rm sp}}{\mathrm dd}
 =
 \frac{
 \displaystyle
 \sum_A r_A\mu_A
 +
 \frac{\mathrm d\ln C}{\mathrm dd}
 }{
 p_{\rm eff}+2
 }.
 \label{eq:appendix_joint_species_rate}
\end{equation}
For locally constant \(C\), this reduces to
\begin{equation}
 \lambda_{\rm sp}
 =
 \frac{\sum_A r_A\mu_A}{p_{\rm eff}+2}.
 \label{eq:appendix_joint_species_rate_constant_C}
\end{equation}
Thus jointly charged states can change not only the normalization and location of a species wall but also its field-space gradient.

An additive form
\begin{equation}
 {\cal N}_{\rm tot}
 \simeq
 \sum_A
 \left(
 \frac{\Lambda}{M_A}
 \right)^{p_A}
 \label{eq:appendix_restricted_additive_species}
\end{equation}
may still be used when the contributing towers are genuinely distinct sectors and there is no parametrically important population of states jointly charged under several tower directions.  It should not be used as the generic description of a multidimensional KK lattice.

Finally, a formal solution of Eq.~\eqref{eq:appendix_species_implicit_equation} is only a candidate stopping surface for a reduced stochastic EFT.  If individual tower states, string states, or other degrees of freedom become dynamically relevant before the formal species/Hubble equality is reached, the reduced process must stop at that earlier loss of validity unless the new degrees of freedom
are included explicitly.

Independent loss mechanisms may instead define a local multi-boundary survival problem:
\begin{equation}
 \mathcal M_{\rm EFT}
 =
 \{
 \phi\in\mathcal M:
 F_A(\phi)>0,
 \quad
 A=1,\ldots,m
 \}.
\end{equation}
If the local survival probability factorizes near a transverse intersection as
\begin{equation}
 h(\phi,\tau)
 \sim
 C(\sigma,\tau)
 \prod_A s_A^{\nu_A},
 \qquad
 \nu_A>0,
 \label{eq:product_survival}
\end{equation}
then
\begin{equation}
 \Delta b^i_{\rm Doob}
 \simeq
 2D^{ij}
 \sum_A
 \frac{\nu_A n_j^{(A)}}{s_A}
 +
 \mathcal O(1),
 \label{eq:boundary_web_drift}
\end{equation}
or, in regular boundary-function form,
\begin{equation}
 \Delta b^i_{\rm Doob}
 \simeq
 2D^{ij}
 \sum_A
 \nu_A
 \frac{\nabla_jF_A}{F_A}
 +
 \mathcal O(1).
 \label{eq:boundary_web_function_drift}
\end{equation}
With
\begin{equation}
 \mathsf D_{AB}
 =
 n_i^{(A)}D^{ij}n_j^{(B)},
 \label{eq:normal_coupling_matrix}
\end{equation}
the projection onto the \(A\)-th boundary is
\begin{equation}
 \Delta b_{\perp A}
 \equiv
 n_i^{(A)}\Delta b^i_{\rm Doob}
 \simeq
 2\sum_B
 \frac{
 \nu_B\,\mathsf D_{AB}
 }{s_B}
 +
 \mathcal O(1).
 \label{eq:boundary_web_projection}
\end{equation}
This product ansatz is meant only as an illustrative local normal form for transverse, approximately independent absorbing faces; it is not assumed to hold for a generic corner of the EFT-control domain. The factorized form is a local assumption; nonseparable corners require the full survival action.

The distinction between a combined species wall and several independent absorbing walls is therefore physical rather than notational.  States that jointly contribute to a single gravitational species count determine one scalar function \(\Lambda_{\rm sp}(\phi)\).  By contrast, genuinely independent mechanisms for loss of control define separate stopping conditions \(F_A=0\), whose first occurrence determines the hard-loss event.

\subsection{Soft killing and finite horizons}
\label{app:soft_finite}

Gradual loss of EFT control may be represented by a nonnegative killing profile
\begin{equation}
 \kappa_{\rm kill}(\phi)\geq0.
 \label{eq:soft_killing_rate_definition}
\end{equation}
For an autonomous stochastic generator and a static controlled domain, the remaining-horizon survival function obeys
\begin{equation}
 \partial_\tau h
 =
 \left(
 \mathcal L-\kappa_{\rm kill}
 \right)h.
 \label{eq:soft_killing_survival}
\end{equation}
The associated survival action is
\begin{equation}
 \mathcal S_{\rm surv}
 \equiv
 -\ln h,
 \label{eq:soft_killing_survival_action}
\end{equation}
and the conditioned drift retains the standard Doob form
\begin{equation}
 \Delta b_{\rm Doob}^{i}
 =
 2D^{ij}\nabla_j\ln h
 =
 -2D^{ij}\nabla_j\mathcal S_{\rm surv}.
 \label{eq:soft_killing_doob}
\end{equation}
Soft killing changes the boundary-value problem through \(\kappa_{\rm kill}\), but it does not alter the exact gradient structure of the Doob correction. At each fixed remaining horizon,
\begin{equation}
 \frac12
 \left(D^{-1}\right)_{ij}
 \Delta b_{\rm Doob}^{j}
 =
 \nabla_i\ln h
 \label{eq:soft_killing_exactness}
\end{equation}
whenever the diffusion tensor is invertible on the stochastic state space.

A finite-width killing profile generally smooths the response across a model-dependent transition region. Unlike a regular Dirichlet hard wall, it does not by itself imply
\begin{equation}
 h\propto s
 \qquad
 \text{or}
 \qquad
 \Delta b_\perp^{\rm Doob}
 \simeq
 \frac{2D_\perp}{s}.
 \label{eq:soft_killing_no_universal_wall_law}
\end{equation}
Accordingly, no universal \(1/s\) coefficient should be assigned to a soft loss channel unless a separate limiting argument shows that the killing profile converges to a regular absorbing boundary.

This distinction is particularly relevant when the microscopic endpoint of EFT control is uncertain over a finite field interval.  Varying a hard-wall parameter while simultaneously retuning the physical initial distances does not test this uncertainty.  A direct comparison instead keeps the underlying physical model fixed and varies either the hard stopping prescription or a finite-width \(\kappa_{\rm kill}\) profile describing the transition out of control.

For an autonomous process without soft killing on a static controlled domain \(\mathcal M_{\rm EFT}\), finite-duration control is encoded by
\begin{equation}
 h(\phi,\tau)
 =
 \mathbb P_\phi
 \left[
 T_{\partial\mathcal M_{\rm EFT}}>\tau
 \right],
 \label{eq:finite_horizon_survival}
\end{equation}
where \(\tau\) is the prescribed remaining duration. Suppose that one subsequently evaluates this autonomous survival solution at a state-dependent horizon,
\begin{equation}
 h_\tau(\phi)
 \equiv
 h\!\left(
 \phi,\tau(\phi)
 \right).
 \label{eq:state_dependent_horizon_composition}
\end{equation}
The field-space gradient then satisfies
\begin{equation}
 \nabla_i\ln h_\tau
 =
 \left(
 \nabla_i\ln h
 \right)_\tau
 +
 \partial_\tau\ln h\,
 \nabla_i\tau,
 \label{eq:finite_horizon_chain_rule}
\end{equation}
where \(\left(\nabla_i\ln h\right)_\tau\) denotes the field-space derivative taken at fixed remaining horizon. The corresponding composed Doob response is
\begin{equation}
 \Delta b_{\rm Doob,\tau}^{i}
 =
 2D^{ij}
 \left[
 \left(
 \nabla_j\ln h
 \right)_\tau
 +
 \partial_\tau\ln h\,
 \nabla_j\tau
 \right].
 \label{eq:finite_horizon_doob_chain}
\end{equation}

This chain rule is a statement about evaluating an autonomous solution at a field-dependent duration. It is not, by itself, a general treatment of an explicitly moving absorbing boundary, and an arbitrary function \(\tau(\phi)\) does not automatically define a dynamically consistent conditioned process. Nor does such a composition, by itself, represent a path-dependent completion constraint such as the Trans-Planckian Censorship Conjecture. If the generator, killing profile, or controlled domain depends explicitly on e-fold time, one must instead use the two-time survival function
\begin{equation}
 H(\phi,N;N_f)
 =
 \mathbb E_{\phi,N}\!\left[
 e^{-\int_N^{N_f}\kappa_{\rm kill}(\phi(N'),N')\,\mathrm dN'}
 \mathbf 1_{\{T_{\rm exit}>N_f\}}\right]
 \label{eq:appendix_nonautonomous_survival}
\end{equation}
introduced in Eq.~\eqref{eq:nonautonomous_survival_revision}. It obeys
\begin{equation}
 \left(
 \partial_N
 +
 \mathcal L_N
 -
 \kappa_{\rm kill}(\phi,N)
 \right)H
 =
 0
 \label{eq:appendix_nonautonomous_backward}
\end{equation}
with the time-dependent absorbing and terminal conditions given in Eqs.~\eqref{eq:nonautonomous_absorbing_condition_revision} and \eqref{eq:nonautonomous_terminal_condition_revision}. Its conditioned drift is
\begin{equation}
 b_{\rm surv}^{i}(\phi,N;N_f)
 =
 b_N^{i}(\phi,N)
 +
 2D_N^{ij}(\phi,N)
 \nabla_j\ln H(\phi,N;N_f).
 \label{eq:appendix_nonautonomous_doob}
\end{equation}

\subsubsection{Competing completion and TCC first passage}
\label{app:tcc_competing_first_passage}

The TCC constrains the accumulated expansion of a cosmological history, rather than survival against an unrelated EFT-loss wall for a state-dependent duration.  A stochastic formulation must therefore keep track of completion and TCC violation as competing path events \cite{Bedroya:2019snp,Bedroya:2019tba,Bedroya:2020rmd,Brahma:2019vpl,Guleryuz:2021zik,Brandenberger:2021pzy}.

Let \(T_{\rm end}\) denote the first time at which the inflationary history reaches its completion surface, let \(T_{\rm EFT}\) denote the first loss of the separately specified EFT-control domain, and let \(T_{\rm TCC}\) denote the first time at which the TCC condition is violated.  The probability relevant for a history that both remains under EFT control and completes before TCC violation is
\begin{equation}
 P_{\rm comp}^{\rm TCC}
 =
 \mathbb P
 \left[
 T_{\rm end}
 <
 T_{\rm EFT},
 \;
 T_{\rm end}
 <
 T_{\rm TCC}
 \right].
 \label{eq:tcc_competing_completion_probability}
\end{equation}
Equivalently,
\begin{equation}
 P_{\rm comp}^{\rm TCC}
 =
 \mathbb P
 \left[
 T_{\rm end}
 <
 T_{\rm EFT}\wedge T_{\rm TCC}
 \right].
 \label{eq:tcc_competing_completion_probability_compact}
\end{equation}
This is a competing first-passage problem.  It is not equivalent to asking for the probability \(\mathbb P[T_{\rm EFT}>N_{\rm TCC}]\): a model with no EFT-loss wall would make the latter probability unity even if the inflationary history itself continued beyond the TCC limit.

For a fixed initial UV scale \(\Lambda_i\), a convenient accumulated TCC variable in e-fold time is
\begin{equation}
 Y(N)
 \equiv
 (N-N_i)
 +
 \ln\!\left[
 \frac{H(N)}{\Lambda_i}
 \right].
 \label{eq:tcc_accumulated_clock}
\end{equation}
With this convention, the controlled side of the TCC condition is
\begin{equation}
 Y(N)<0,
 \label{eq:tcc_clock_controlled_side}
\end{equation}
and the TCC-violation time is the first hitting time
\begin{equation}
 T_{\rm TCC}
 \equiv
 \inf
 \left\{
 N\geq N_i:
 Y(N)\geq0
 \right\}.
 \label{eq:tcc_first_violation_time}
\end{equation}
For constant \(H\), this reduces to the familiar deterministic duration
\begin{equation}
 N_{\rm TCC}-N_i
 =
 \ln\!\left(
 \frac{\Lambda_i}{H}
 \right),
 \label{eq:tcc_constant_H_duration}
\end{equation}
and \(\Lambda_i=M_{\rm Pl}\) gives the usual \(\ln(M_{\rm Pl}/H)\) expression.

When \(H\) fluctuates or evolves together with other stochastic variables, \(Y\) is part of the path-dependent state.  One may therefore enlarge the Markov state from \(X\) to
\begin{equation}
 \widehat X
 =
 (X,Y),
 \label{eq:tcc_augmented_state}
\end{equation}
with augmented generator \(\widehat{\cal L}\). For example, when \(H=H(X,N)>0\) is a smooth state function, let \(\ell=\ln[H(X,N)/\Lambda_i]\), \(\mathrm dX^i=a^i\mathrm dN+\Sigma^i{}_a\mathrm dW^a\), and \(D^{ij}=\tfrac12\Sigma^i{}_a\Sigma^j{}_a\). It\^o's formula gives
\begin{equation}
 \mathrm dY=\bigl[1+(\partial_N+\mathcal L_N)\ell\bigr]\mathrm dN
 +\partial_i\ell\,\Sigma^i{}_a\mathrm dW^a.
 \label{eq:tcc_clock_ito_revision}
\end{equation}
Thus, in a local coordinate chart,
\begin{equation}
 \begin{aligned}
 \widehat{\mathcal L}_N={}&\mathcal L_N
 +\bigl[1+(\partial_N+\mathcal L_N)\ell\bigr]\partial_Y\\
 &+2D^{ij}\partial_j\ell\,\partial_i\partial_Y
 +D^{ij}\partial_i\ell\partial_j\ell\,\partial_Y^2 .
 \end{aligned}
 \label{eq:tcc_augmented_generator_revision}
\end{equation}
The clock shares the original noise; it is not an independent Brownian coordinate. For deterministic \(H(N)\), this reduces to \(\mathrm dY=(1-\epsilon_H)\mathrm dN\). If \(\Sigma_{\rm end}\), \(\Sigma_{\rm EFT}\), and \(Y=0\) denote respectively the completion, EFT-loss, and TCC-violation surfaces, then the completion probability
\begin{equation}
 h_{\rm comp}^{\rm TCC}(\widehat X)
 \equiv
 \mathbb P_{\widehat X}
 \left[
 T_{\rm end}
 <
 T_{\rm EFT}\wedge T_{\rm TCC}
 \right]
 \label{eq:tcc_completion_h_function}
\end{equation}
satisfies the corresponding backward first-passage equation
\begin{equation}
 \widehat{\cal L}
 h_{\rm comp}^{\rm TCC}
 =
 0
 \label{eq:tcc_completion_backward_equation}
\end{equation}
for the autonomous augmented process, or its explicitly time-dependent counterpart when the generator is nonautonomous. On an unbounded domain the event defines the minimal nonnegative solution; histories that never complete are not successes. The clock monitors the specified interval starting at \(N_i\) with fixed \(\Lambda_i\); a varying UV scale or a requirement over all possible initial times needs its own path event.

The boundary data distinguish success from the two competing failures:
\begin{equation}
 h_{\rm comp}^{\rm TCC}
 =
 1
 \quad
 \text{on }\Sigma_{\rm end}\cap\{Y<0\}\text{ before EFT loss},
 \label{eq:tcc_completion_success_boundary}
\end{equation}
whereas
\begin{equation}
 h_{\rm comp}^{\rm TCC}
 =
 0
 \quad
 \text{on }
 \Sigma_{\rm EFT}
 \text{ and on }
 Y=0
 \label{eq:tcc_completion_failure_boundaries}
\end{equation}
away from any coincident-boundary convention.  The resulting Doob transform, if one wishes to condition on successful completion, is generated by \(h_{\rm comp}^{\rm TCC}\), not by an autonomous EFT-survival function evaluated at a substituted horizon.

This formulation also allows a finite failure tolerance.  For an initial state \(\widehat X_i\), a probabilistic TCC requirement may be written as
\begin{equation}
 1-
 h_{\rm comp}^{\rm TCC}(\widehat X_i)
 \leq
 p_{\max},
 \label{eq:tcc_failure_tolerance}
\end{equation}
provided that the failure event on the left has been defined to include precisely the competing channels under consideration.  If EFT loss and TCC violation are to be reported separately, their individual first-passage probabilities should instead be retained rather than combined into one failure fraction.

The earlier state-dependent expression \(\tau_{\rm TCC}(\phi)\sim\ln[M_{\rm Pl}/H(\phi)]\) can still be useful as a local or deterministic estimate in a regime where \(H\) is effectively fixed and the completion time is specified independently.  Substituting that expression into \(h(\phi,\tau)\) and differentiating with Eq.~\eqref{eq:finite_horizon_chain_rule}, however, constructs only a composed scalar weight.  It does not in general impose the path event Eq.~\eqref{eq:tcc_competing_completion_probability}, and should therefore not be identified with a stochastic implementation of the TCC.

This competing-first-passage viewpoint is closely related to other recent uses of stochastic crossing probabilities to quantify quantum-gravity lifetimes.  In particular, Ref.~\cite{Anchordoqui:2026edd} combines a Distance-Conjecture tower with a generalized Higuchi condition to define a finite field-space loss wall and computes the fixed-confidence time over which stochastic diffusion keeps a prescribed fraction of inflationary branches on its healthy side.  That construction is not itself the TCC event above, but it provides an independent example in which quantum-gravity control data become a stochastic first-passage lifetime problem.

The conceptual distinction is therefore
\begin{equation}
 \text{prescribed duration}
 \;\neq\;
 \text{EFT first-exit time}
 \;\neq\;
 \text{completion before TCC violation}.
 \label{eq:tcc_event_distinction_summary}
\end{equation}
All three can be treated within the same backward-probability framework, but they require different events, state variables, and boundary data.

\bibliography{refs}
\end{document}